\pdfoutput=1

\documentclass[%
 reprint,
 superscriptaddress,
 nobibnotes,
 amsmath,amssymb,
 aps,
 floatfix,
 showkeys,
 longbibliography,
]{revtex4-2}

\usepackage{graphicx}
\usepackage{dcolumn}
\usepackage{bm}
\usepackage{chemformula}
\usepackage[separate-uncertainty=true,separate-uncertainty,multi-part-units=single]{siunitx}
\usepackage{hyperref}
\hypersetup{colorlinks=true,allcolors=magenta}

\usepackage{subcaption}
\usepackage{xspace}
\usepackage{placeins}

\usepackage{textgreek}

\newcommand\stringCommand[1]{
  \expandafter\newcommand\csname #1\endcsname[0]}

\stringCommand{ptc}{K$_2$CO$_3$\xspace}
\stringCommand{func}{CS1000a$_{\text{f}}$\xspace}
\stringCommand{nofunc}{CS1000a$_{\text{nf}}$\xspace}
\stringCommand{carbon}{CO$_2$\xspace}
\stringCommand{water}{H$_2$O\xspace}
\stringCommand{Kptc}{K$_{\text{K}_2\text{CO}_3}^{+}$\xspace}
\stringCommand{Cptc}{C$_{\text{K}_2\text{CO}_3}^{4+}$\xspace}
\stringCommand{Optc}{O$_{\text{K}_2\text{CO}_3}^{2-}$\xspace}
\stringCommand{Hwater}{H$_{\text{H}_2\text{O}}$\xspace}
\stringCommand{Owater}{O$_{\text{H}_2\text{O}}$\xspace}
\stringCommand{Ccarbon}{C$_{\text{CO}_2}$\xspace}
\stringCommand{Ocarbon}{O$_{\text{CO}_2}$\xspace}

\stringCommand{HoApth}{HoA/forced_scale/lin}
\stringCommand{rdfpth}{rdf/adsorbates_swap/show_loading}

\begin{document}

\title{Enhancing Direct Air Capture through Potassium Carbonate Doping of Activated Carbons}

\author{N. van Dongen}
    \affiliation{Materials Simulation \& Modelling, Department of Applied Physics and Science Education, Eindhoven University of Technology, 5600 MB, Eindhoven, The Netherlands}
\author{A. J. F. van Hoof}
    \affiliation{Carbyon B.V., High Tech Campus 32, 5656 AE, Eindhoven, The Netherlands}
\author{S. Calero}
    \affiliation{Materials Simulation \& Modelling, Department of Applied Physics, Eindhoven University of Technology, 5600 MB, Eindhoven, The Netherlands}
    \affiliation{Eindhoven Institute for Renewable Energy Systems (EIRES), Eindhoven University of Technology, Eindhoven 5600 MB, The Netherlands}
\author{J. M. Vicent-Luna}
    \email[Corresponding author: ]{j.vicent.luna@tue.nl}
    \affiliation{Materials Simulation \& Modelling, Department of Applied Physics, Eindhoven University of Technology, 5600 MB, Eindhoven, The Netherlands}
    \affiliation{Eindhoven Institute for Renewable Energy Systems (EIRES), Eindhoven University of Technology, Eindhoven 5600 MB, The Netherlands}


\begin{abstract}

Direct air capture of carbon dioxide (\carbon) is one of the most promising strategies to mitigate rising atmospheric \carbon levels. Among various techniques, adsorption using porous materials is a viable method for extracting \carbon from air, even under humid conditions. However, identifying optimal adsorbent materials remains a significant challenge. Moreover, the performance of existing materials can be improved by doping with active species that boost gas capture, a relatively unexplored field. In this study, we perform atomistic simulations to investigate the adsorption, structural, and energetic properties of \carbon and water in realistic models of activated carbons. We first analyze the impact of explicitly considering surfaces containing functional groups, which aims to imitate the chemical environment of experimental samples. Additionally, we introduce potassium carbonate within the pores of the adsorbent to evaluate its effect on \carbon and water adsorption. Our results demonstrate that both functional groups and potassium carbonate enhance adsorption, primarily by shifting the adsorption onset pressures to lower values. Specifically, potassium carbonate clusters act as extra adsorption sites for \carbon and water, facilitating the nucleation of water molecules and promoting the formation of a hydrogen bond network within the activated carbon pores.

\end{abstract}

\keywords{Carbon capture, activated carbons, potassium carbonate, Monte Carlo simulations}

\maketitle


\section{Introduction}\label{sec:intro}

The ever increasing levels of carbon dioxide in the atmosphere of our planet is one of the largest environmental concerns of our time.\cite{Dowell2017} \carbon is one of the major contributing greenhouse gasses, and its rise is strongly correlated to global warming and anthropogenic climate change.\cite{Raupach2007} This has led to, and is increasingly leading to, many adverse effects such as: increases in droughts, desertification, melting of icecaps and permafrost, rising of sea levels, increase in floods, increasing probabilities of extreme weather events, and destruction of ecosystems.\cite{Yu2008, Yang2016, LU2009} However, our dependency on fossil-fuels and other processes emitting CO$_2$ as a byproduct are not feasible to be replaced easily. Emissions are expected to continue to increase due to growing demands, while alternatives cannot keep up or are economically infeasible.\cite{LU2009, conti2016international, SanzPerez2016} Current trends show that atmospheric concentration is heading to an average of 420 ppm in 2024, which is up from about. 280 ppm in pre-industrial times, and is expected to exceed 550 ppm by 2050.\cite{Lan2023, Raupach2007} To be able to reduce our emissions to acceptable levels according to climate goals, `negative emissions' from capturing and storing CO$_2$ from the atmosphere are required on top of proposed reductions.\cite{SanzPerez2016, ipcc2013climate} Among various methods of carbon capture and storage technologies being worked on for this purpose, our focus is on Direct Air Capture (DAC).

A promising method for DAC is the use of solid regenerable sorbents made of alkali-metal carbonates. This material can capture \carbon by a reaction with \ptc in the presence of water at room temperature, creating potassium bicarbonate:
\[
    \text{K$_2$CO$_3$} + \text{H$_2$O} + \text{CO$_2$} \rightleftarrows 2\text{HKCO$_3$}.
\] A complete reverse reaction can be achieved at temperatures of $150$ $^\circ$C, which is relatively low.\cite{Meis2013} Although the actual process is slightly more complex due to the formation of hydrates with \ptc, the adsorbing and heating process to desorb forms the basis of carbon capture. This process can match the energy efficiency of current commercial liquid amine processes in flue gas treatment. Compared to them, solid \ptc is practically beneficial, is thermally stable, can capture CO$_2$ at atmospheric concentrations, and is more environmentally friendly in use and disposal.\cite{Zhao2011, Wang2011, Wang2014, Meis2013, Masoud2021} Other solid sorbents that utilize amine functionalities exist; however, the anticipated limited lifetime of these materials makes them less interesting.

The \ptc must be deposited on some porous support structure to increase the surface area of the reaction, to prevent attrition, and to improve regeneration.\cite{Yang2016, Masoud2021} The choice of this support is very influential; the pore structure, pore size distribution (PSD), and surface groups are all important to determine the capture capacity and kinetics.\cite{Wang2014} These properties are important to create a fast CO$_2$ adsorption technology in which the sorbent can be saturated quickly, which is desired for large-scale industrial applications. Activated carbons have become a popular material type choice for this owing to the tunability of both these aspects. Moreover, carbons are cheap because they are created from organic waste material, are thermally stable, and have a high surface area--all of which are desirable properties.\cite{Wang2011} Finding a perfect carbon is difficult for a variety of reasons. By far, one of the greatest challenges is that of excessive adsorption of water over CO$_2$ in these doped carbons. Potassium carbonate is naturally hygroscopic, and surface functional groups additionally increase water adsorption. This is a major hindrance in the viability of a sorbent; Desorbing excess water requires significant energy, which is detrimental to the technology commercially and reduces the net amount of CO$_2$ captured considering what is exhausted generating energy for the process. Lowering the H$_2$O:CO$_2$ ratio is therefore the goal of a sorbent and process combination, for which adsorption and its kinetics must be known to improve this.

The addition of \ptc on activated carbons has been proposed for \carbon capture and separation applications.\cite{QUEREJETA2019208, Ali2017, Kundu2024, Korah2025, GUO2021100011} However, the physical and chemical properties of all these sorbents and the external conditions of temperature and relative humidity affect the adsorption, and its mechanisms are not well understood at the molecular scale. Molecular simulations allow for exploring these types of matter and are an increasingly more prevalent and valuable tool for screening and understanding carbon capture.\cite{Akinola2022, Wadi2023} Relevant to the question at hand, this includes research on improving the selectivity of \carbon over water (and other gases) in many classes of porous materials. \cite{Liu2012, DiBiase2015, Jiang2022} Doping, modifying the surfaces, or exchanging ions in the case of zeolites have seen good success with these techniques.\cite{Babarao2012, Abdelrasoul2017, AkbariBeni2020, Zhang2023, Sadeghi2024} For using \ptc, research explaining its capture mechanisms is proceeding, but, to our best knowledge, none directly focus on its direct inclusion in carbons to study adsorption in realistic environments for carbon capture.\cite{Liu2017, Cai2023} As such, in this paper, we present two methods of including \ptc in carbon models and explore the effects on adsorption and its mechanisms of \carbon and water separately using Monte Carlo simulations. For carbon, a model of the purely microporous activated carbon CS1000a is used. This specific carbon was chosen due to the accuracy of the model and the availability of two versions, which allows for an analysis between the presence and absence of surface functionalized groups.\cite{MaderoCastro2022}

\section{Methodology}\label{sec:methods}

\subsection{Monte Carlo simulation}

We used classical molecular simulations to assess the adsorption properties of \carbon and water in the activated carbon CS1000a, dependent on the presence of surface functionalization and potassium carbonate. For this purpose, Monte Carlo simulations in the grand canonical ensemble (GCMC) are used to compute single adsorbate isotherms and heats of adsorption at various loadings.\cite{Dubbeldam2013} Specifically, we employ configurational bias Monte Carlo using the RASPA software package.\cite{Dubbeldam2016RASPA} Unless otherwise specified, all components have translation moves; and if they are also molecules, they have rotation moves. The adsorbates additionally have moves for their complete reinsertion and moves to either add or remove them from the system. Moves are always chosen with equal probability.

Equilibration and sampling proved challenging because of the complexity of adsorbate loading in the system. This is mainly due to water, which is known to form clusters at active sites.\cite{Brennan2001, Monge2001} This occurs in simulation, which causes a wide variance in the cycles needed to equilibrate and properly average over cases with potentially slow and large fluctuations in the sizes of these clusters. Therefore, an approach was used where we monitored the loading status for each simulation after a short initial equilibration run of $2\times10^{4}$ cycles. The cycles were run repeatedly in blocks in the range of $10^{5}-10^{6}$ cycles, depending on the equilibration of the system in each operating condition. Subsequently, the gathered adsorption isotherms were fitted to a multisite version of the Sips model using the RUPTURA software for interpolation and interpretation.\cite{Sips1948, Sharma2023RUPTURA} Further information on this, including fitting parameters for every isotherm, is included in the supplementary materials (\ref{supp:fitting}).

For the implementation of the force fields, the two models of CS1000a with differing surface chemistry were implemented as they were by Madero-Castro et al.\cite{MaderoCastro2022} The structures were originally obtained using a hybrid reverse Monte Carlo (HRMC) technique by Jain et al., based on the experimental structure of an activated carbon from pyrolyzing sucrose at $1000$ $^{\circ}$C.\cite{Jain2006, Peng2018} From this method, no oxygen-containing surface groups are present that are responsible for the partial hydrophilicity of the realistic activated carbon.\cite{Billemont2013} To this end, a version of the structure was created using the method by Peng et al. by outfitting it with carboxylic acid (-COOH) and alcohol groups (-OH) to closely match experimental $\text{O/C}$ and $\text{H/C}$ ratios of $0.0087$ and $0.091$, respectively.\cite{Peng2017, Peng2020, Jain2006} This version with functional groups, henceforth referred to as \func, is compared to the version with no surface functionalization, from now on referred to as \nofunc. In modeling, both structures are considered rigid frameworks and form cubic unit cells of systems with edges of $50$ \r{A}.

The water model used is the extended simple point charge (SPC/E) model by Berendsen et al.\cite{Berendsen1987} It has been used in a previous study on CS1000a for its ability to predict the structure and thermodynamics of bulk liquid water at ambient temperatures well.\cite{Billemont2013} We used the force field for \carbon reported by Garc\'ia-S\'anchez et al.,\cite{Garcia2009} and for potassium carbonate, we used a model by Jo and Banerjee, based on the CVFF force field.\cite{Jo2015} It describes the \ptc as a carbonate CO$_3^{2-}$ polyatomic ion and two separate K$^+$ ions. This force field was chosen since it was created to be applied in the modeling of interactions with carbon nanotubes\cite{Jo2015} We verified its implementation and use for more general applicability by computing radial distribution functions (RDF) at 300 K. We compared the RDFs created by the model with the RDF results of Ding et al., which proved agreeable\cite{Ding2018} These RDFs and more information on this verification have been provided in the supplementary materials \ref{supp:rdf melt}. For nonbonded interactions, all models use Lennard-Jones potentials to represent the van der Waals interactions and Coulombic potentials with partial charges centered on the atoms to represent the electrostatic interactions. Lennard-Jones cross-interaction parameters are determined with Lorentz-Berthelot mixing rules.\cite{Boda2008} These interactions have a cutoff of 12 \r{A}, and do not use any form of tail corrections. For longer-range electrostatic interactions, Ewald summations are used.\cite{Ewald1921}to assess behavior and stability.

\subsection{Modelling CS1000a with potassium carbonate}
For the studied mechanism of carbon capture, potassium carbonate is needed in the pores. As far as we are aware, there is no precedent in the literature on doping activated carbons with potassium carbonate in computational studies. Hence, we explore two different methods of adding \ptc within the pores of ACs, and we explore the effects on adsorption and its mechanisms, both for \carbon and for water separately. 

The easiest method to add \ptc is by simply arbitrarily adding several molecules of it into the pores of the model. However, this method does not result in configurations that one would expect when considering experimental considerations. That is, during impregnation in porous structures, the potassium carbonate deposited on the surfaces can be left in the form of crystal clusters.\cite{Cai2020, Masoud2022} As such, it is of interest to create structures with small clusters of \ptc and compare the results with those when randomly adding molecules. Furthermore, considering the effects of the quantity of \ptc is also insightful. To this end, starting from each structure, we prepared four new initial configurations: 5 randomly inserted \ptc, 10 randomly inserted \ptc, a single cluster made up of 5 \ptc, and two clusters with 5 \ptc. For reference, the mass percentages of \ptc are summarized in table \ref{tab:ptc mass percent}. 

First, we generated a framework with randomly distributed \ptc where the CO$_3^{2-}$ and K$^+$ ions were first arbitrarily inserted in a 1:2 ratio. To create these molecules, a MC simulation was run, where both components were allowed to keep their regular set of MC moves, but K$^+$ was additionally allowed to undergo random translations. This ensured a quick formation of molecules and an opportunity to get into equilibrium with its environment. Alternatively, we inserted the \ptc ions forming small clusters. To create a realistic structure of small clusters, the AMS driver program from SCM was used.\cite{AMSdriver} First, a cluster of 5 molecules was selected from the lattice of a potassium carbonate crystal. This cluster was then optimized under vacuum conditions using the Machine Learning potential M3GNet.\cite{Chen2022}

Subsequently, the optimized clusters were inserted into the pores and considered to be completely rigid. Using only non-bonded interactions as defined by the \ptc model, they were relaxed with a 5000 cycle MC simulation. During this, the clusters were allowed to translate and rotate, enabling them to settle next to the walls. For the frameworks in which two clusters were inserted, the clusters were ensured to end up far apart from each other to minimize their interactions. Hereafter, the rigid bonds from the clusters were removed, switching definitions to the initially defined model of \ptc. The local environment of the carbon and a change in the model have definite effects on the shape of the just decoupled clusters. Therefore, another round of equilibration is required for them to settle properly. For this, MC simulations of $5\times10^5$ steps were performed using the normal set of MC moves. 

\begin{table}[h!]
    \caption{Mass percentages of \ptc in the simulated structures.}
    \label{tab:ptc mass percent}
    \begin{tabular}{lll}
        & \func  & \nofunc  \\ \cline{2-3} 
        \multicolumn{1}{l|}{5 \ptc} & 1.23\% & \multicolumn{1}{l|}{1.25\%} \\
        \multicolumn{1}{l|}{10 \ptc} & 2.43\% & \multicolumn{1}{l|}{2.46\%} \\ \cline{2-3} 
    \end{tabular}
\end{table}

\begin{figure}[!ht]
\centering
    \begin{subfigure}[b]{\columnwidth}
        \includegraphics[width=0.8\textwidth]{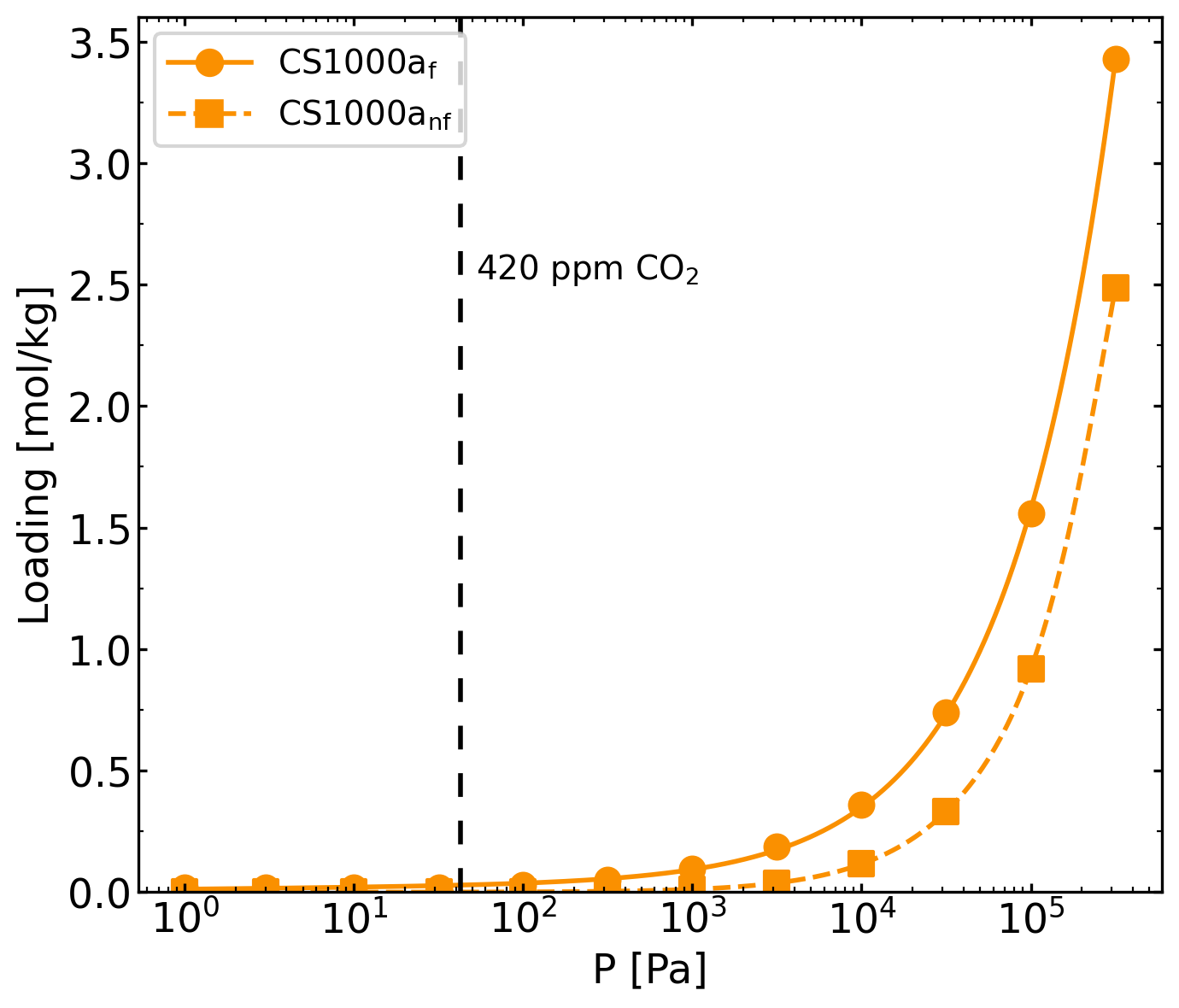}
        \caption{}\label{fig:CO2 300.15}
    \end{subfigure}
    
    \vspace{1em} 
    
    \begin{subfigure}[b]{\columnwidth}
        \includegraphics[width=0.8\textwidth]{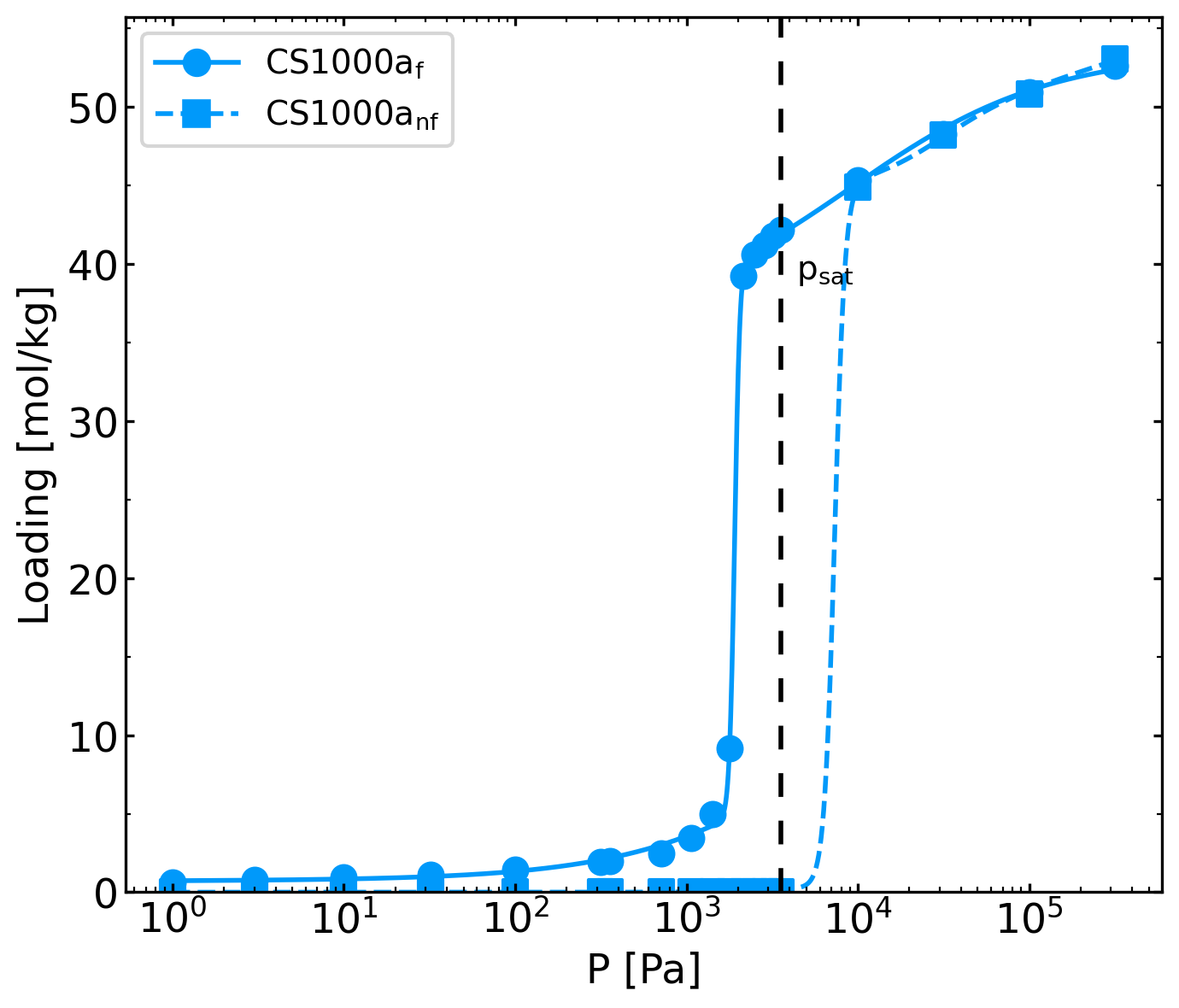}
        \caption{}\label{fig:H2O 300.15}
    \end{subfigure}
    
    \caption{Comparisons of computed adsorption isotherms of CO$_2$ (a) and \water (b) in \func (solid lines) and \nofunc (dashed lines) at $300.15$ K. Lines are fits through the data points obtained using RUPTURA.\cite{Sharma2023RUPTURA} The vertical line in (a) indicates the partial pressure of \carbon equivalent to 420 ppm \carbon (at 1 atm), its current approximate atmospheric concentration.\cite{Lan2023} The vertical line in (b) represents the saturation pressure at the simulation temperature.}
    \label{fig:iso no K2CO3 300.15}
\end{figure}

\section{Results and Discussion}\label{sec:results}

\begin{figure*}[!ht]
\centering
    \begin{subfigure}[b]{0.25\textwidth}
        \includegraphics[width=\textwidth]{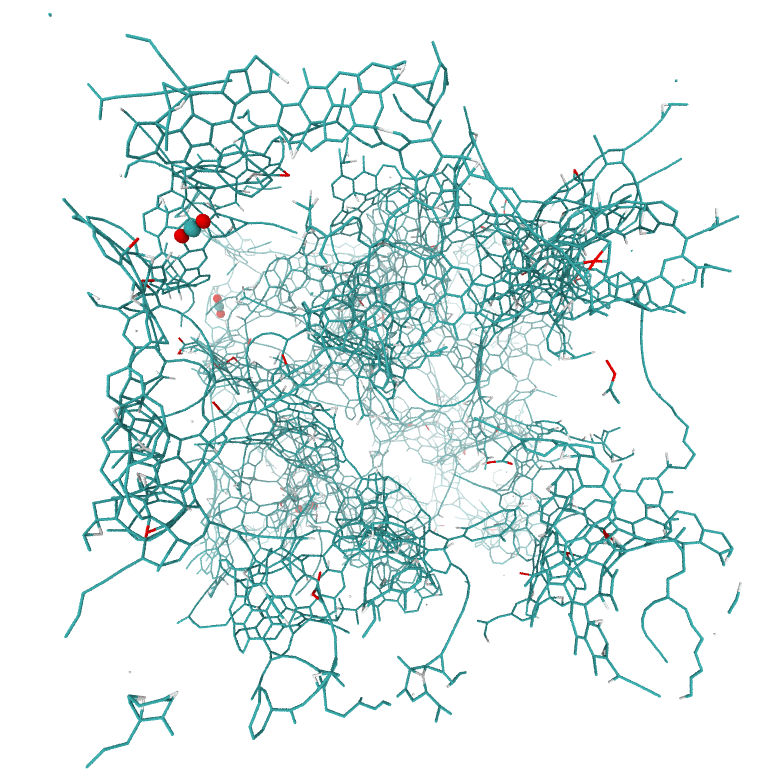}
        \caption{}\label{fig:CO2 f 316}
    \end{subfigure}
    \begin{subfigure}[b]{0.25\textwidth}
        \includegraphics[width=\textwidth]{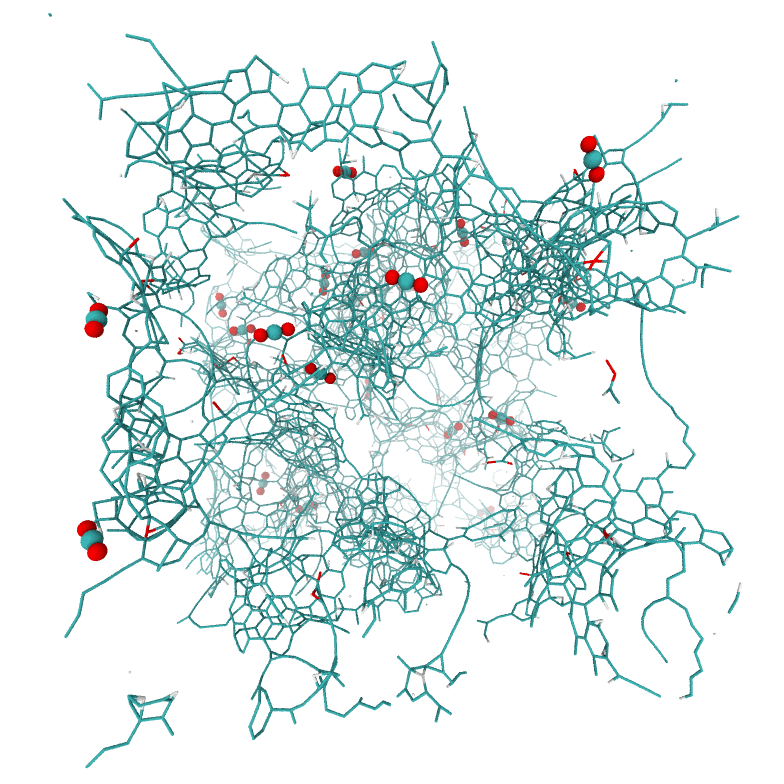}
        \caption{}\label{fig:CO2 f 10000}
    \end{subfigure}
    \begin{subfigure}[b]{0.25\textwidth}
        \includegraphics[width=\textwidth]{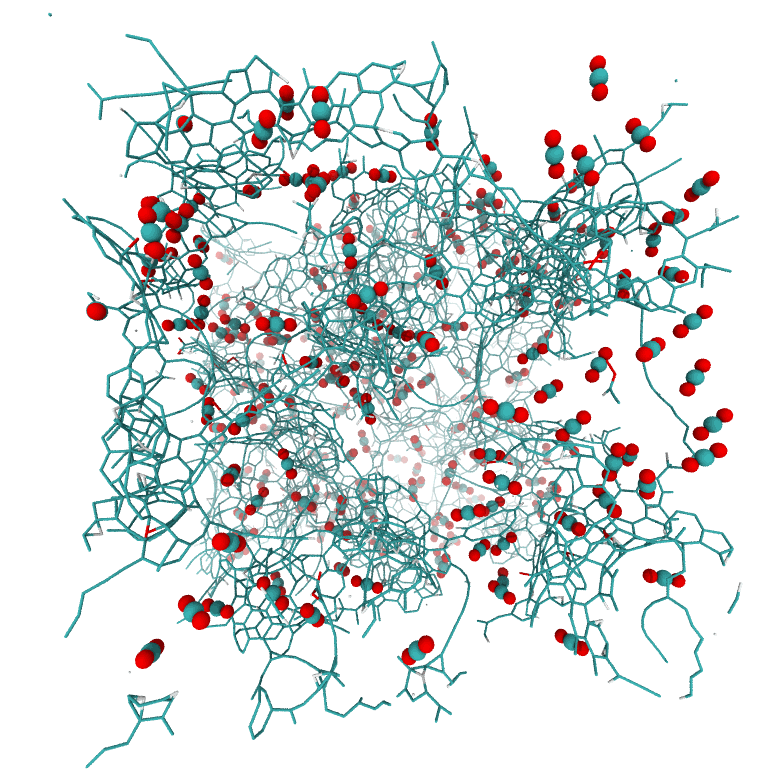}
        \caption{}\label{fig:CO2 f 316228}
    \end{subfigure}
    \begin{subfigure}[b]{0.25\textwidth}
        \includegraphics[width=\textwidth]{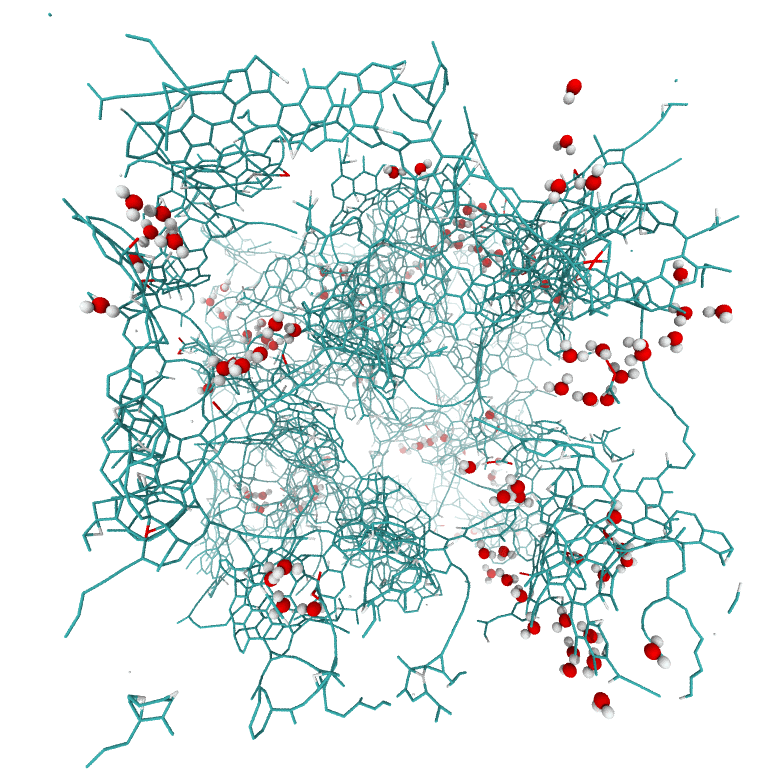}
        \caption{}\label{fig:H2O f 357}
    \end{subfigure}
    \begin{subfigure}[b]{0.25\textwidth}
        \includegraphics[width=\textwidth]{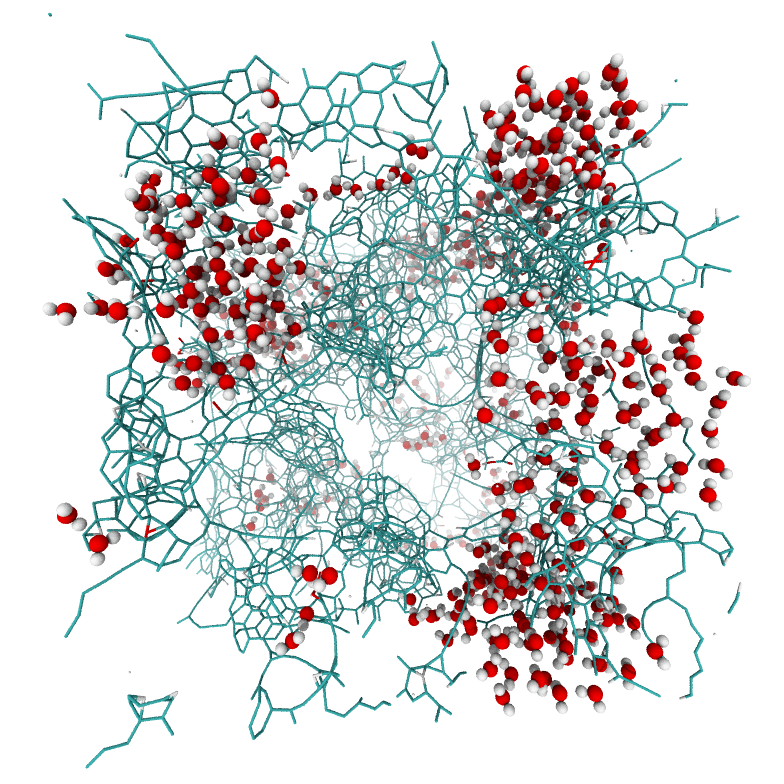}
        \caption{}\label{fig:H2O f 1784}
    \end{subfigure}
    \begin{subfigure}[b]{0.25\textwidth}
        \includegraphics[width=\textwidth]{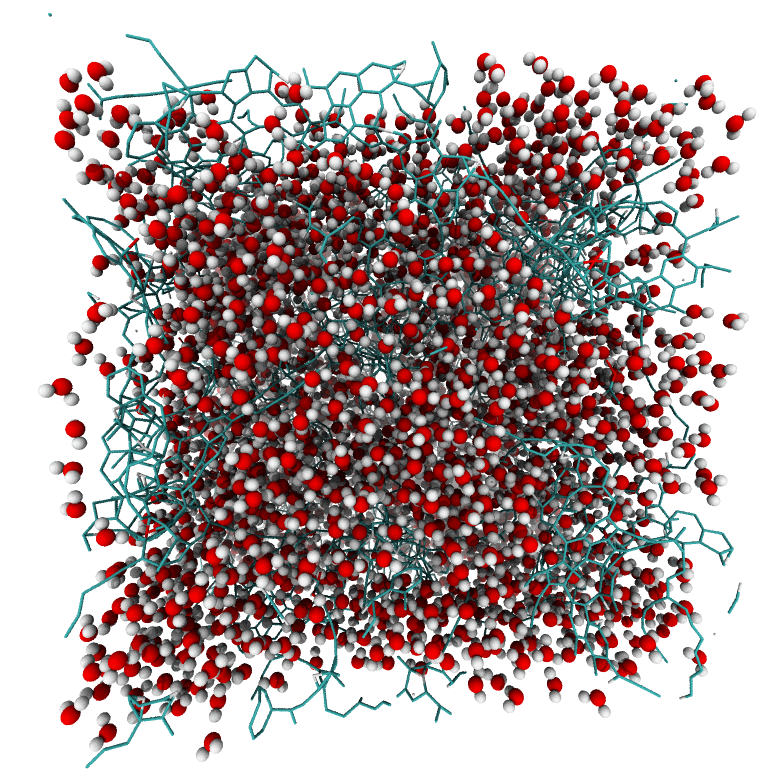}
        \caption{}\label{fig:H2O f 316228}
    \end{subfigure}
    \caption{Snapshots showcasing adsorption in \func at $300.15$ K under various partial pressure conditions. (a-c) Adsorption of \carbon at $316$, $10000$, and $316228$ Pa respectively. (e-f) Adsorption of \water at $357$, $1784$, and $316228$ Pa respectively. These conditions for water correspond to $10\%$ relative humidity, $50\%$ relative humidity, and saturation.}\label{fig:renders func no K2CO3}
\end{figure*}

\subsection{Effects of surface chemistry}
Before adding the effects of potassium carbonate, it is worthwhile to analyze the adsorption in pristine carbon models. This is to understand and review the mechanisms of water and \carbon adsorption by comparing their dependence on the surface functionality without added species. Figure \ref{fig:iso no K2CO3 300.15} shows a comparison of the computed adsorption isotherms on \func and \nofunc for \carbon (a) or water (b) at 300.15 K. Here, symbols indicate the data obtained from the MC simulations, while the lines are fits through these data created using RUPTURA.\cite{Sharma2023RUPTURA} For \carbon, both structures show the same behavior where adsorption increases continuously over the computed pressure range while never reaching saturation. Evidently, the loading in \func is significantly greater than \nofunc, specially at carbon capture working conditions. Taking as reference $420$ ppm atmospheric \carbon, at atmospheric pressure this equates to partial pressure conditions of $42.55$ Pa.\cite{Lan2023} Using the fitted functions for interpolation, \func ($2.937\times10^{-2}$ mol/kg) shows about a two-order-of-magnitude increase in adsorption over \nofunc ($5.251\times10^{-4}$ mol/kg). For reference, this pressure has been marked in Fig. \ref{fig:CO2 300.15} with a vertical line. However, the concentration of adsorbed \carbon under these conditions is still too low to be used in practical applications, and this emphasizes the need to combine activated carbons with sorbents such as potassium carbonate to enhance \carbon capture. 

The results clearly indicate that the functionalized groups create more favorable adsorption conditions. For visualization of adsorption behavior, renderings were made using VMD software at different stages in the adsorption process.\cite{HUMP96, STON1998} Figures \ref{fig:renders func no K2CO3}a-c display representative snapshots of \carbon adsorption in \func at low, medium and high pressure values, respectively. There is relatively little adsorption compared to water (Figures \ref{fig:renders func no K2CO3}d-f), and the molecules appear to be randomly distributed. The effects of the functionalized groups are most clear when compared with the renders of \nofunc, which have been provided in Figures \ref{fig:renders nofunc no K2CO3}a-c. At matching pressures, these carbons show considerably less \carbon adsorption.

\begin{figure}[!ht]
\centering
    \begin{subfigure}[b]{\columnwidth}
        \includegraphics[width=0.8\textwidth]{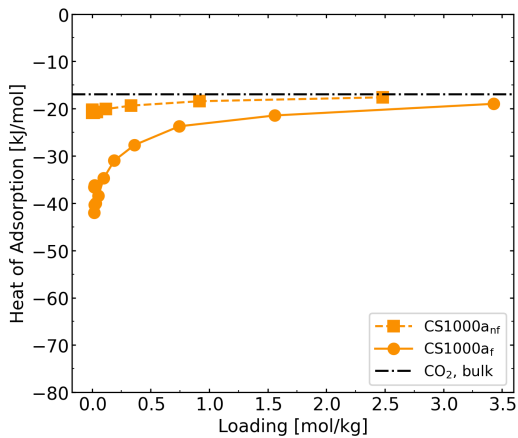}
        \caption{}\label{fig:hoa CO2 300.15}
    \end{subfigure}
    
    \vspace{1em} 

    \begin{subfigure}[b]{\columnwidth}
        \includegraphics[width=0.8\textwidth]{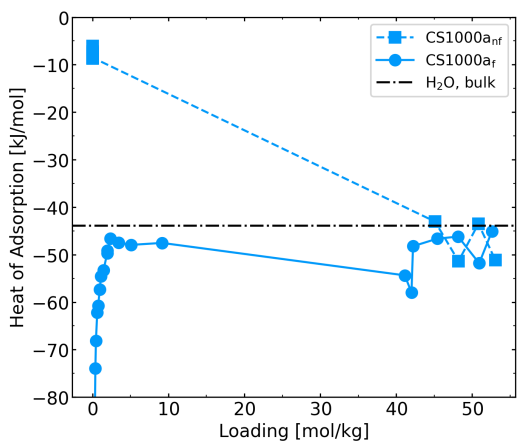}
        \caption{}\label{fig:hoa H2O 300.15}
    \end{subfigure}
    
    \caption{Comparisons of computed heats of adsorption of CO$_2$ (a) and \water (b) in \func (solid lines) and \nofunc (dashed lines) at $300.15$ K. The horizontal dash-dotted lines represent the heats of vaporization of the respective components.}
    \label{fig:HoA no K2CO3 300.15}
\end{figure}
For water, the computed isotherms in Figure \ref{fig:H2O 300.15} differ considerably between \func and \nofunc. Their shapes and differences can be understood through the generally accepted water filling mechanism in microporous carbons. 
Starting at low pressure, water adsorbs at the functional groups due to their polarity. Subsequently, water starts adsorbing on the already adsorbed water through hydrogen bonding. This forms clusters that grow with increasing pressure, filling up the pores. After this, the loading increases continuously with increasing pressure.\cite{Brennan2001, Monge2001, Billemont2013} These mechanisms are very clearly visualized in the renders of \func going through Figures \ref{fig:renders func no K2CO3}d-f. A few molecules adsorbed at $10\%$ relative humidity (RH), as described, grow large clusters concentrated at $50\%$ RH. At the highest simulated RH conditions, the structure is fully saturated.

Without functional groups, the adsorbent is completely hydrophobic, so water has no place to form the initial hydrogen bonds to start nucleating. Due to the pore size, no nucleation can start either.\cite{Brennan2001, Monge2001, Billemont2013} This results in no adsorption of water until well above the saturation pressure for water, the pressure of which has been indicated for the used simulation temperature in the Fig. \ref{fig:H2O 300.15} with a vertical line. After some point, the pores practically instantly fill, and the loading becomes equal to that of \func. This lack of adsorption into a jump is also visible in renders of the structure in Figures \ref{fig:renders nofunc no K2CO3}d-f. For carbon capture, the relevant range of conditions is up to 100\% RH as this constitutes the conditions where water remains as vapour in the air. The adsorption of water below its vapour saturation pressure in \nofunc is negligible, the amount of water captured by \func is significant under the same conditions. An analogous study was presented by Billemont et al.\cite{Billemont2013} Though they use different operating conditions and a different molecular model for \carbon, our results and conclusions align with their work. In this context, this paper provides a more direct discussion with regard to DAC and will be used as a baseline for when \ptc is included.

Observations of an increase in adsorption strength due to functionalized groups are confirmed by the heat of adsorption curves depicted in Figure \ref{fig:HoA no K2CO3 300.15}. Here, the horizontal dash-dotted lines represent the heat of vaporization of the respective components. At 300.15 K these are 17.0 kJ/mol for \carbon and 43.9 kJ/mol for water.\cite{Stephenson1987, etoolbox2010h2o_vap} At low coverage, both \carbon and water show heats of adsorption of a greater magnitude compared to \nofunc. The trends for \nofunc differ between the components. For \carbon, due to van der Waals interactions with the walls, slightly more favorable heats of adsorption are observed for adsorption. For water, due to the hydrophobicity of the surface of \nofunc, it instead stays above -10 kJ/mol until a sudden jump toward the heat of vaporization at high loading. The other trends show a curve towards their respective heats of vaporization. It is worth mentioning that the noise observed in the heats of adsorption of water at high loading is expected because of the difficulty of inserting additional molecules within the hydrogen bond network created by the adsorbed water molecules. However, the results clearly indicate that, on average, the heats of adsorption of water in the large pores of CS1000a converge to the vaporization enthalpy of water.\cite{Peng2019}. For additional validation of this finding, a qualitative comparison can be made with the heats of adsorption of methanol in CS1000a, a polar molecule that behaves similarly to water. Madero et al. \cite{MaderoCastro2022} compared the experimental and computed heats of adsorption of methanol in CS1000a, showing how similar to water the heat of adsorption converges to the enthalpy of vaporization of this molecule.

The effects of temperature were also investigated for the pristine carbon models at 288.15 and 313.15 K. For reference, results for the isotherms and heats of adsorption are available in Figures \ref{fig:iso no K2CO3 allT} and \ref{fig:HoA no K2CO3 allT} respectively in the supplementary materials. As expected, we observe that the isotherms shift properly when the temperature is modified, and the heats of adsorption show mostly overlapping trends, tending toward the heats of vaporization.

\subsection{Effects of potassium carbonate}

\begin{figure}[!ht]
    \centering
    \includegraphics[width=\columnwidth]{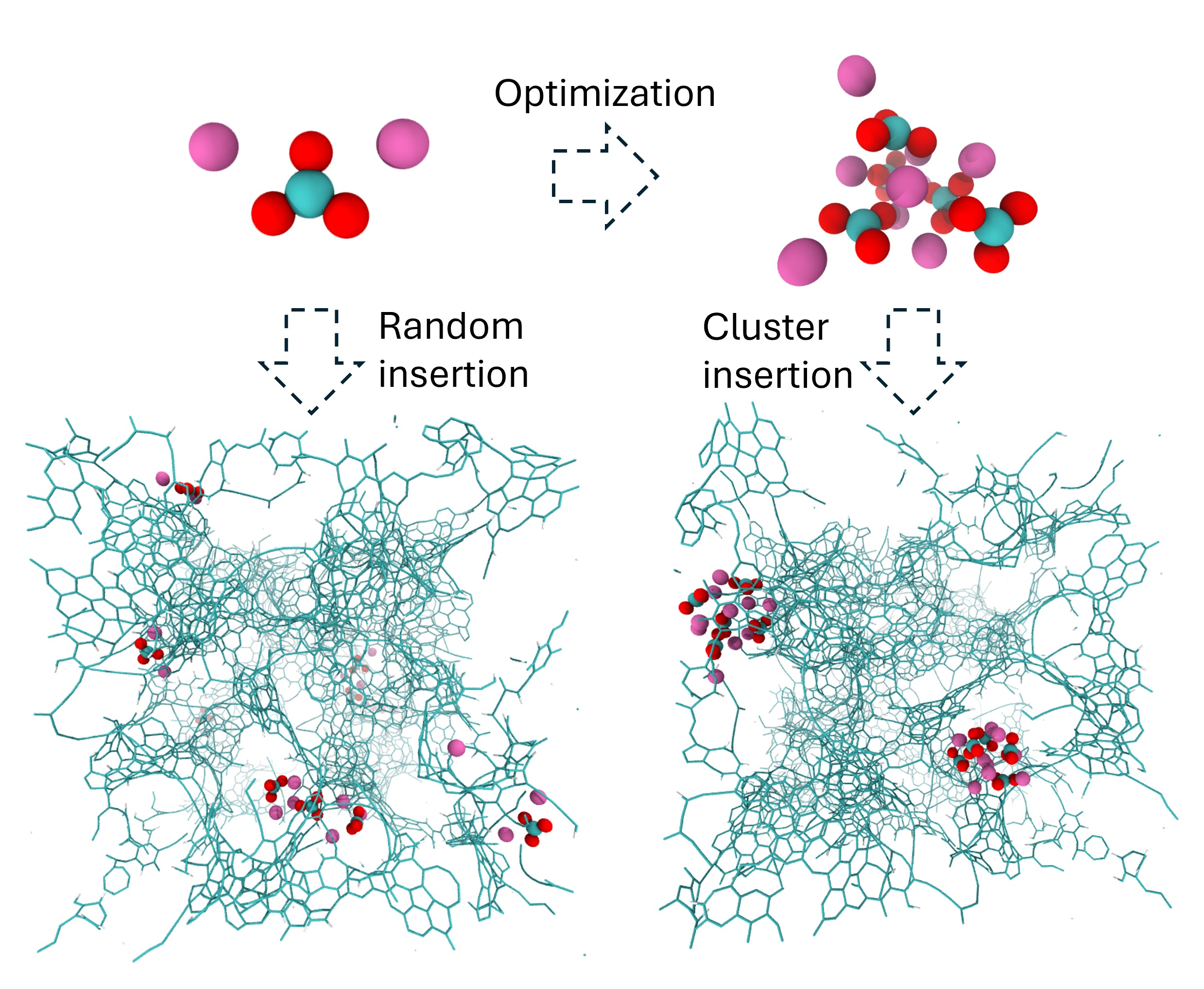}
    \caption{Schematic explanation of the two routes to create structures containing \ptc. In the first case, the \ptc units are randomly added to an empty structure. In the second case, previously optimized small clusters of $5$ \ptc were initially added to the pores. Then, the ions of \ptc are relaxed within the structure to accommodate the clusters next to a wall correctly.}
    \label{fig:k2co3 creation}
\end{figure}

\begin{figure*}[!ht]
\centering
    \begin{subfigure}[b]{0.25\textwidth}
        \includegraphics[width=\textwidth]{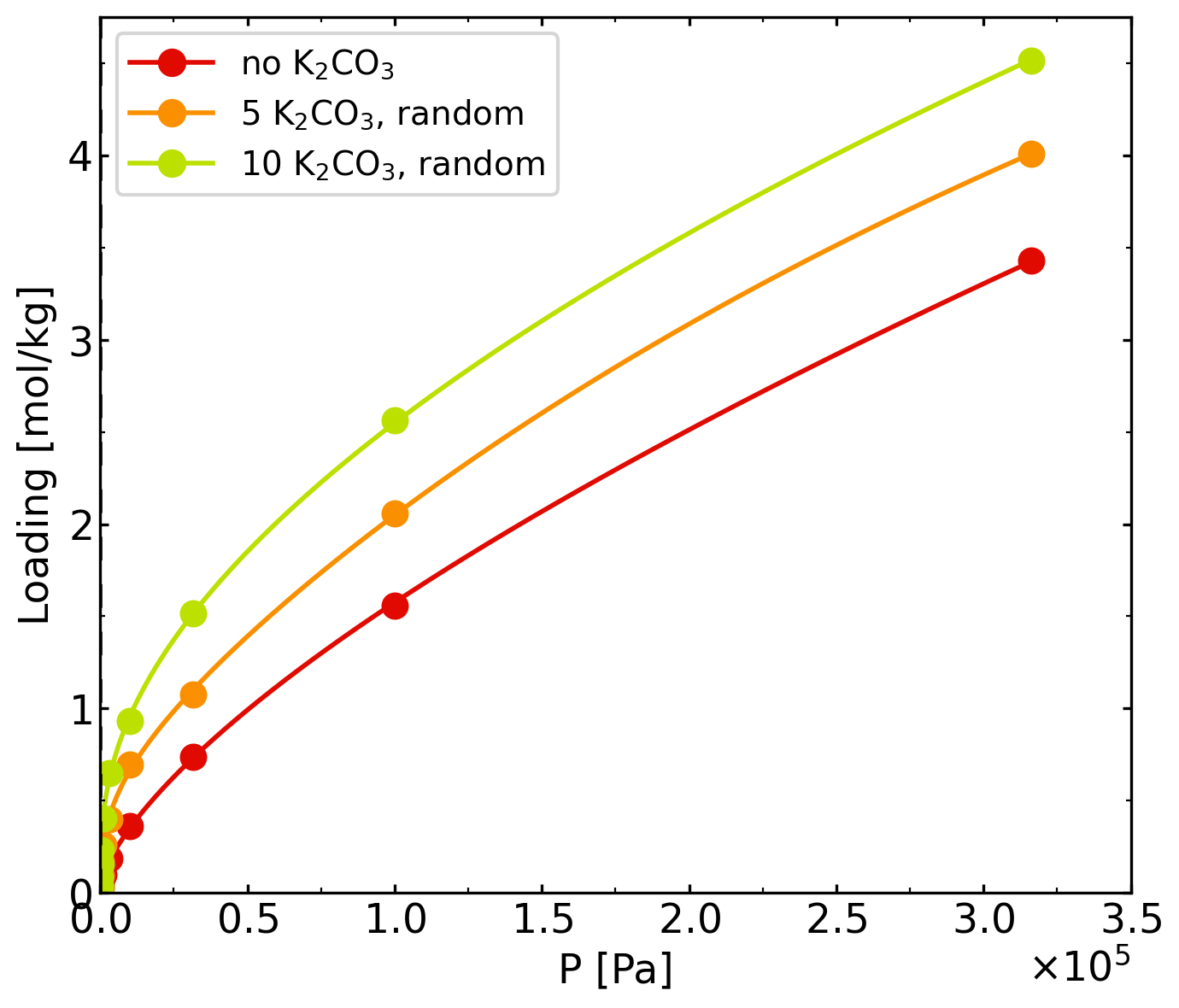}
        \caption{}\label{fig:CO2 f random}
    \end{subfigure}
    \begin{subfigure}[b]{0.25\textwidth}
        \includegraphics[width=\textwidth]{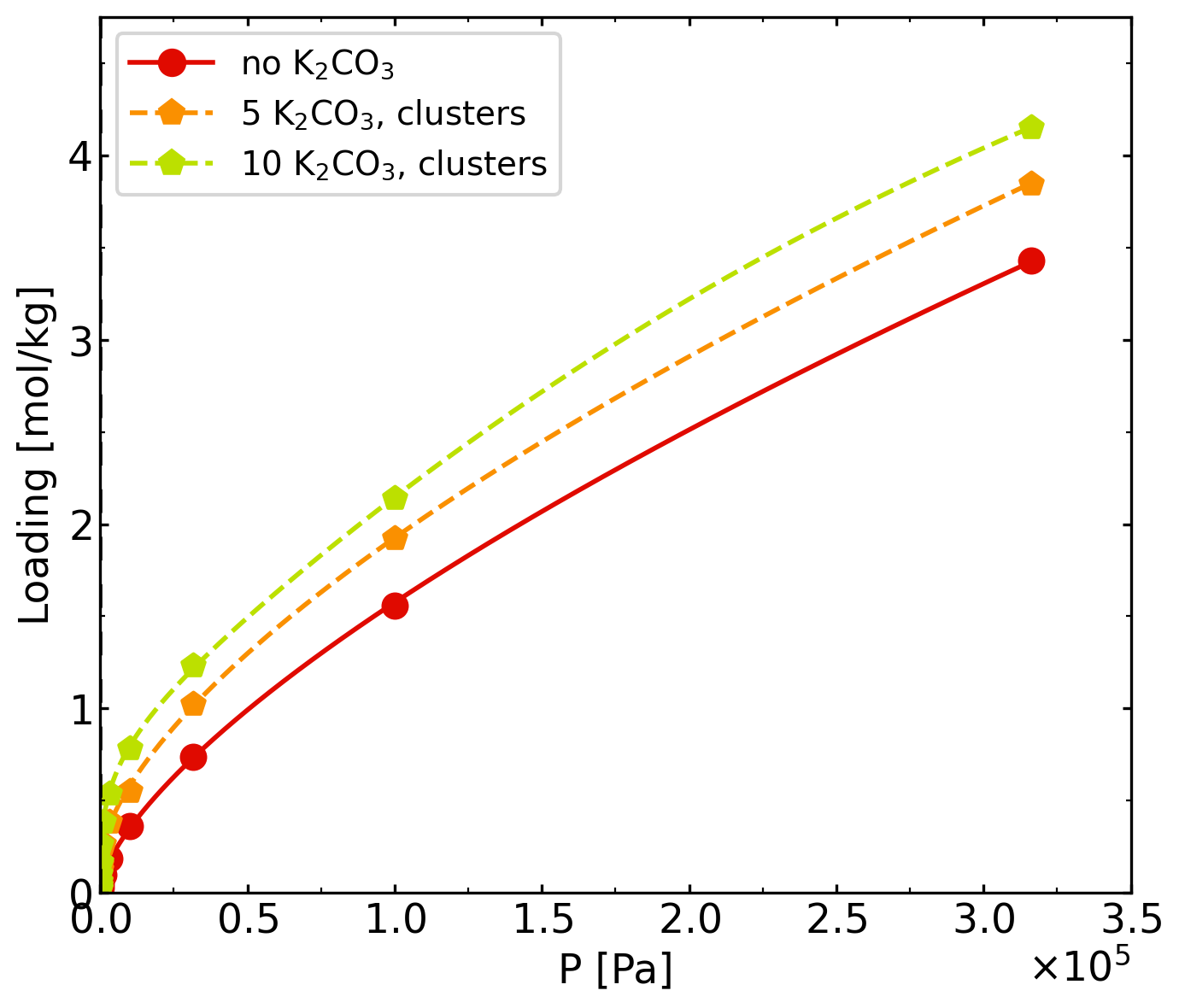}
        \caption{}\label{fig:CO2 f clusters}
    \end{subfigure}
    \begin{subfigure}[b]{0.25\textwidth}
        \includegraphics[width=\textwidth]{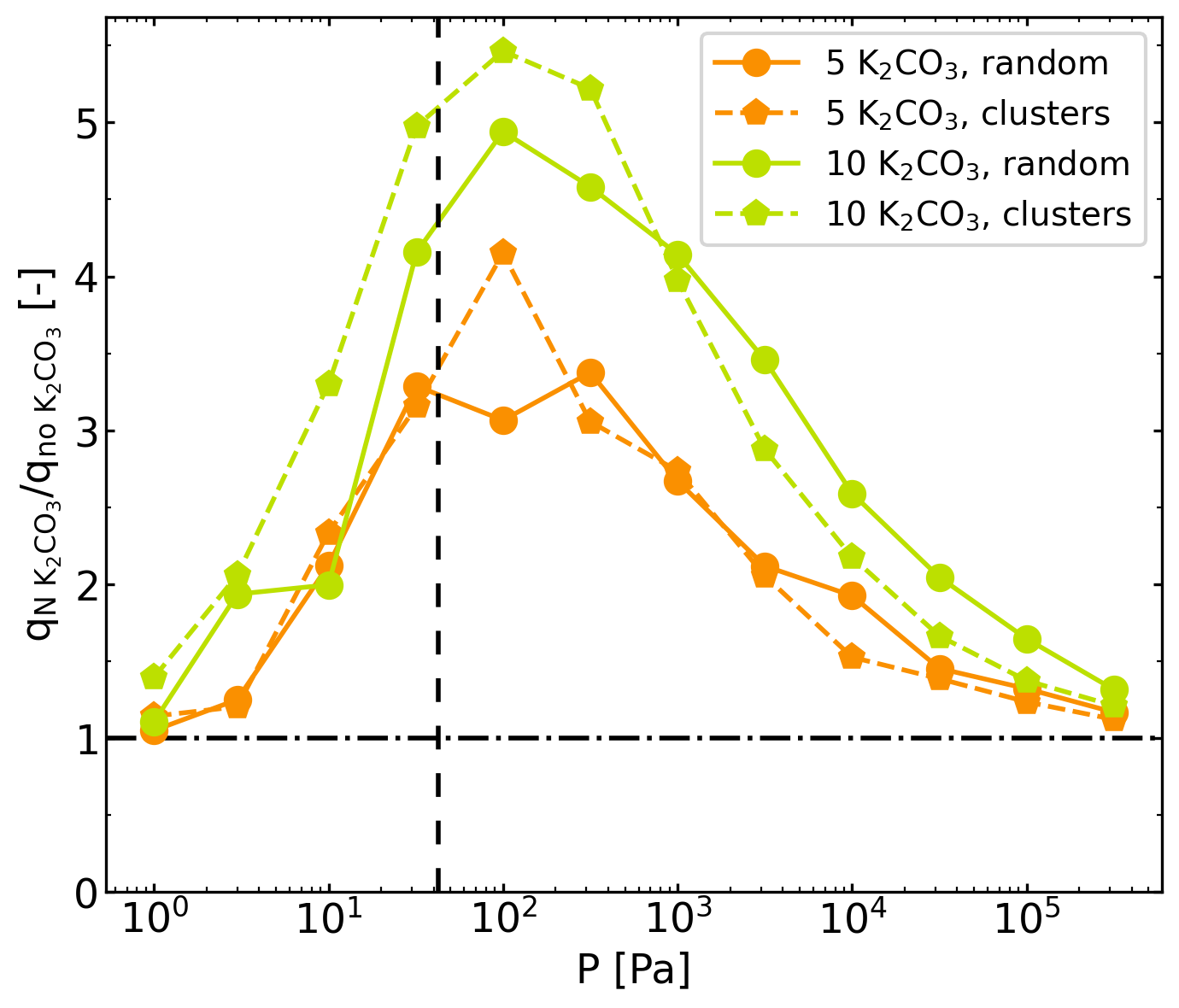}
        \caption{}\label{fig:CO2 f 10 K2CO3 scaled}
    \end{subfigure}
    \begin{subfigure}[b]{0.25\textwidth}
        \includegraphics[width=\textwidth]{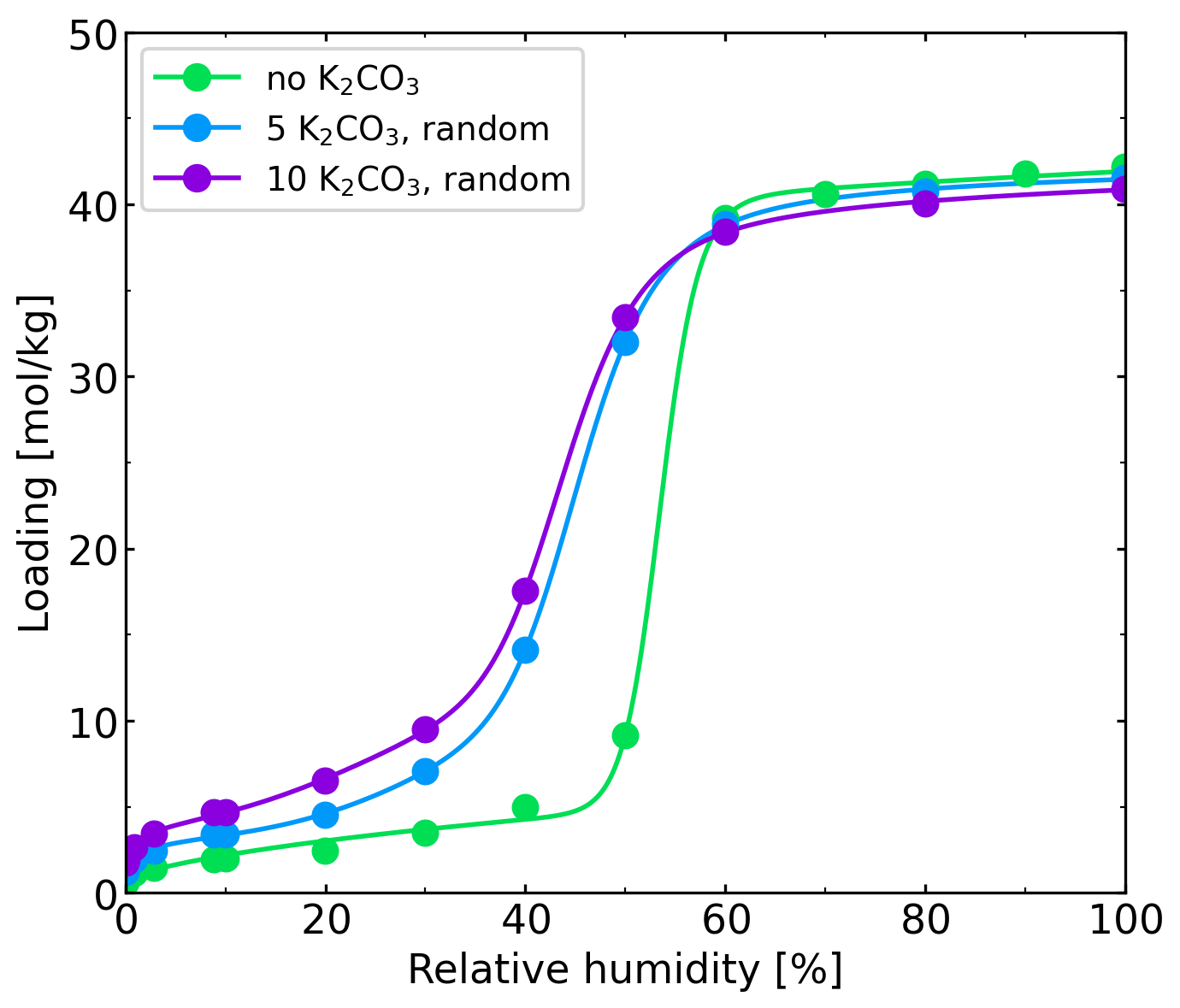}
        \caption{}\label{fig:H2O f random}
    \end{subfigure}
    \begin{subfigure}[b]{0.25\textwidth}
        \includegraphics[width=\textwidth]{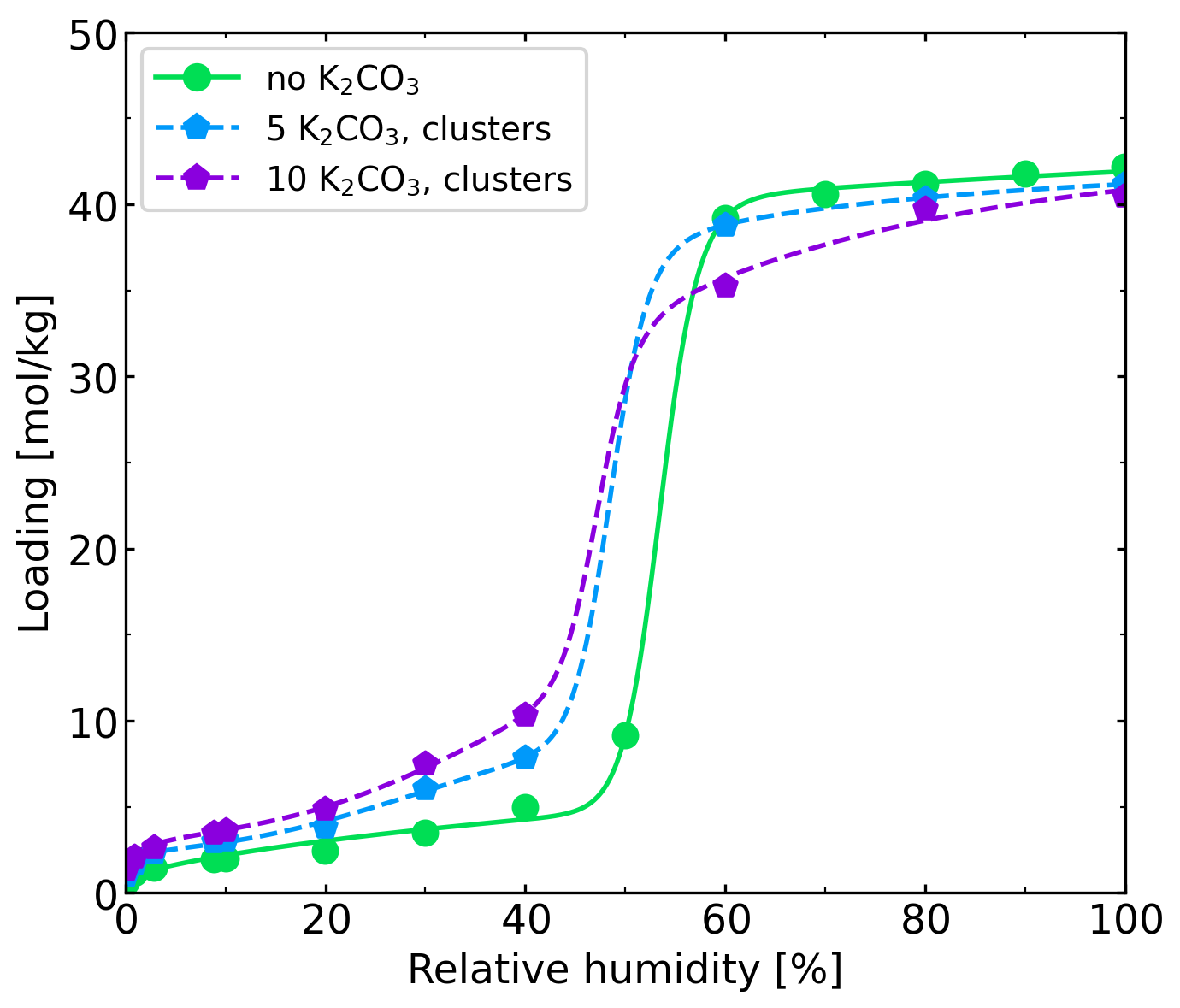}
        \caption{}\label{fig:H2O f clusters}
    \end{subfigure}
    \begin{subfigure}[b]{0.25\textwidth}
        \includegraphics[width=\textwidth]{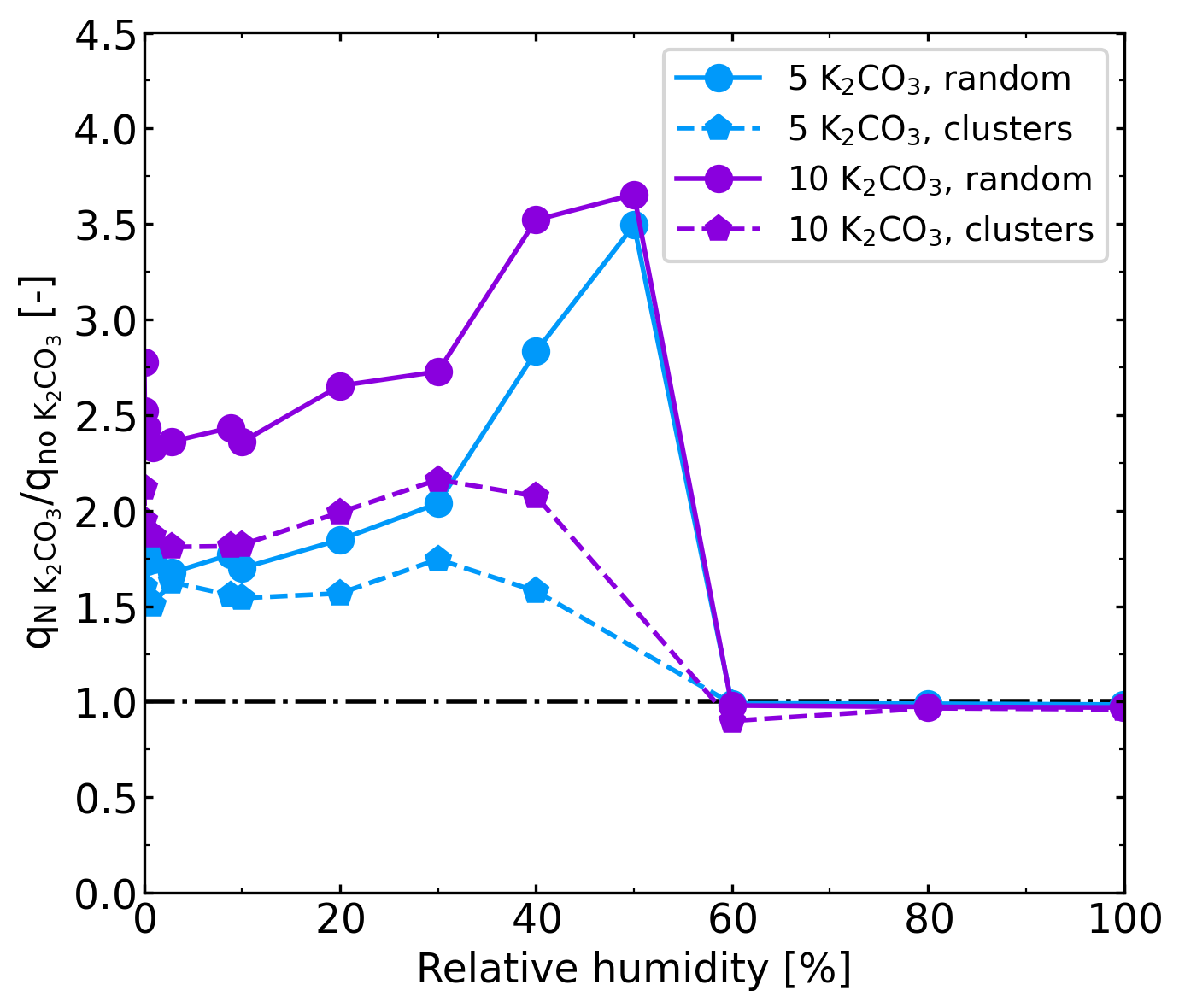}
        \caption{}\label{fig:H2O f 10 K2CO3 scaled}
    \end{subfigure}
    \caption{Comparisons of computed adsorption isotherms of \carbon (a-c) and \water (d-f) in \func at $300.15$ K for varying amounts of \ptc. Figures (a,d) show isotherms in which the \ptc was added randomly into the carbon, while (b,e) show isotherms in which the \ptc was added in clusters of $5$. Lines in these Figures are fits through the data points obtained using RUPTURA.\cite{Sharma2023RUPTURA} In Figures (c,f), isotherms using both methods of \ptc inclusion are shown, scaled by the isotherm without \ptc. The black vertical line in the \carbon isotherms indicates the partial pressure equivalent to 420 ppm (at 1 atm), its current approximate atmospheric concentration.\cite{Lan2023}}\label{fig:iso func K2CO3}
\end{figure*}

\begin{figure*}[!ht]
\centering
    \begin{subfigure}[b]{0.2\textwidth}
        \includegraphics[width=\textwidth]{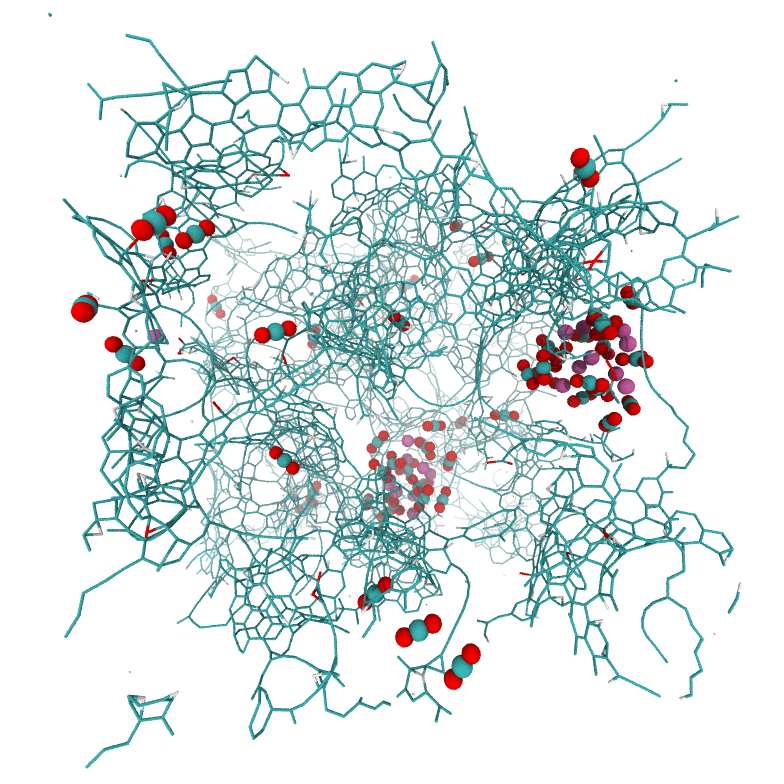}
        \caption{}\label{fig:CO2 f 10000 cluster}
    \end{subfigure}
    \begin{subfigure}[b]{0.2\textwidth}
        \includegraphics[width=\textwidth]{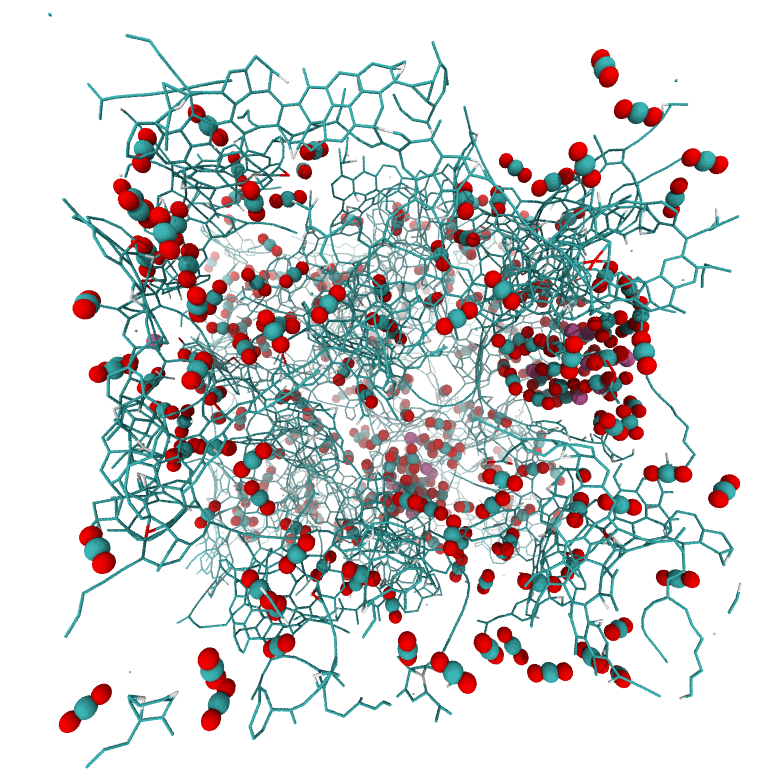}
        \caption{}\label{fig:CO2 f 316228 cluster}
    \end{subfigure}
    \begin{subfigure}[b]{0.2\textwidth}
        \includegraphics[width=\textwidth]{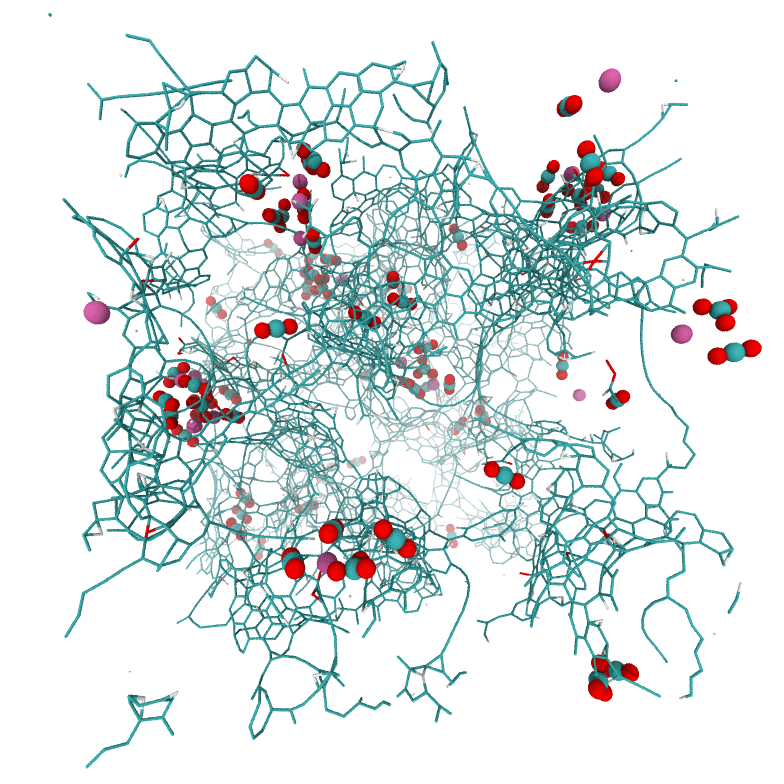}
        \caption{}\label{fig:CO2 f 10000 random}
    \end{subfigure}
    \begin{subfigure}[b]{0.2\textwidth}
        \includegraphics[width=\textwidth]{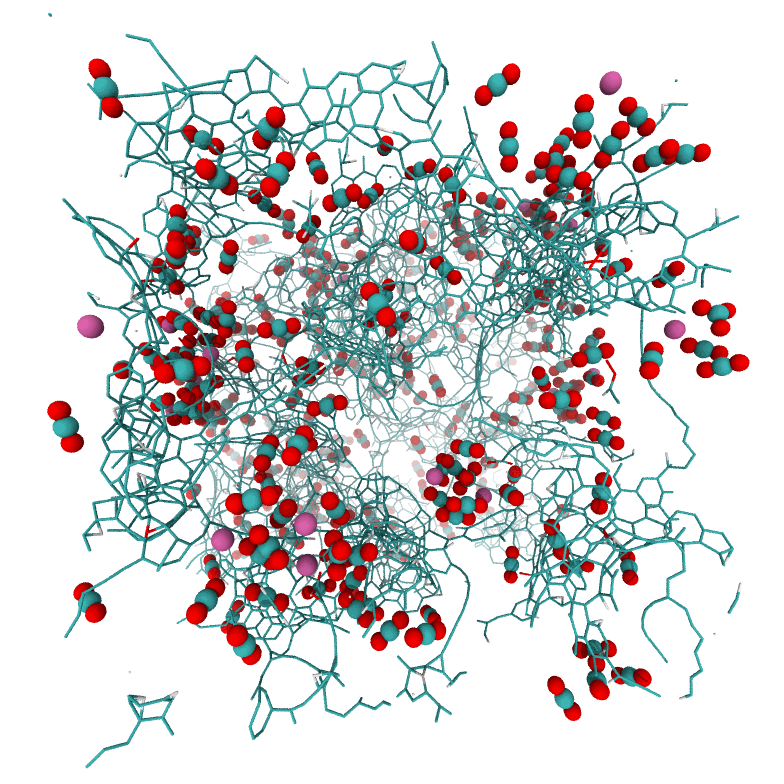}
        \caption{}\label{fig:CO2 f 316228 random}
    \end{subfigure}
    \begin{subfigure}[b]{0.2\textwidth}
        \includegraphics[width=\textwidth]{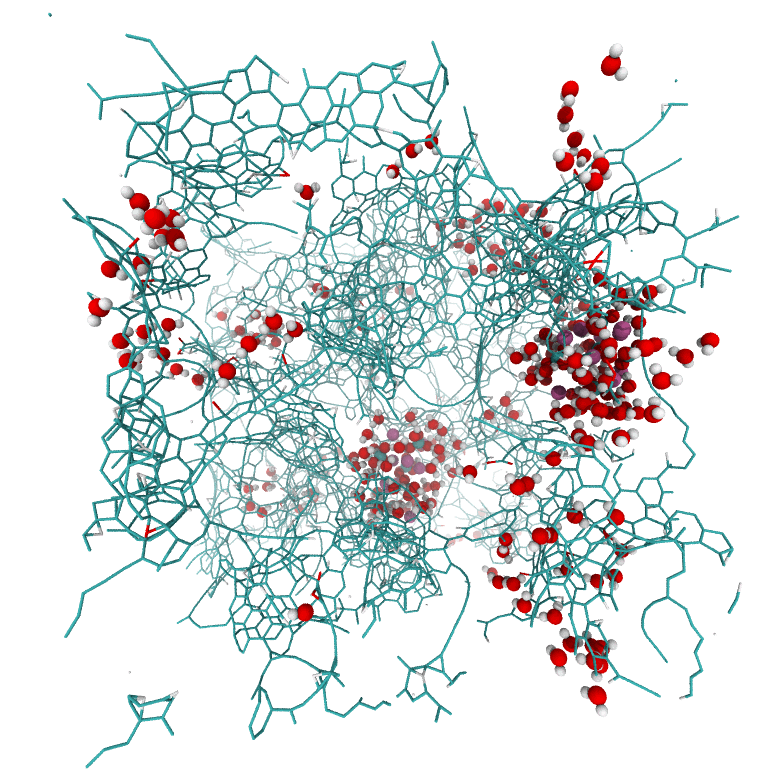}
        \caption{}\label{fig:H2O f 357 cluster}
    \end{subfigure}
    \begin{subfigure}[b]{0.2\textwidth}
        \includegraphics[width=\textwidth]{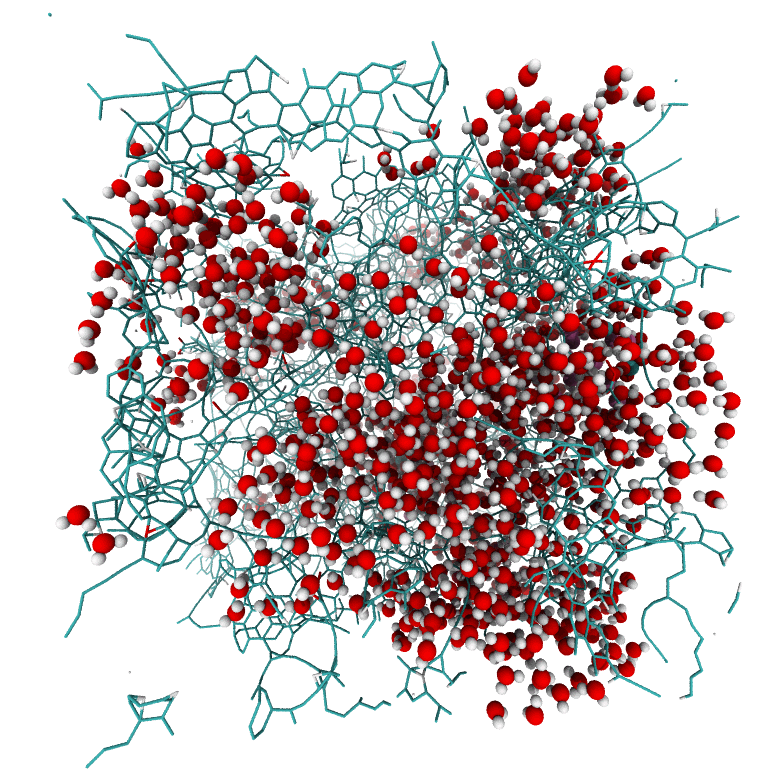}
        \caption{}\label{fig:H2O f 1784 cluster}
    \end{subfigure}
    \begin{subfigure}[b]{0.2\textwidth}
        \includegraphics[width=\textwidth]{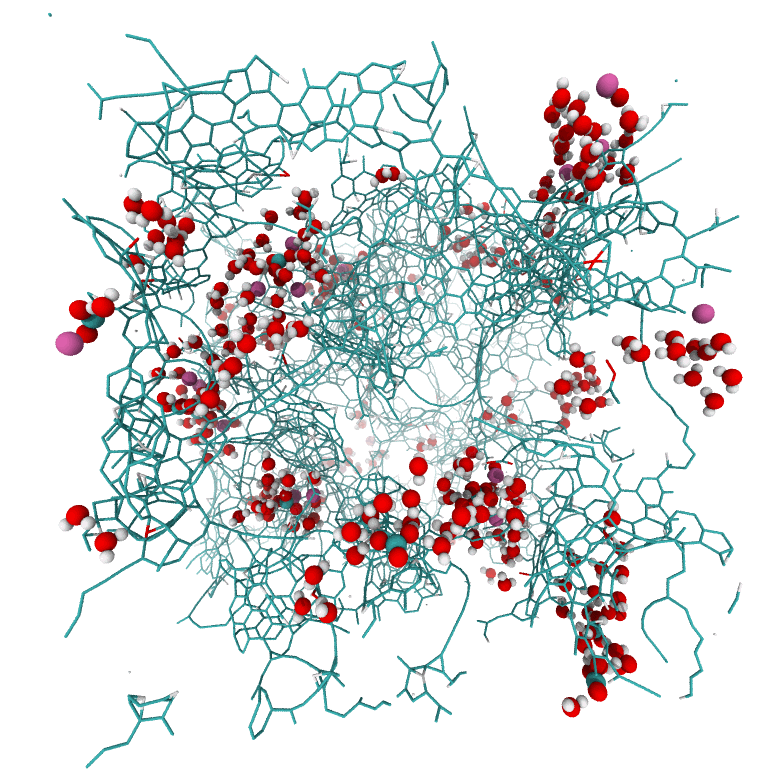}
        \caption{}\label{fig:H2O f 357 random}
    \end{subfigure}
    \begin{subfigure}[b]{0.2\textwidth}
        \includegraphics[width=\textwidth]{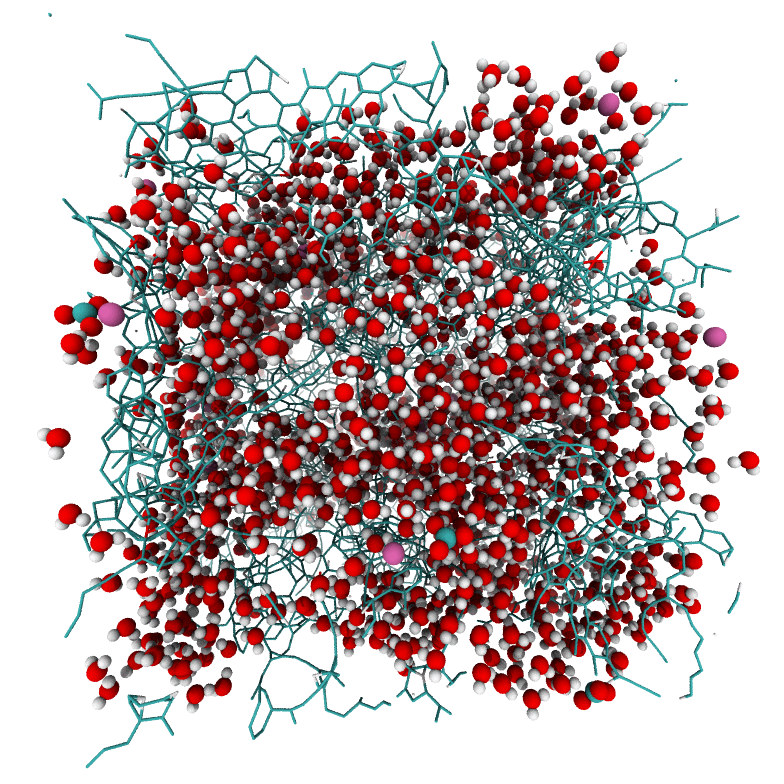}
        \caption{}\label{fig:H2O f 1784 random}
    \end{subfigure}
    \caption{Snapshots showcasing adsorption of \carbon (a-d) and \water (e-h) in \func at $300.15$ K with $10$ \ptc added in clusters (a,b,e,f) or randomly (c,d,g,h) under various partial pressure conditions. \carbon is shown at $10000$ Pa in (a,c) and at $316228$ Pa in (b,d). \water is shown at $357$ Pa ($10\%$ relative humidity) in (e,g) and at 1784 Pa ($50\%$ relative humidity) in (f,h).}\label{fig:renders func 10 K2CO3}
\end{figure*}

After the analysis of the \carbon and water adsorption in the pristine adsorbents, we study the effect of doping the ACs with \ptc species. We used two approaches to dope the adsorbents with \ptc as described in the methodology. Figure \ref{fig:k2co3 creation} schematically recaps the insertion of random \ptc ions and their insertion after forming small clusters. After the \ptc ions are relaxed within the ACs pores, we computed the adsorption isotherms for \carbon and water $300.15$ K. Figure \ref{fig:iso func K2CO3} shows the results alongside comparisons with the pristine \func carbon. As realistic carbons contain surface functionalization, the results for \nofunc are shown in the supplementary materials Fig. \ref{fig:iso nofunc K2CO3}. They are still useful as they isolate the effects of the potassium carbonate, because there are no other active adsorption sites within the pores as in the case of the structures with functional groups.  
The computed water isotherms this time were limited to the $0-100\%$ relative humidity range, covering the pressure range relevant to atmospheric conditions.

Figures \ref{fig:CO2 f random} and \ref{fig:CO2 f clusters} show the \carbon adsorption isotherms with \ptc randomly added or in cluster form, respectively, compared to the isotherm of the pristine carbon. Regardless of the insertion method, increasing the \ptc content enhances adsorption. Quantitatively, increases in \ptc lead to consistent increases in adsorption regardless of the presence of functional groups. This shows evidence that \ptc is able to serve properly as an additional active sites for \carbon. The renderings of the adsorption behavior in Fig. \ref{fig:renders func 10 K2CO3}a-d confirm this finding. These images visualize the \carbon adsorption in \func with 10 \ptc at various pressures. They indeed show significant quantities of \carbon congregating around the \ptc surfaces. The influence of the functional groups is still relevant as clearly more \carbon is present compared to the counterparts without these in Figures \ref{fig:renders nofunc no K2CO3}a-d.

To make a comparison of the effectiveness of \ptc in increasing adsorption, Fig. \ref{fig:CO2 f 10 K2CO3 scaled} scales all \carbon isotherms from the carbons with \ptc with the \carbon isotherm of the pristine carbon. No matter the amount of \ptc or its manner of insertion, the resultant shape is qualitatively similar. All trends show a peak around the $10^{2}$ Pa range, indicating that this is where \ptc is most effective in increasing \carbon adsorption as a general trend. This is probably related to the covering of most of the surface of \ptc with a monolayer of \carbon. After this, higher pressures are needed for continued filling, which becomes increasingly more pronounced as less \carbon is able to set near the surfaces. The atmospheric conditions of carbon indicated by the vertical line are within the broader peak of increased effectiveness, thus clearly benefiting the application.

Figure \ref{fig:CO2 f 10 K2CO3 scaled} additionally provides a direct comparison between the methods of inserting \ptc randomly or in clusters. Although the differences in effectiveness are not major, a minimal but definite difference in trends can be found. This is most pronounced for 10 \ptc, which below $10^3$ Pa shows \func with clusters adsorbing a slightly higher amount of \carbon. Above this pressure, the structure with randomly added \ptc adsorbs more. Considering that potassium carbonate has a larger surface area when it has been inserted randomly, it is understandable that more is able to adsorb. At low pressures, not enough \carbon is adsorbed for this extra surface to matter, and clustered \ptc molecules instead act as stronger active centers for superior \carbon adsorption. The structure with only 5 \ptc shows minimal effects of surface area differences, and either method is about as effective. Only at the $10^2$ Pa peak does the cluster insertion method show better adsorption, while at higher pressures the random insertion method only has slightly preferential adsorption.

Figures \ref{fig:H2O f random} and \ref{fig:H2O f clusters} show the water adsorption isotherms with \ptc randomly added or in cluster form, respectively, compared to the isotherm of pristine carbon. At lower RH, large quantitative increases in loading are found as more \ptc are present. This also tends towards an earlier onset of saturation. The effects are even more pronounced for \nofunc (Figures \ref{fig:iso nofunc K2CO3}d,e), revealing large gaps in onset to saturation between 5 and 10 \ptc. The \ptc appears to act hygroscopically properly and serves as additional adsorption sites for water to form hydrogen bonds with. This is confirmed by the computed radial distribution functions (RDFs) between water and \Optc as discussed below. This means that they serve as additional locations for the water filling mechanism to form clusters around, which is evident from the structures modeled in Figures \ref{fig:renders func 10 K2CO3} e-h. These show snapshots of water adsorption in \func with 10 \ptc at $10\%$ RH and $50\%$ RH for both methods of \ptc insertion. Large clusters of water are present, including around the locations of \ptc, which grow to near saturation at 50\% RH. The renderings of \nofunc are shown in Figures \ref{fig:renders func 10 K2CO3} e-h, which unambiguously display the adsorption due to the added \ptc.

The adsorption mechanism also leads to large differences in effectiveness between the insertion methods. This is best observed in Fig. \ref{fig:H2O f 10 K2CO3 scaled}, which displays all water isotherms with \ptc, scaled with the water isotherm of the pristine carbon. The extra surface area created by random insertions of \ptc is significant to the point where 5 randomly added \ptc molecules have water adsorption similar to that of 2 clusters below 40\% RH. After this, it even outperforms due to the earlier onset of filling. These differences are also visible in the renders; there is a tangible increase in adsorption in the structures with randomly added \ptc (Figures \ref{fig:renders func 10 K2CO3}g,h) compared to their counterparts which used cluster insertion (Figures \ref{fig:renders func 10 K2CO3}e,f). 

With Fig. \ref{fig:H2O f 10 K2CO3 scaled} also provides insight into the effective increases in adsorption due to the addition of \ptc. Regardless of the amount or method of insertion, the trends are similar. At the lowest pressure conditions, a greater increase in the water uptake is observed, showing preferential adsorption. After this, the relative increase is stable up to the point at which the structure starts to fill rapidly to reach saturation. This filling is earlier than in the pristine carbon, and thus shows up as peaks whose locations and shape are dependent on the amount and method of insertion. In general, the random insertion method leads to a slightly higher water uptake because there are more active adsorption sites that trigger the nucleation of water within the pores. After the point of pristine carbon reaches saturation, all loadings are nearly the same, resulting in a flat line at value one.


Finally, we investigated the microscopic structure of \ptc and surrounding adsorbates in the pores by means of computing radial distribution functions. To this aim, we focused on the structure of \func with 10 \ptc in cluster formation, which can serve as a good representation of a realistic carbon containing both surface groups and potassium carbonate in clusters.\cite{Cai2020, Masoud2022}

\begin{figure}[!t]
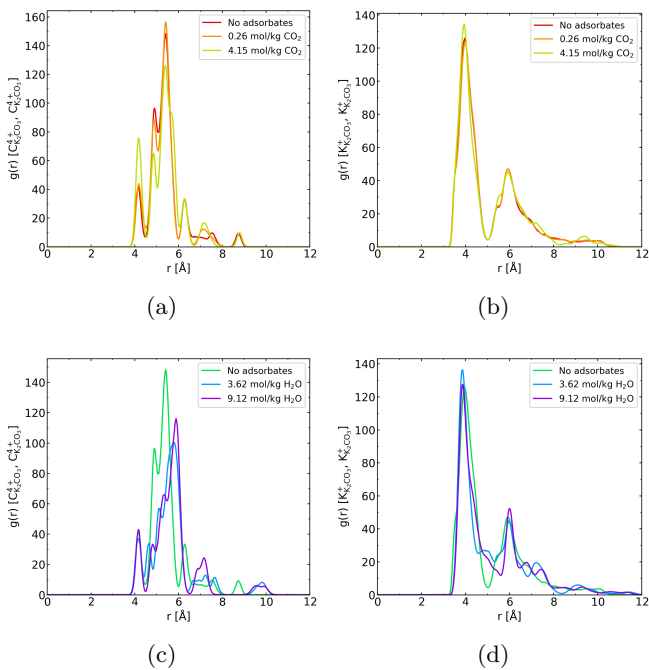

\centering
    \begin{minipage}{\columnwidth}
        \centering
        \begin{subfigure}[b]{0.49\columnwidth}
            \includegraphics[width=\textwidth]{images/\rdfpth/rdf-f-CO2-C_K2CO3_C_K2CO3.png}
            \caption{}\label{fig:rdf CO2 f C - C 2}
        \end{subfigure}
        \hfill
        \begin{subfigure}[b]{0.49\columnwidth}
            \includegraphics[width=\textwidth]{images/\rdfpth/rdf-f-CO2-K_K2CO3_K_K2CO3.png}
            \caption{}\label{fig:rdf CO2 f K - K 2}
        \end{subfigure}
    \end{minipage}
    
    \vspace{1em} 
    
    \begin{minipage}{\columnwidth}
        \centering
        \begin{subfigure}[b]{0.49\columnwidth}
            \includegraphics[width=\textwidth]{images/\rdfpth/rdf-f-H2O-C_K2CO3_C_K2CO3.png}
            \caption{}\label{fig:rdf H2O f C - C 2}
        \end{subfigure}
        \hfill
        \begin{subfigure}[b]{0.49\columnwidth}
            \includegraphics[width=\textwidth]{images/\rdfpth/rdf-f-H2O-K_K2CO3_K_K2CO3.png}
            \caption{}\label{fig:rdf H2O f K - K 2}
        \end{subfigure}
    \end{minipage}
    
    \caption{Computed radial distribution functions (RDFs) with \ptc in \func containing $2$ clusters of $5$ \ptc in the presence of \carbon (a,b) and water (c,d) at $300.15$ K. The RDFs compare the effects of these adsorbates on the structure of the \ptc cluster at different loading. RDFs are shown between $\text{\Cptc} - \text{\Cptc}$ (a,c) and $\text{\Kptc} - \text{\Kptc}$ (b,d).}
    \label{fig:rdfs effect on ptc}
\end{figure}

Following the previously obtained adsorption isotherms, additional runs of $2\times10^5$ cycles were performed after 5000 cycles of equilibration at low and high loadings to obtain the RDFs between ions and adsorbates. Figure \ref{fig:rdfs effect on ptc} shows the RDFs between the ions of the clusters at two different concentrations of adsorbed molecules. Interactions involving $\text{\ptc} - \text{\ptc}$ in systems without adsorbates are also included as a baseline for comparison. Figure \ref{fig:rdf CO2 f C - C 2} shows the RDFs between the carbon atoms of \ptc under the adsorption of \carbon. With increased pressure and loading, only slight changes occur in peak heights and in the position of the peak at $5.3$ \r{A}. RDFs between different ions come to the same conclusions, changing only slightly at higher loading, as exemplified by the RDF between K$^+$ ions in Fig. \ref{fig:rdf CO2 f K - K 2}. These small changes indicate the shape of clusters is not affected by \carbon. This indicates that \ptc plays an important role on enhancing \carbon adsorption, while preserving its microstructure while confined within the ACs pores. Water predictably has a more significant influence on the clusters. This is evident from both Figures \ref{fig:rdf H2O f C - C 2} and \ref{fig:rdf H2O f K - K 2}, which contain the RDFs of $\text{\Cptc} - \text{\Cptc}$ and $\text{\Kptc} - \text{\Kptc}$, respectively under water adsorption. Both figures show that the peaks jump and the shape changes in the RDFs when comparing the baseline without adsorbates to that with 3.62 mol/kg \water. These changes are most pronounced for the \Cptc ions, which reveal a tendency of these ions to move further apart. Changes that occur with increasing loading to 9.12 mol/kg \water are noticeable, but not as extreme. Under current conditions, although there are changes, these results do not contain major shifts, indicating cluster destabilization and breakup.

\begin{figure*}[!ht]
\centering
    \begin{subfigure}[b]{0.28\textwidth}
        \includegraphics[width=\textwidth]{images/\rdfpth/rdf-f-CO2-C_co2_O_K2CO3.png}
        \caption{}\label{fig:rdf CO2 f C_CO2 - O}
    \end{subfigure}
    \begin{subfigure}[b]{0.28\textwidth}
        \includegraphics[width=\textwidth]{images/\rdfpth/rdf-f-CO2-O_co2_O_K2CO3.png}
        \caption{}\label{fig:rdf CO2 f O_CO2 - O}
    \end{subfigure}
    \begin{subfigure}[b]{0.25\textwidth}
        \includegraphics[width=\textwidth]{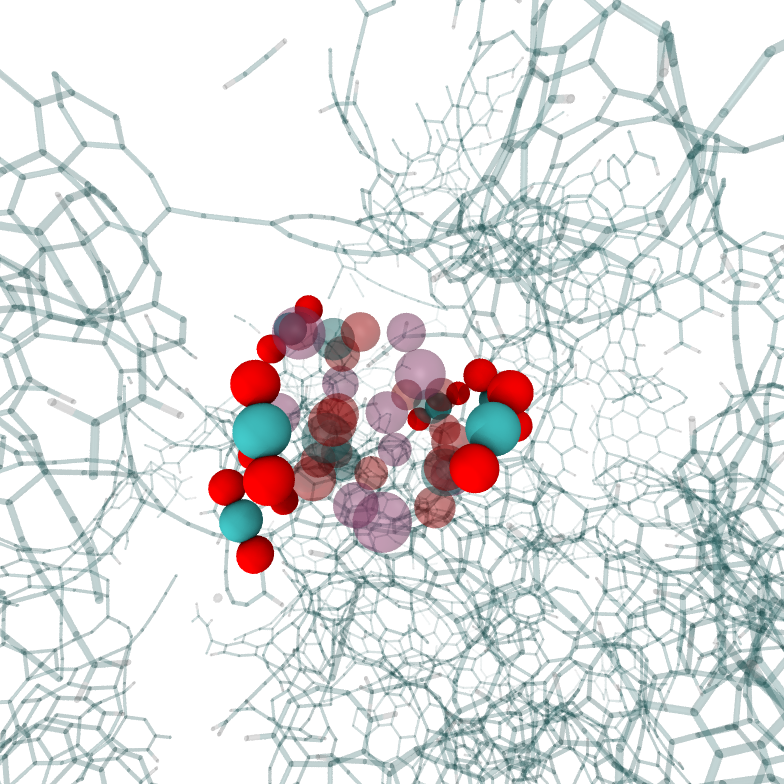}
        \caption{}\label{fig:closeup CO2}
    \end{subfigure}
    \begin{subfigure}[b]{0.28\textwidth}
        \includegraphics[width=\textwidth]{images/\rdfpth/rdf-f-H2O-Hw1_O_K2CO3.png}
        \caption{}\label{fig:rdf H2O f H_H2O - O}
    \end{subfigure}
    \begin{subfigure}[b]{0.28\textwidth}
        \includegraphics[width=\textwidth]{images/\rdfpth/rdf-f-H2O-Ow1_O_K2CO3.png}
        \caption{}\label{fig:rdf H2O f O_H2O - O}
    \end{subfigure}
    \begin{subfigure}[b]{0.25\textwidth}
        \includegraphics[width=\textwidth]{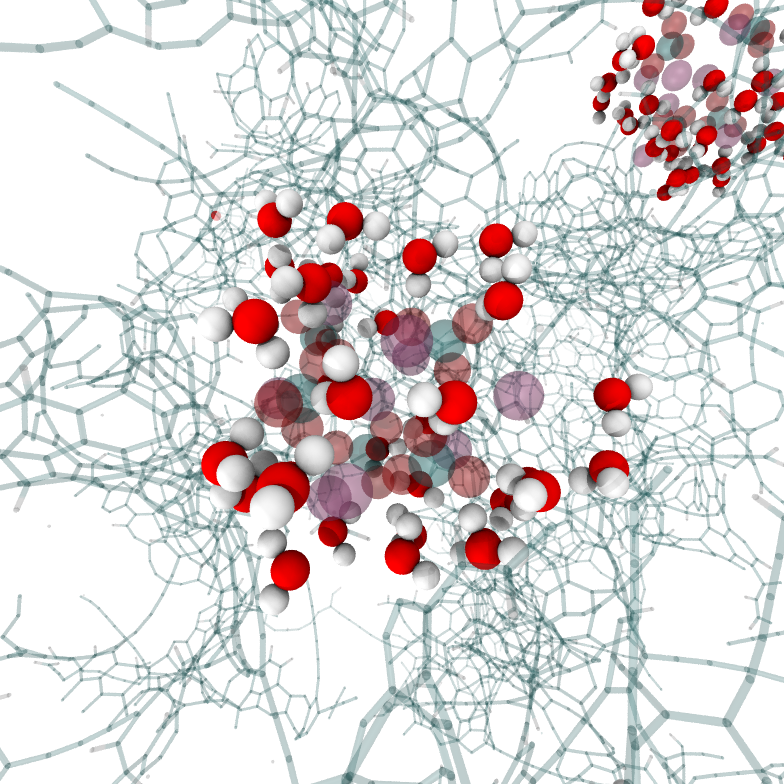}
        \caption{}\label{fig:closeup H2O}
    \end{subfigure}
    \caption{Computed radial distribution functions (RDFs) between \ptc and adsorbates in \func containing $2$ clusters of $5$ \ptc in the presence of \carbon (a,b) and water (d,e) at $300.15$ K. For \carbon, RDFs are shown between $\text{\Ccarbon} - \text{\Optc}$ (a) and $\text{\Ocarbon} - \text{\Optc}$ (b). For water, RDFs are shown between $\text{\Hwater} - \text{\Optc}$ (d) and $\text{\Owater} - \text{\Optc}$ (e). Snapshots of closeups are shown for \carbon (c) and water (f) in \nofunc with 2 clusters to visualize the loading behavior clearly.}\label{fig:rdfs ptc adsorbates}
\end{figure*}

Figure \ref{fig:rdfs ptc adsorbates} shows the RDFs between the adsorbates and the cluster ions with increasing loading. The same trend is observed in all cases. With increasing pressure, the peaks shrink, but the locations of all peaks remain the same. This scaling effect is explained simply by the amount of adsorbates present in the system, which makes only a smaller proportion able to be near the clusters. We thus conclude that there are no significant changes in the distribution of atoms around the clusters in either case.

The RDFs between the adsorbates and \ptc themselves show a predominant first peak at distances lower than 4 \r{A} that covers the first coordination shell of the added salt, even at very low adsorbate concentrations. This confirms the role of \ptc ions being the most active adsorption sites for \carbon and water in doped ACs. For \carbon, the distribution is fairly homogeneous around the clusters. Here, RDFs show that distances between \carbon and \Optc (Figures \ref{fig:rdfs ptc adsorbates}a,b) are slightly smaller than between \carbon and \Kptc or \Cptc (Figures \ref{fig:rdfs ptc adsorbates cont}a-d). The water RDFs show that \water acts as a hydrogen bond donor and \Optc acts as a hydrogen bond acceptor. Peaks between \water and \Optc in \ref{fig:rdfs ptc adsorbates}d,e clearly have a much larger amplitude at lower distances than those between \water and \Kptc or \Cptc in Figures \ref{fig:rdfs ptc adsorbates cont}e-h. Further RDFs between the \water atoms themselves in Fig. \ref{fig:H2O H2O mutual HB} show the hydrogen bonding between water molecules, which occurs when water clusters around active sites. Snapshots with close-ups of \carbon and water molecules adsorbing around the \ptc clusters are shown in Figures \ref{fig:closeup CO2} and \ref{fig:closeup H2O} to illustrate the observed behaviors of the RDFs. For visual clarity, these snapshots were made in \nofunc, which isolates adsorption to the two clusters of \ptc present.

\section{Conclusion}\label{sec:conclusions}

In this work, we investigated the adsorption properties of activated carbon sorbents doped with potassium carbonate (\ptc) for direct air capture (DAC) applications. This process captures \carbon through a reversible reaction with water and \ptc, releasing the stored compounds at elevated temperatures. The amounts of adsorbed \carbon and water in the carbon pores under specific operating conditions are critical to the performance of this reaction. To gain insights at the molecular level, we employed Monte Carlo simulations to study \carbon and water adsorption in a representative activated carbon, CS1000a.

To assess the effect of pore surface chemistry, we considered two models of CS1000a: one with and one without surface functional groups. The presence of functional groups enhances the adsorption of both \carbon and water. While \carbon uptake increases slightly across the entire pressure range—without altering the overall shape of the isotherm—the impact on water adsorption is more pronounced. Functional groups serve as nucleation sites for water, facilitating pore filling at pressures below water’s saturation pressure.

We also explored two methods of incorporating \ptc into the pores: random insertion of CO$_3^{2-}$ and K$^+$ ions, and the addition of small \ptc clusters. These approaches resulted in subtle differences in adsorption mechanisms. Random insertion creates more adsorption sites throughout the pore volume, while clusters act as strong interaction centers, particularly at low pressures. Higher \ptc loadings enhance the adsorption of both adsorbates. Radial distribution functions (RDFs) and structural analyses revealed that \carbon tends to form monolayers on \ptc surfaces. In the carbon model with functional groups, \carbon uptake increases by a factor of $3$–$5\times$ compared to the pristine carbon under DAC conditions (420 ppm \carbon at atmospheric pressure).

Water interactions with \ptc are even stronger than those of \carbon, due to water’s polar nature. Water forms extended hydrogen-bonded clusters around adsorption sites. As a result, the additional surface area from randomly inserted \ptc significantly promotes water adsorption, which leads to earlier pore filling. Water nucleation through hydrogen bonding is the primary mechanism of adsorption, though interactions with \ptc also slightly alter its structural arrangement, as indicated by the RDFs. Depending on the \ptc insertion method and concentration, water uptake increases by a factor of $1.5$–$3\times$ until reaching saturation.

These findings clarify the molecular mechanisms governing \carbon and water adsorption in activated carbons with large pores, emphasizing the roles of surface functional groups and embedded \ptc. Further work is needed to investigate the chemical reactions involving \carbon, water, and \ptc. In addition, activated carbons with different pore structures and surface chemistries should be examined to optimize \carbon capture under ambient conditions and limit excessive water uptake that may block the pores. Achieving the right balance between \carbon and water adsorption is crucial for enhancing the reactive capture process and improving current direct air capture technologies.

\FloatBarrier

\section*{Supporting Information}

The supporting information contains the necessary information to recreate all fits and the verification of the potassium carbonate model. Additional figures included are computed isotherms and heats of adsorption at different temperatures, isotherms for \nofunc with \ptc, snapshots for \nofunc, more RDFs between \ptc and adsorbates in \func with \ptc clusters, and RDFs between \water atoms in \func with \ptc clusters.





\section*{Competing interests}

The authors declare no competing interests.

\bibliography{CO2_doped-ACs}

\renewcommand{\figurename}{Figure.}
\renewcommand{\thetable}{S\arabic{table}}  
\renewcommand{\thefigure}{S\arabic{figure}} 
\setcounter{figure}{0}
\renewcommand{\thesection}{S\arabic{section}} 
\setcounter{section}{0}

\newpage
\onecolumngrid 
\FloatBarrier
\newpage

\begin{center}
    \textbf{\Huge{Supporting Information}}

    \vspace{1.0cm}

    \large{for}

    \vspace{1.0cm}

    \textbf{\Large{Enhancing Direct Air Capture through Potassium Carbonate Doping of Activated Carbons}}
\end{center}

\section{Fit information}\label{supp:fitting}
For interpolation of the isotherm data, the isotherms are fitted using the RUPTURA software.\cite{Sharma2023RUPTURA} With it, we fit to a multi-site version of the Sips model.\cite{Sips1948} The equation for this isotherm is given by
\begin{equation}
    q(p)=\sum_{i=1}^{n}q_i^{\text{sat}}\frac{(b_i p)^{1/\nu_i}}{1+(b_i p)^{1/\nu_i}},
\end{equation}
where $n$ is the number of sites for the fit. For site $i$, $q_i^{\text{sat}}$ is the saturation loading, $\nu_i$ is a heterogeneity factor, and $b_i$ is an equilibrium constant. The fitting parameters obtained for all isotherms in the study are provided in table \ref{tab:fitted params}.

As may be noted, there are places where the fits do not have the appropriate data points available in order to fully predict the correct behavior of the isotherms. This is especially pertinent for the water isotherms, where various data points are unavailable due to the lengthy equilibration process. This leaves many possible fits, primarily for the pressures range where the carbon fills quickly. However, all fits were carefully created such that the unequilibrated points lay slightly beneath the lines shown.

\newpage

\begin{table}
\centering
\resizebox{\columnwidth}{!}{
\begin{tabular}{clccc|lllllllll}
\hline
Adsorbate & Framework & T [K] & K$_2$CO$_3$ & State & q$_{\text{sat,0}}$ & b$_{\text{0}}$ & $\nu_{\text{0}}$ & q$_{\text{sat,1}}$ & b$_{\text{1}}$ & $\nu_{\text{1}}$ & q$_{\text{sat,2}}$ & b$_{\text{2}}$ & $\nu_{\text{2}}$ \\
\hline
CO$_2$ & CS1000a$_\text{f}$ & $288.15$ & $0$ & - & $268.828$ & $1.38801\times10^{-8}$ & $1.22793$ & $84.8083$ & $8.6869\times10^{-11}$ & $2.43502$ & - & - & - \\
CO$_2$ & CS1000a$_\text{f}$ & $300.15$ & $0$ & - & $13596100000.0$ & $2.84265\times10^{-20}$ & $1.46273$ & $0.03125$ & $0.250001$ & $2.87783$ & - & - & - \\
CO$_2$ & CS1000a$_\text{f}$ & $313.15$ & $0$ & - & $0.0179896$ & $520320.0$ & $0.369437$ & $12855500000.0$ & $4.57488\times10^{-20}$ & $1.43076$ & - & - & - \\
CO$_2$ & CS1000a$_\text{nf}$ & $288.15$ & $0$ & - & $1.7031$ & $1.01808\times10^{-5}$ & $1.00196$ & $2.61121$ & $4.64167\times10^{-6}$ & $0.404941$ & - & - & - \\
CO$_2$ & CS1000a$_\text{nf}$ & $300.15$ & $0$ & - & $1.40375$ & $8.81056\times10^{-6}$ & $0.999719$ & $2.67864$ & $3.42886\times10^{-6}$ & $0.480607$ & - & - & - \\
CO$_2$ & CS1000a$_\text{nf}$ & $313.15$ & $0$ & - & $0.946109$ & $4.47819\times10^{-6}$ & $0.320378$ & $2.20404$ & $3.9838\times10^{-6}$ & $0.99978$ & - & - & - \\
H$_2$O & CS1000a$_\text{f}$ & $288.15$ & $0$ & - & $9.99289$ & $0.000956674$ & $0.374432$ & $31.2495$ & $0.00102457$ & $0.0445515$ & $19.9972$ & $9.57408\times10^{-6}$ & $2.93968$ \\
H$_2$O & CS1000a$_\text{f}$ & $300.15$ & $0$ & - & $1.38188$ & $8.39528$ & $32.3697$ & $35.3644$ & $0.000523475$ & $0.0351599$ & $17.2498$ & $0.00011119$ & $1.32998$ \\
H$_2$O & CS1000a$_\text{f}$ & $313.15$ & $0$ & - & $5.56278$ & $5.18107\times10^{-10}$ & $9.78818$ & $33.4967$ & $0.000245162$ & $0.0421983$ & $16.0021$ & $8.42108\times10^{-5}$ & $0.999371$ \\
H$_2$O & CS1000a$_\text{nf}$ & $288.15$ & $0$ & - & $1.77099$ & $7.47617\times10^{-5}$ & $0.292032$ & $47.2963$ & $0.000136922$ & $0.0726278$ & $4.44898$ & $1.36593\times10^{-5}$ & $0.604888$ \\
H$_2$O & CS1000a$_\text{nf}$ & $300.15$ & $0$ & - & $3.9837$ & $3.22697\times10^{-6}$ & $0.77107$ & $44.5314$ & $0.000136447$ & $0.0699203$ & $6.49791$ & $3.268\times10^{-5}$ & $0.577798$ \\
H$_2$O & CS1000a$_\text{nf}$ & $313.15$ & $0$ & - & $46.2306$ & $3.88966\times10^{-5}$ & $0.0541569$ & $4.81713$ & $1.06071\times10^{-5}$ & $0.377755$ & - & - & - \\
CO$_2$ & CS1000a$_\text{f}$ & $300.15$ & $5$ & random & $9.21091$ & $1.55896\times10^{-6}$ & $0.95973$ & $0.00706898$ & $4.36171$ & $0.497889$ & $1.3026$ & $4.66398\times10^{-5}$ & $2.09134$ \\
CO$_2$ & CS1000a$_\text{f}$ & $300.15$ & $5$ & clusters & $0.0147388$ & $461549.0$ & $0.227748$ & $0.185653$ & $0.0071673$ & $1.29103$ & $132849.0$ & $3.06468\times10^{-13}$ & $1.53769$ \\
CO$_2$ & CS1000a$_\text{f}$ & $300.15$ & $10$ & random & $898.99$ & $2.80833\times10^{-10}$ & $1.72671$ & $0.492482$ & $0.00103027$ & $1.64134$ & $0.00994535$ & $15717500.0$ & $0.582258$ \\
CO$_2$ & CS1000a$_\text{f}$ & $300.15$ & $10$ & clusters & $0.054664$ & $0.0109155$ & $0.823911$ & $3.33096$ & $3.35559\times10^{-6}$ & $2.56939$ & $5.37919$ & $2.71963\times10^{-6}$ & $0.730417$ \\
CO$_2$ & CS1000a$_\text{nf}$ & $300.15$ & $5$ & random & $1.19784$ & $3.54875\times10^{-5}$ & $1.49932$ & $2.87602$ & $5.09154\times10^{-6}$ & $0.52059$ & $0.0294278$ & $0.0327476$ & $0.26237$ \\
CO$_2$ & CS1000a$_\text{nf}$ & $300.15$ & $5$ & clusters & $1.76909$ & $7.86441\times10^{-6}$ & $1.49079$ & $2.34509$ & $4.7797\times10^{-6}$ & $0.441332$ & $0.104001$ & $0.0165566$ & $0.983537$ \\
CO$_2$ & CS1000a$_\text{nf}$ & $300.15$ & $10$ & random & $0.101647$ & $0.0330879$ & $0.299067$ & $1.03646$ & $0.000164145$ & $1.57597$ & $4.76986$ & $3.56738\times10^{-6}$ & $0.715113$ \\
CO$_2$ & CS1000a$_\text{nf}$ & $300.15$ & $10$ & clusters & $3.95357$ & $3.80472\times10^{-6}$ & $0.62092$ & $0.245232$ & $0.00403509$ & $1.18782$ & $1.11226$ & $2.32895\times10^{-5}$ & $1.95332$ \\
H$_2$O & CS1000a$_\text{f}$ & $300.15$ & $5$ & random & $10.0559$ & $0.000728463$ & $0.288344$ & $26.9863$ & $0.000616398$ & $0.0815502$ & $17.9595$ & $2.3479\times10^{-6}$ & $4.6894$ \\
H$_2$O & CS1000a$_\text{f}$ & $300.15$ & $5$ & clusters & $3.205$ & $0.081493$ & $2.16001$ & $10.8897$ & $0.000636007$ & $0.409244$ & $28.5783$ & $0.000577098$ & $0.0384296$ \\
H$_2$O & CS1000a$_\text{f}$ & $300.15$ & $10$ & random & $24.844$ & $0.000637759$ & $0.0843347$ & $138.659$ & $1.27692\times10^{-11}$ & $5.60692$ & $10.0312$ & $0.000793509$ & $0.358133$ \\
H$_2$O & CS1000a$_\text{f}$ & $300.15$ & $10$ & clusters & $18.0159$ & $0.000543396$ & $0.345302$ & $20.2467$ & $0.00059274$ & $0.0382297$ & $8.94785$ & $0.000540827$ & $3.68507$ \\
H$_2$O & CS1000a$_\text{nf}$ & $300.15$ & $5$ & random & $27.1104$ & $0.000319319$ & $0.0456701$ & $13.6949$ & $0.00039854$ & $0.143852$ & $7.92201$ & $2.83328\times10^{-6}$ & $4.73668$ \\
H$_2$O & CS1000a$_\text{nf}$ & $300.15$ & $5$ & clusters & $16333.2$ & $4.73563\times10^{-6}$ & $0.461951$ & $36.2897$ & $0.000303782$ & $0.0311494$ & $6.60539$ & $1.97649\times10^{-7}$ & $5.14148$ \\
H$_2$O & CS1000a$_\text{nf}$ & $300.15$ & $10$ & random & $10.5485$ & $6.81628\times10^{-6}$ & $5.46445$ & $7.58773$ & $0.000472452$ & $0.38534$ & $30.1085$ & $0.000494304$ & $0.0387433$ \\
H$_2$O & CS1000a$_\text{nf}$ & $300.15$ & $10$ & clusters & $2.14465$ & $0.000484455$ & $0.26035$ & $6.55638$ & $6.40015\times10^{-6}$ & $5.26029$ & $36.1044$ & $0.000336004$ & $0.0439293$ \\
\hline
\end{tabular}}
\caption{Multi-site Langmuir Sips fitting parameters for pure component adsorption of carbon dioxide and water in CS1000a.}
\label{tab:fitted params}
\end{table}

\section{RDF potassium carbonate verification}\label{supp:rdf melt}
We computed RDFs of bulk \ptc at 300 K with MC using the model in the work of Jo and Banerjee.\cite{Jo2015} We compared these with the solid state results from Ding et al. who simulated RDFs with MD with using a different model.\cite{Ding2018} First, before computing the RDFs, a simulation box was created with initial sides of 60 \r{A} containing 192 CO$_3^{2-}$ and 384 K$^+$. The system was relaxed with an NPT simulation at 300 K and atmospheric pressure for $5\times10^4$ cycles. Then the RDFs were measured under NVT conditions for $5\times10^4$ cycles after 5000 equilibration cycles under the now fixed volume conditions.

\begin{figure*}[!ht]
\centering
    \begin{subfigure}[b]{0.325\textwidth}
        \includegraphics[width=\textwidth]{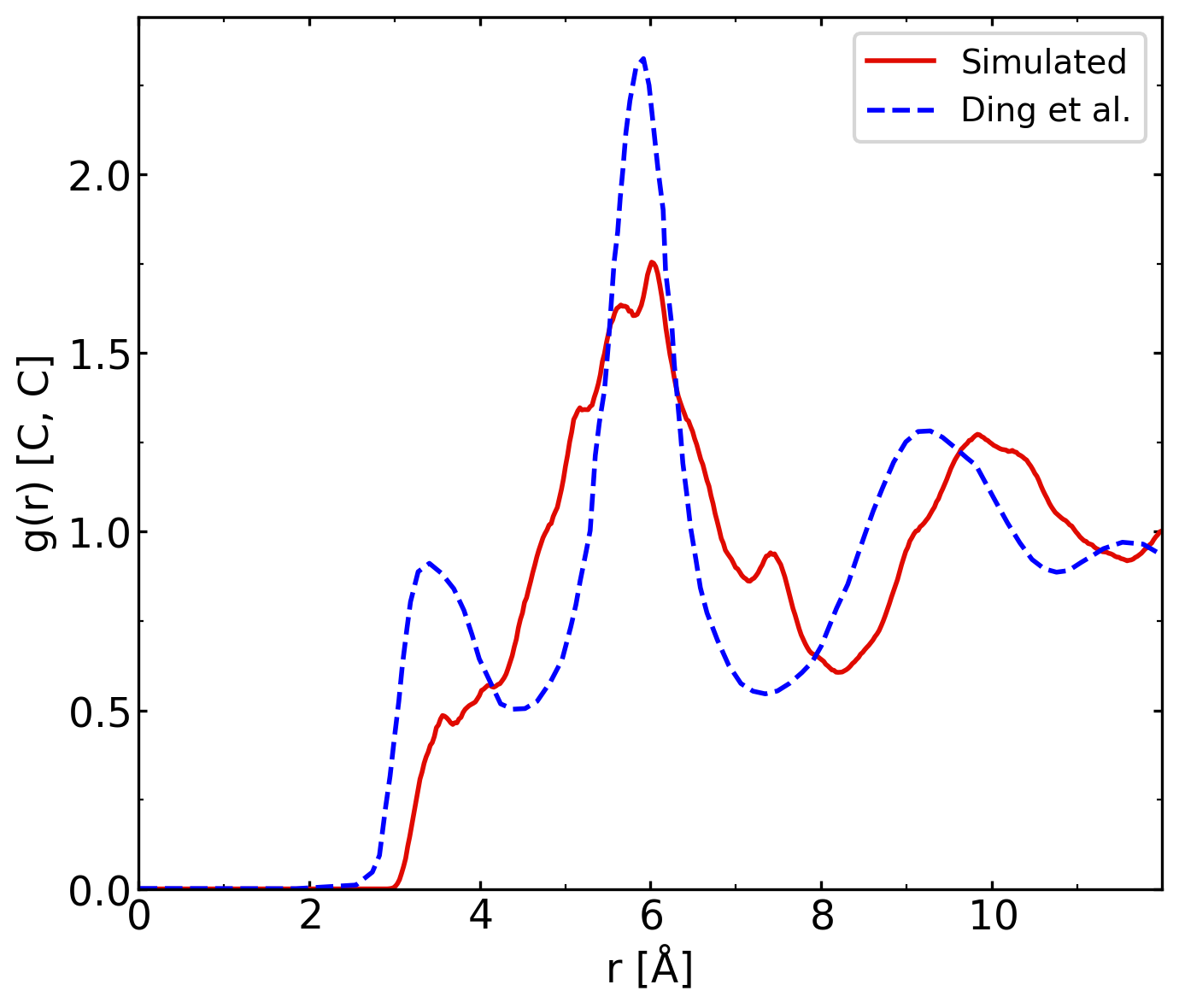}
        \caption{}\label{fig:rdf valid CC}
    \end{subfigure}
    \begin{subfigure}[b]{0.325\textwidth}
        \includegraphics[width=\textwidth]{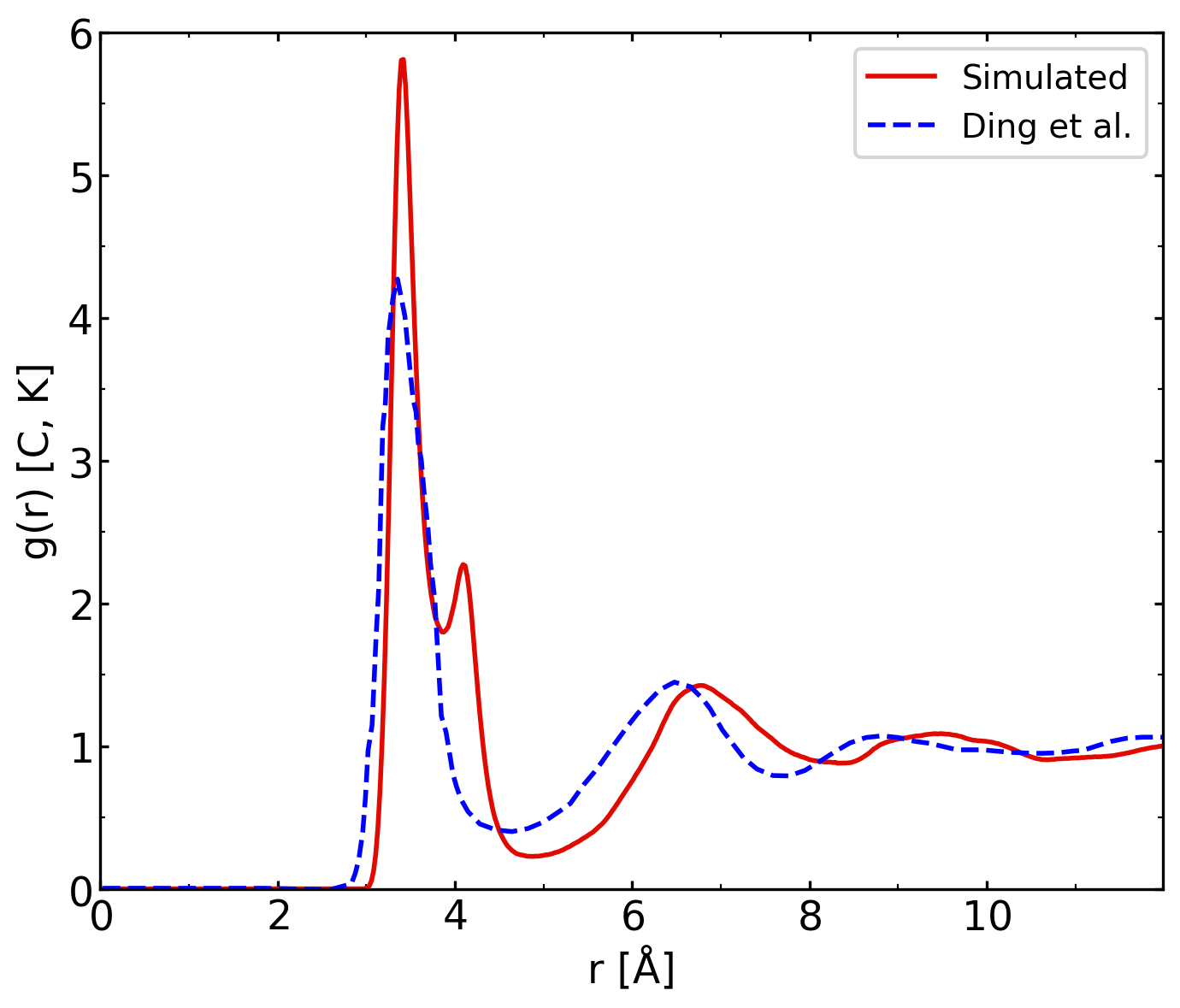}
        \caption{}\label{fig:rdf valid CK}
    \end{subfigure}
    \begin{subfigure}[b]{0.325\textwidth}
        \includegraphics[width=\textwidth]{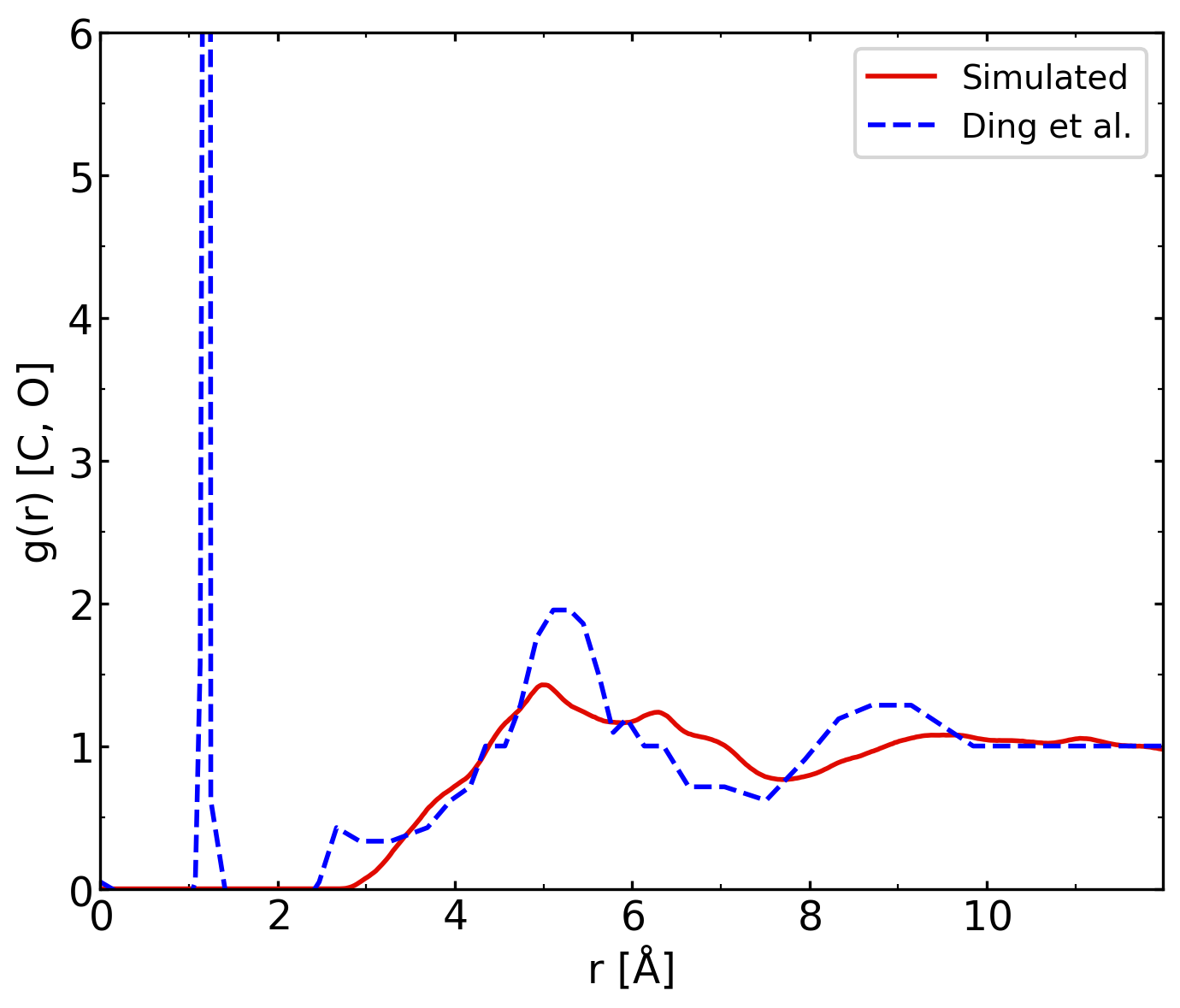}
        \caption{}\label{fig:rdf valid CO}
    \end{subfigure}
    \begin{subfigure}[b]{0.325\textwidth}
        \includegraphics[width=\textwidth]{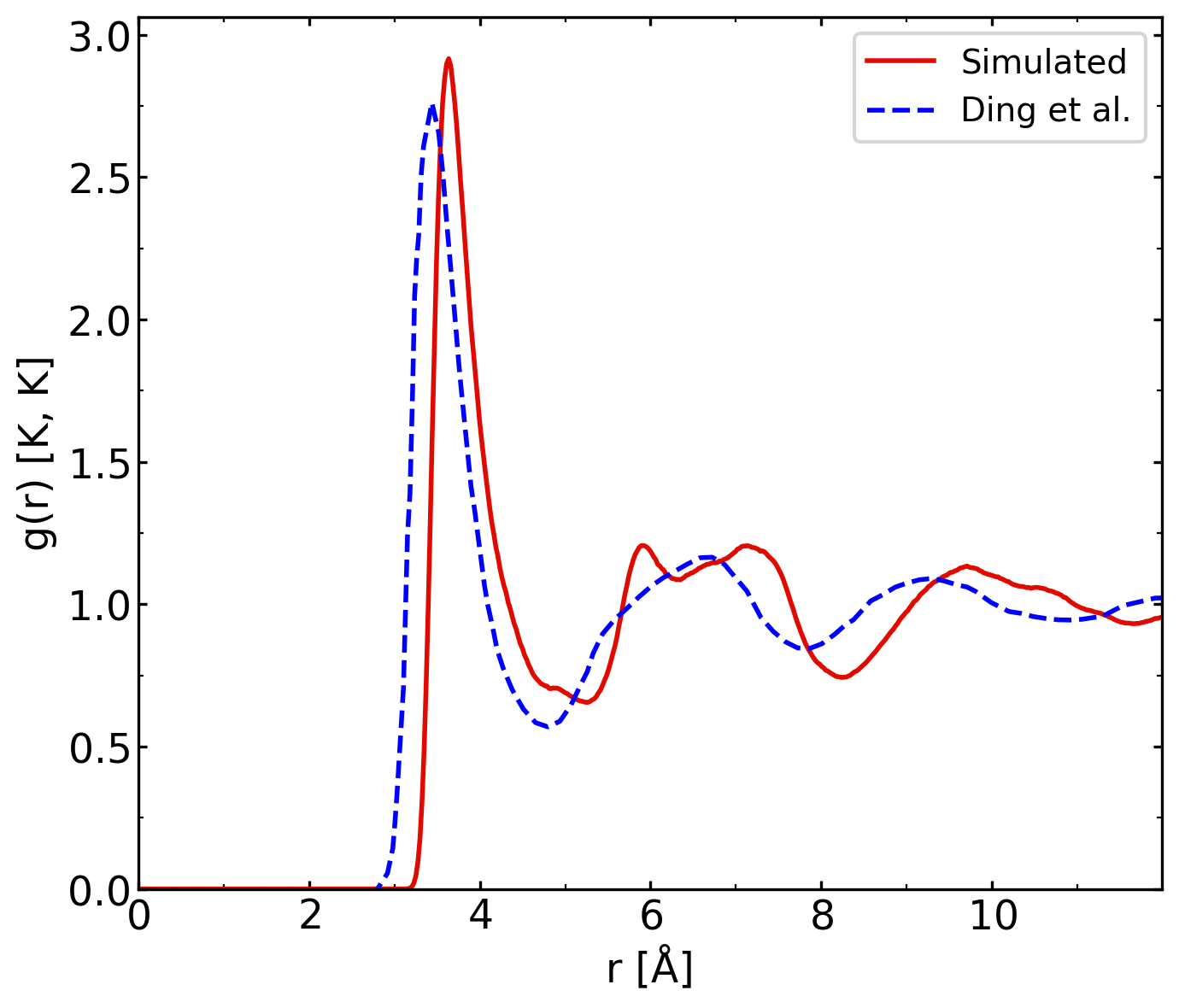}
        \caption{}\label{fig:rdf valid KK}
    \end{subfigure}
    \begin{subfigure}[b]{0.325\textwidth}
        \includegraphics[width=\textwidth]{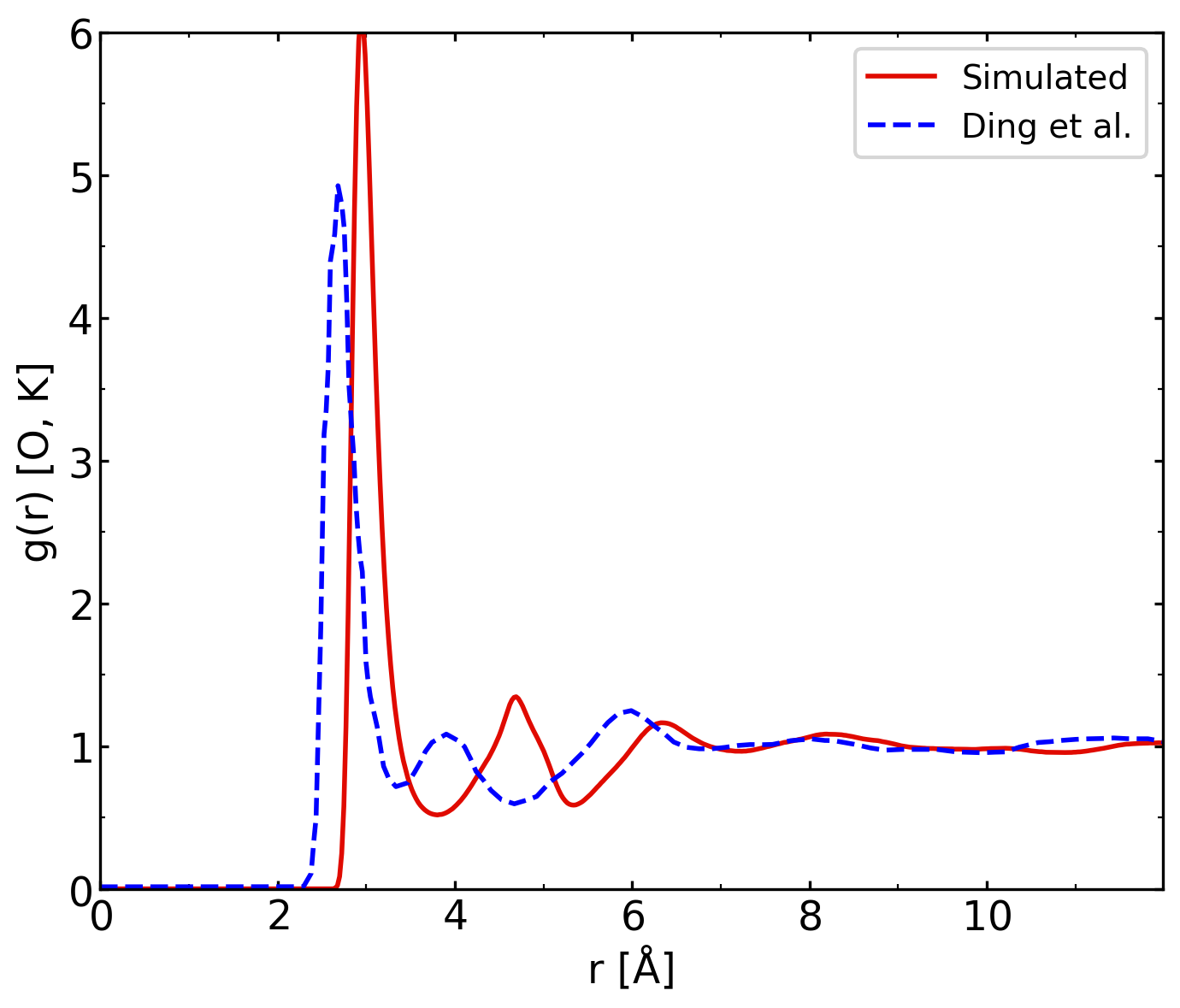}
        \caption{}\label{fig:rdf valid OK}
    \end{subfigure}
    \begin{subfigure}[b]{0.325\textwidth}
        \includegraphics[width=\textwidth]{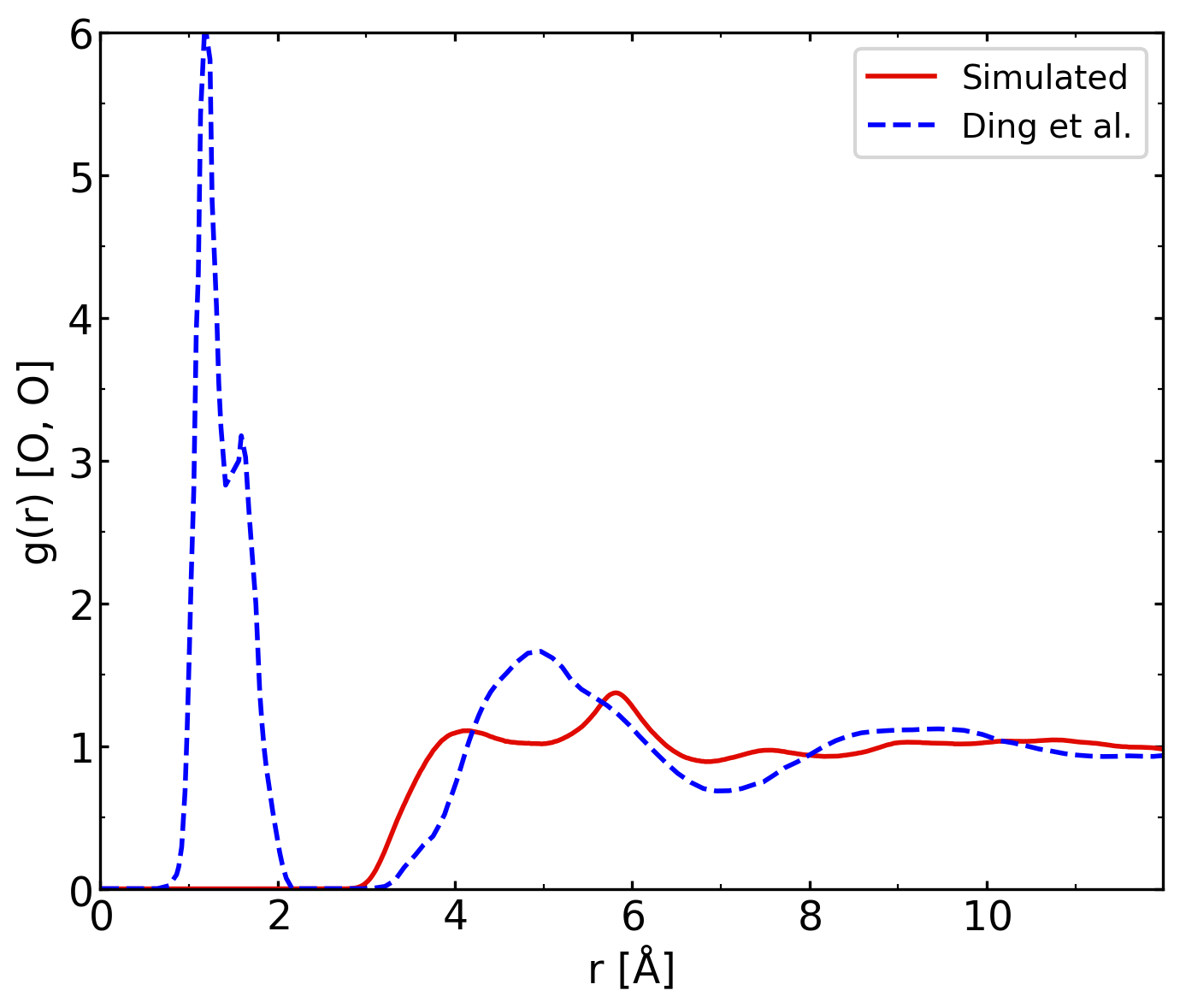}
        \caption{}\label{fig:rdf valid OO}
    \end{subfigure}
    \caption{Computed radial distribution functions (RDFs) between elements of \ptc at $300$ K. The solid red lines are RDFS computed with Monte Carlo simulations using the model in the work of Jo and Banerjee.\cite{Jo2015} The dashed blue lines are results taken directly from the Ding et al., who performed Molecular Dynamics simulations using a different model.\cite{Ding2018} Shown here are RDFs for (a) $\text{C}-\text{C}$, (b) $\text{C}-\text{K}$, (c) $\text{C}-\text{O}$, (d) $\text{K}-\text{K}$, (e) $\text{O}-\text{K}$, and (f) $\text{O}-\text{O}$.}\label{fig:rdf melt validation}
\end{figure*}

The resultant RDFs are shown together with those traced from Ding et al. in figure \ref{fig:rdf melt validation}.\cite{Ding2018} Overall, comparable peaks are observed. Differences observed occur mostly with our results being shifted right as if the sizes of the ions are larger. In several places, the large peaks also show smaller peaks/perturbations not present in the reference. We not in the two Figures \ref{fig:rdf valid CO} and \ref{fig:rdf valid OO} containing the RDFs between $\text{C}-\text{O}$ and $\text{O}-\text{O}$ respectively, the first peaks are completely absent. This is merely because RASPA does not compute the RDF within a defined molecular unit, so in this case within the polyatomic ion CO$_3^{2-}$ the first strong peaks are not taken into account. To also not, due to the height of the $\text{C}-\text{O}$ peak in the original work of Ding et al., the fidelity of the data is low and hence not as accurate. Disregarding the matters based on these points, we conclude the behavior is acceptable.


\FloatBarrier
\section{Figures}\label{supp:figs}
\begin{figure*}[!ht]
\centering
    \begin{subfigure}[b]{0.495\textwidth}
        \includegraphics[width=\textwidth]{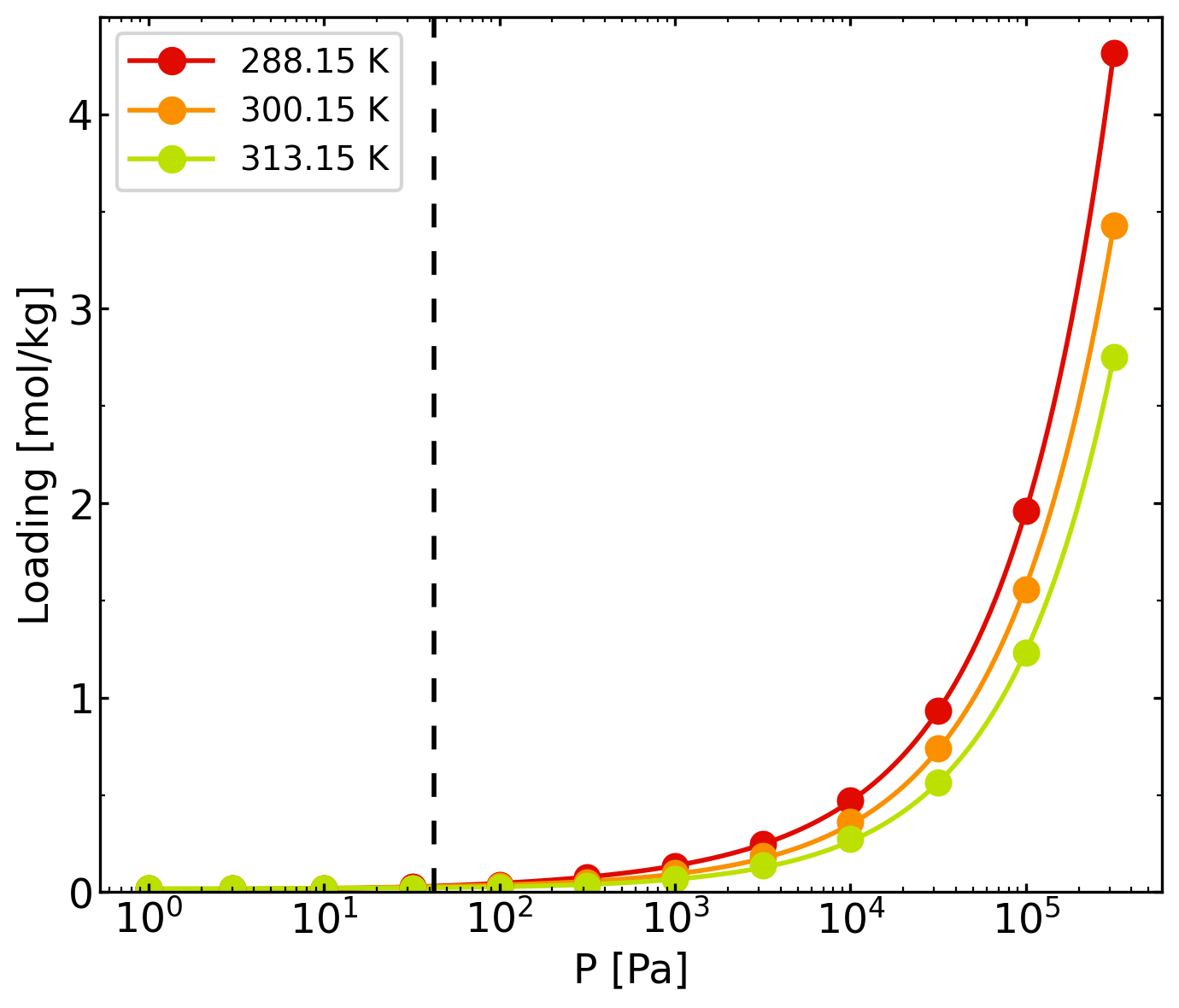}
        \caption{}\label{fig:CO2 f allT}
    \end{subfigure}
    \begin{subfigure}[b]{0.495\textwidth}
        \includegraphics[width=\textwidth]{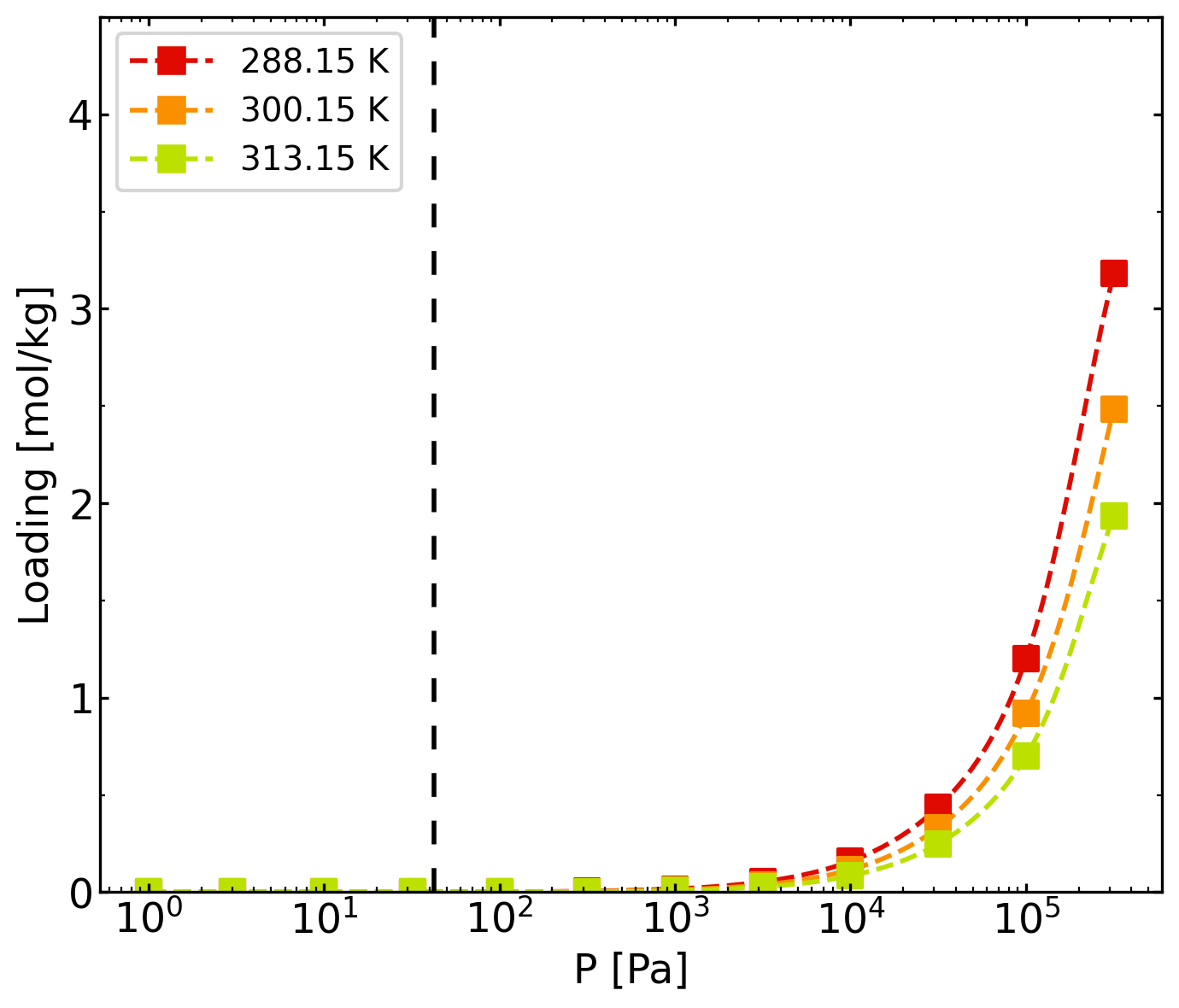}
        \caption{}\label{fig:CO2 nf allT}
    \end{subfigure}
    \begin{subfigure}[b]{0.495\textwidth}
        \includegraphics[width=\textwidth]{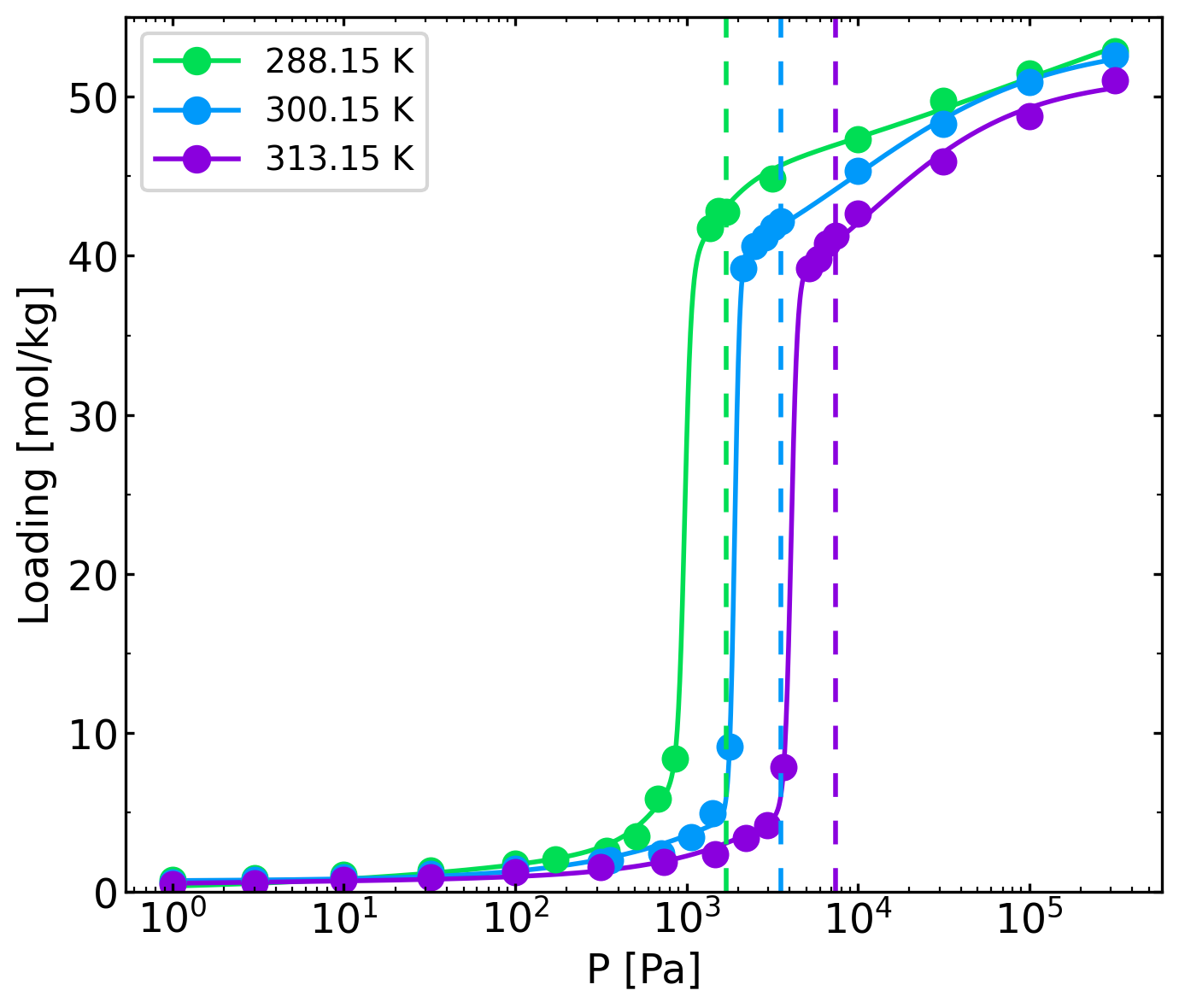}
        \caption{}\label{fig:H2O f allT}
    \end{subfigure}
    \begin{subfigure}[b]{0.495\textwidth}
        \includegraphics[width=\textwidth]{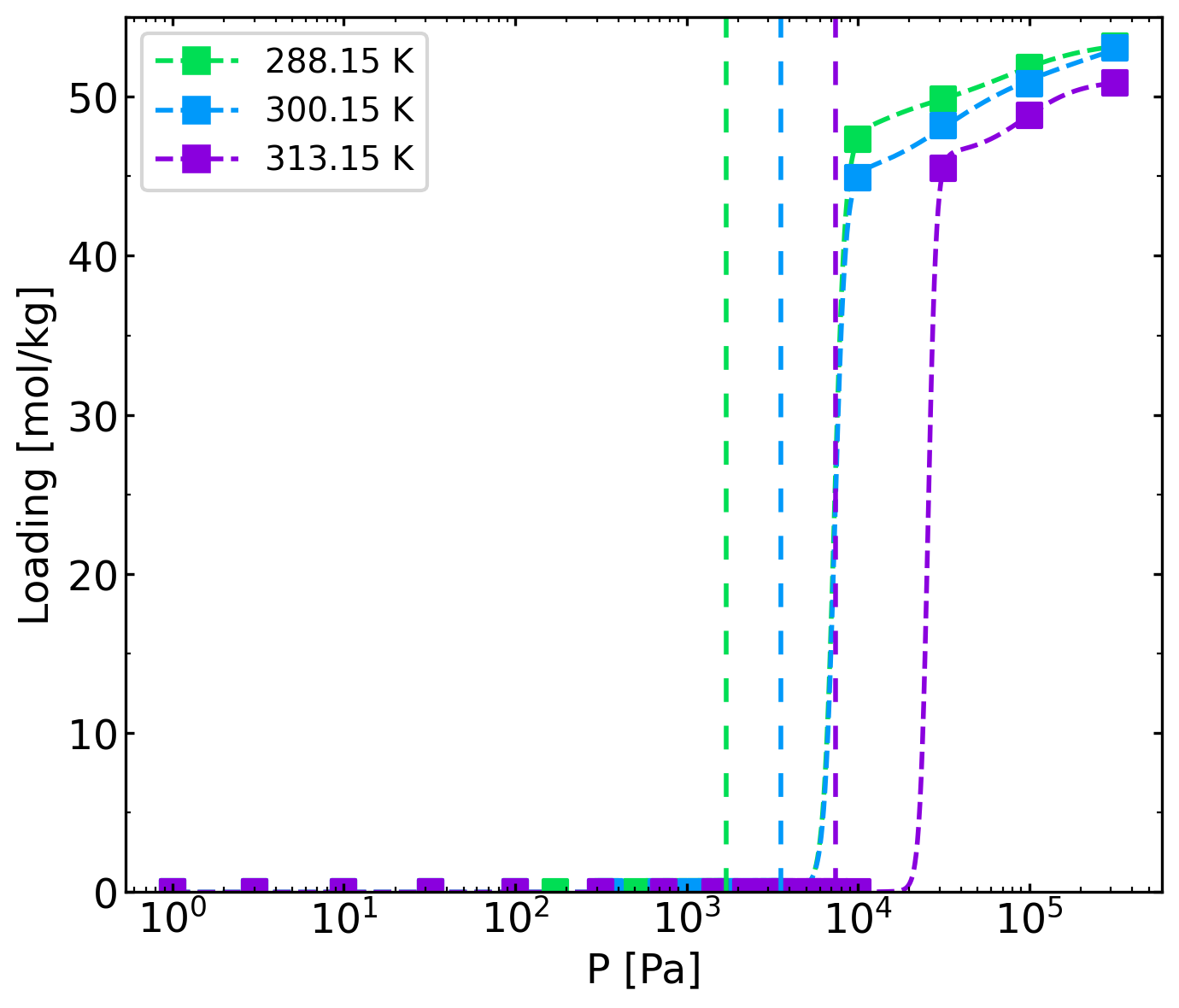}
        \caption{}\label{fig:H2O nf allT}
    \end{subfigure}
    \caption{Comparisons of computed adsorption isotherms for differing temperatures ($288.15$, $300.15$, and $313.15$ K). Figures (a,b) present results for \carbon, (c,d) for \water (c,d), with (a,c) being in \func, and (b,d) in \nofunc. Lines are fits through the data points obtained using RUPTURA.\cite{Sharma2023RUPTURA} The black vertical line in the \carbon isotherms indicate the partial pressure equivalent to 420 ppm (at 1 atm), its current approximate atmospheric concentration.\cite{Lan2023} The vertical lines in the \water isotherms represent the saturation pressures corresponding to their colors.}\label{fig:iso no K2CO3 allT}
\end{figure*}

\begin{figure*}[!ht]
\centering
    \begin{subfigure}[b]{0.495\textwidth}
        \includegraphics[width=\textwidth]{images/\HoApth/HoA-CO2-f-all_T.png}
        \caption{}\label{fig:hoa CO2 f allT}
    \end{subfigure}
    \begin{subfigure}[b]{0.495\textwidth}
        \includegraphics[width=\textwidth]{images/\HoApth/HoA-CO2-nf-all_T.png}
        \caption{}\label{fig:hoa CO2 nf allT}
    \end{subfigure}
    \begin{subfigure}[b]{0.495\textwidth}
        \includegraphics[width=\textwidth]{images/\HoApth/HoA-H2O-f-all_T.png}
        \caption{}\label{fig:hoa H2O f allT}
    \end{subfigure}
    \begin{subfigure}[b]{0.495\textwidth}
        \includegraphics[width=\textwidth]{images/\HoApth/HoA-H2O-nf-all_T.png}
        \caption{}\label{fig:hoa H2O nf allT}
    \end{subfigure}
    \caption{Comparisons of computed heats of adsorption for differing temperatures ($288.15$, $300.15$, and $313.15$ K). Figures (a,b) present results for \carbon, (c,d) for \water (c,d), with (a,c) being in \func, and (b,d) in \nofunc. The horizontal dash-dotted lines represent the heats of vaporation of the respective components at 300.15 K. Heats of evaporation at the other temperatures do not change appreciably, and are therefore not shown.}\label{fig:HoA no K2CO3 allT}
\end{figure*}

\begin{figure*}[!ht]
\centering
    \begin{subfigure}[b]{0.325\textwidth}
        \includegraphics[width=\textwidth]{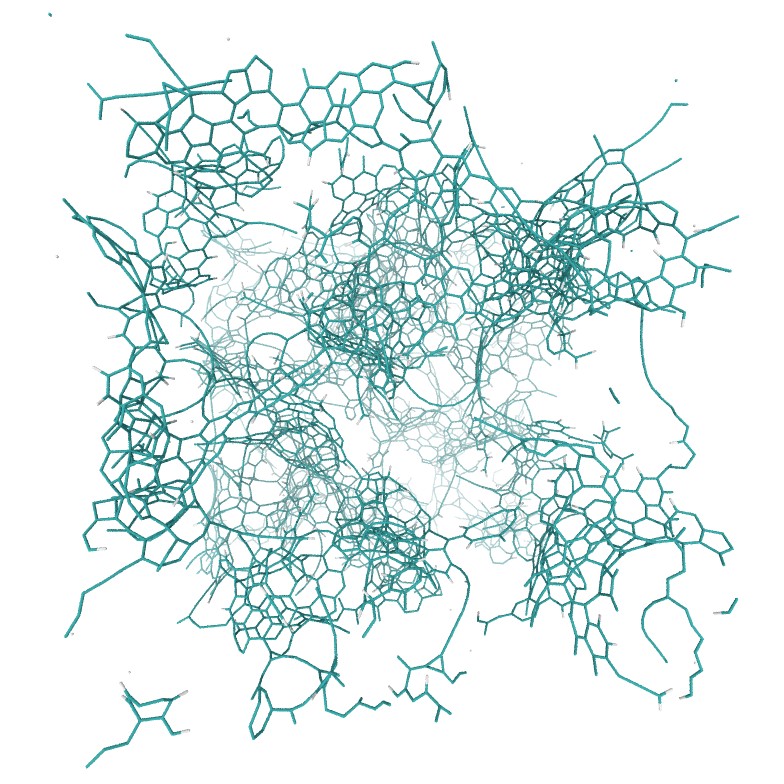}
        \caption{}\label{fig:CO2 nf 316}
    \end{subfigure}
    \begin{subfigure}[b]{0.325\textwidth}
        \includegraphics[width=\textwidth]{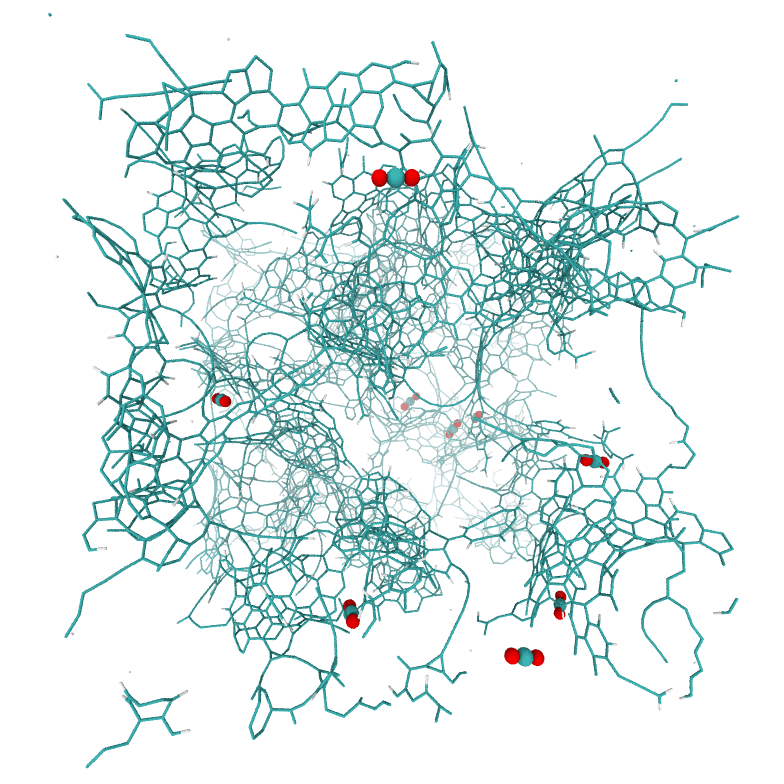}
        \caption{}\label{fig:CO2 nf 10000}
    \end{subfigure}
    \begin{subfigure}[b]{0.325\textwidth}
        \includegraphics[width=\textwidth]{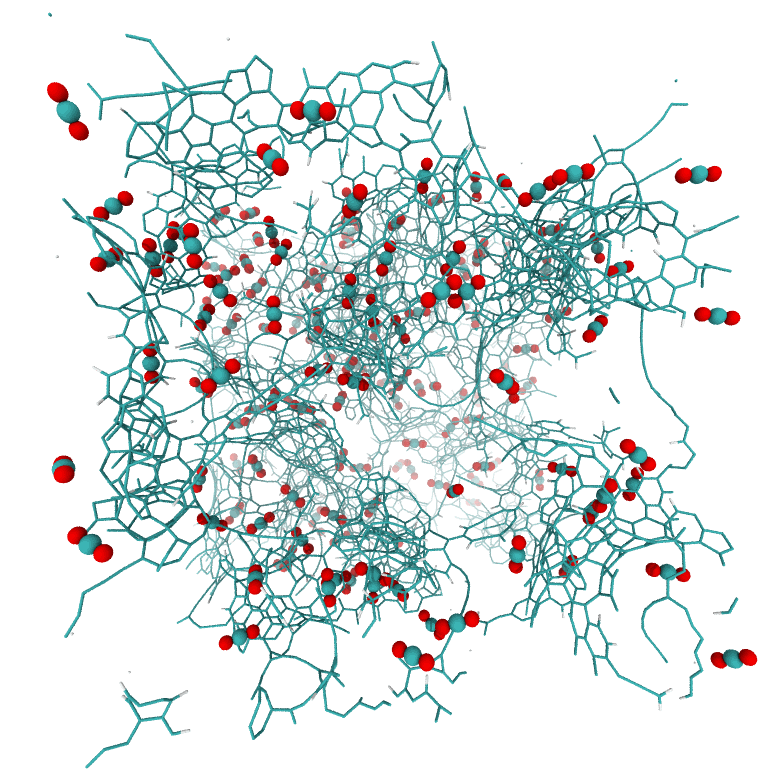}
        \caption{}\label{fig:CO2 nf 316228}
    \end{subfigure}
    \begin{subfigure}[b]{0.325\textwidth}
        \includegraphics[width=\textwidth]{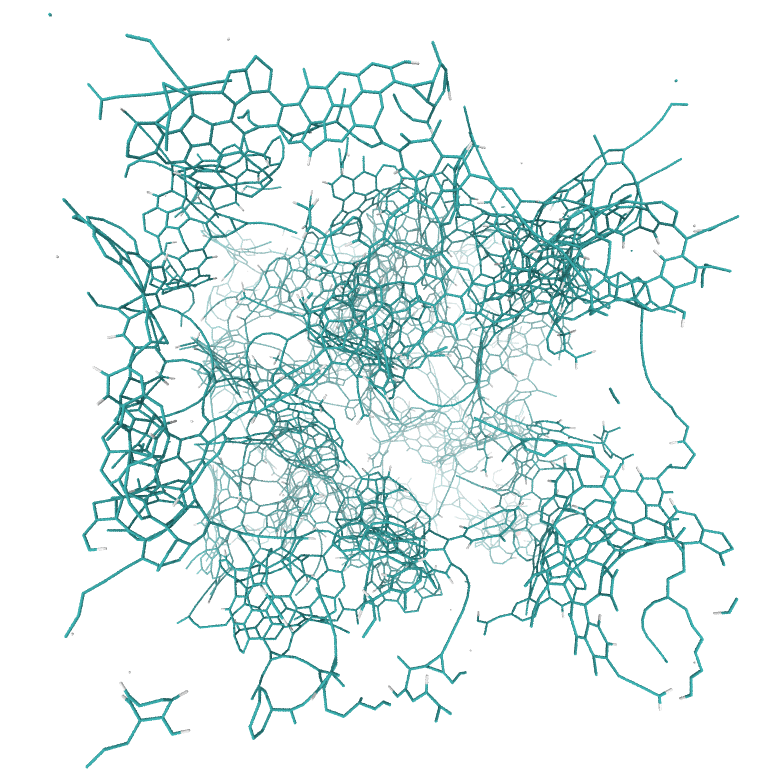}
        \caption{}\label{fig:H2O nf 357}
    \end{subfigure}
    \begin{subfigure}[b]{0.325\textwidth}
        \includegraphics[width=\textwidth]{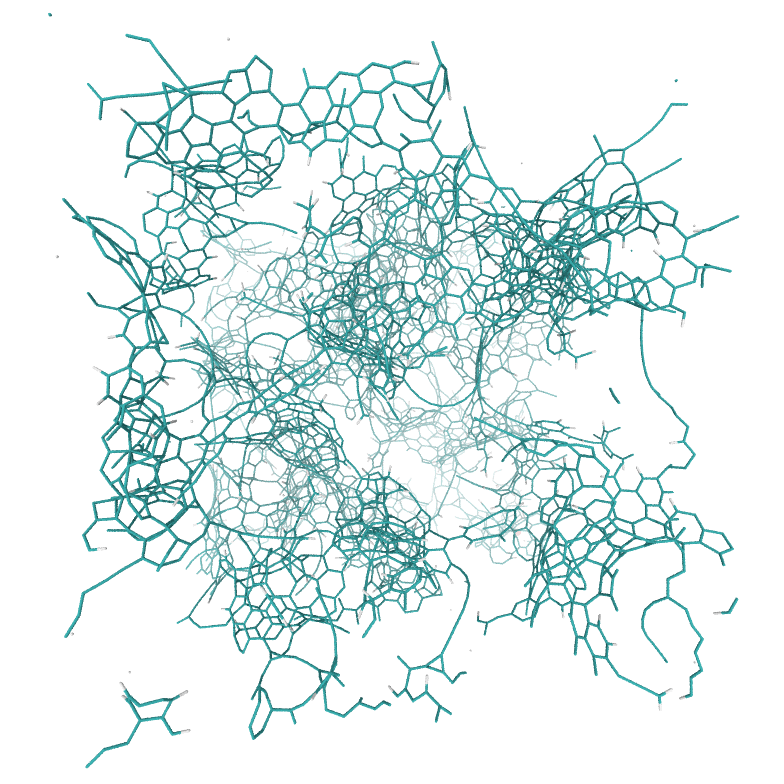}
        \caption{}\label{fig:H2O nf 1784}
    \end{subfigure}
    \begin{subfigure}[b]{0.325\textwidth}
        \includegraphics[width=\textwidth]{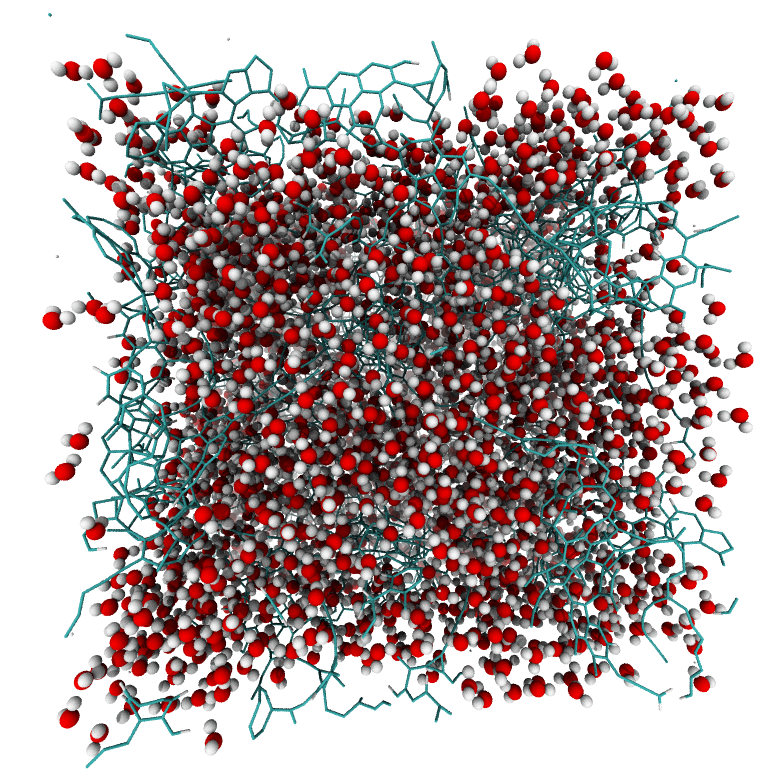}
        \caption{}\label{fig:H2O nf 316228}
    \end{subfigure}
    \caption{Snapshots showcasing adsorption in \nofunc at $300.15$ K under various partial pressure conditions. (a-c) Adsorption of \carbon at $316$, $10000$, and $316228$ Pa respectively. (e-f) Adsorption of \water at $357$, $1784$, and $316228$ Pa respectively. These conditions for water correspond to $10\%$ relative humidity, $50\%$ relative humidity, and saturation.}\label{fig:renders nofunc no K2CO3}
\end{figure*}

\begin{figure*}[!ht]
\centering
    \begin{subfigure}[b]{0.325\textwidth}
        \includegraphics[width=\textwidth]{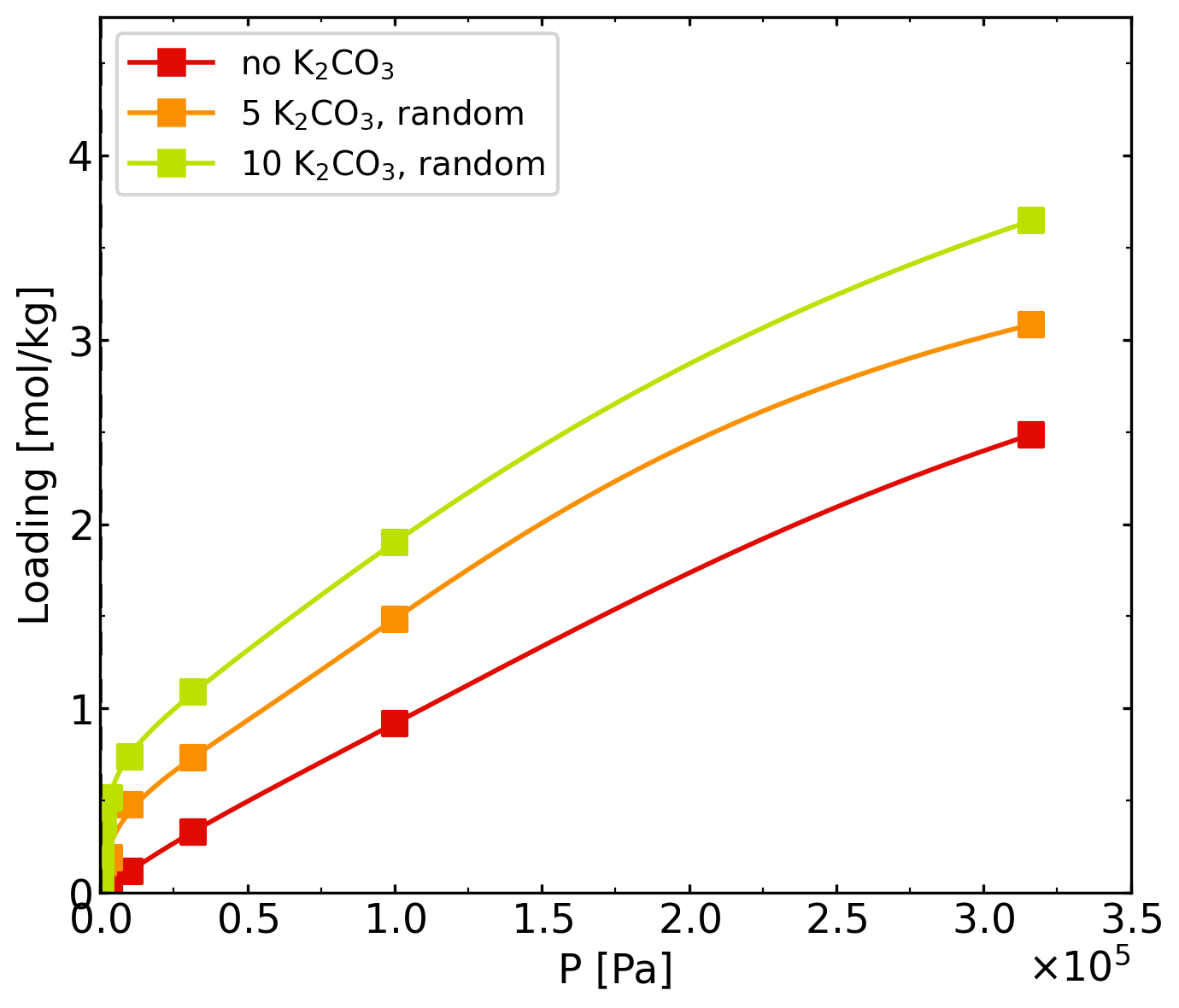}
        \caption{}\label{fig:CO2 nf random}
    \end{subfigure}
    \begin{subfigure}[b]{0.325\textwidth}
        \includegraphics[width=\textwidth]{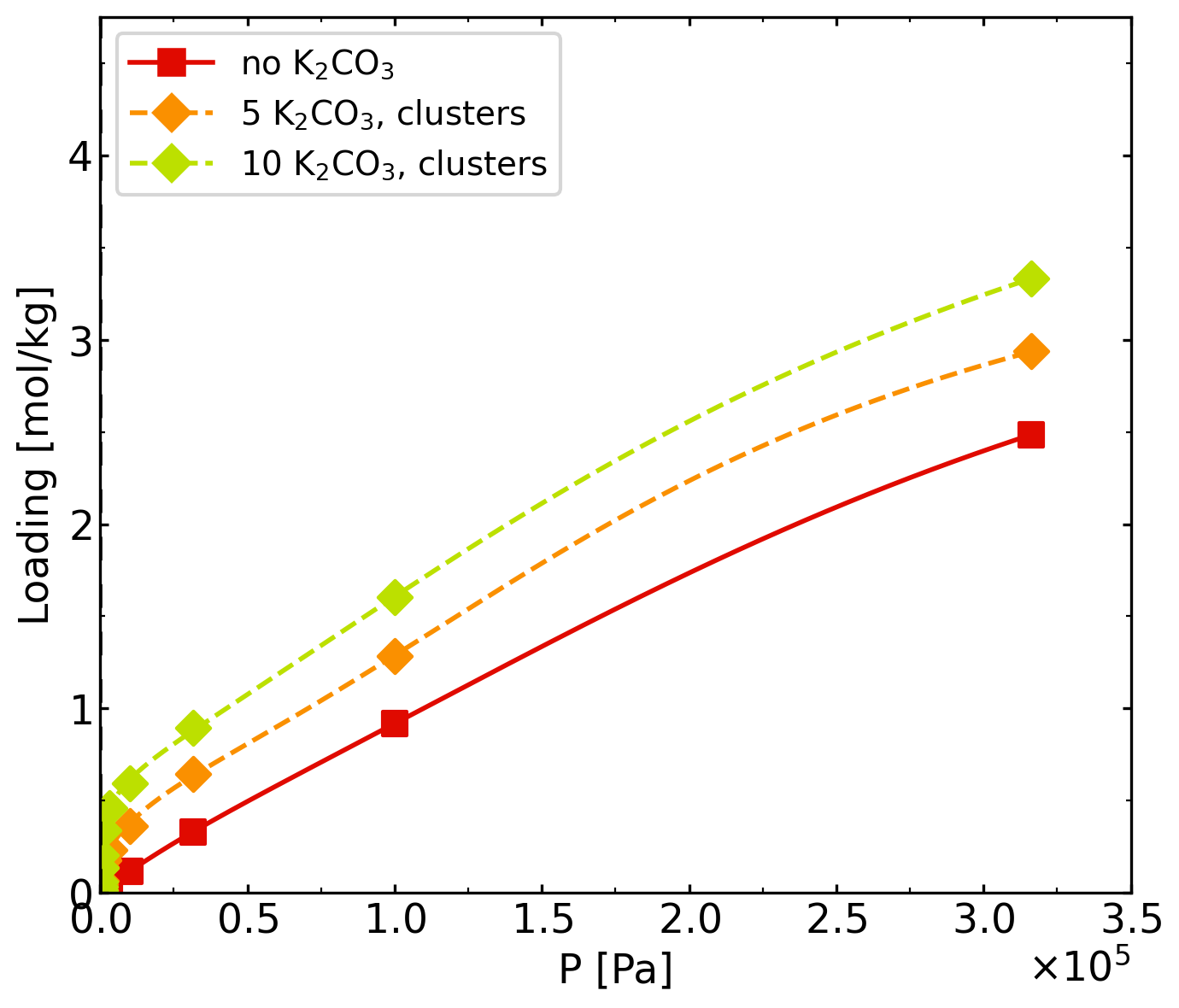}
        \caption{}\label{fig:CO2 nf clusters}
    \end{subfigure}
    \begin{subfigure}[b]{0.325\textwidth}
        \includegraphics[width=\textwidth]{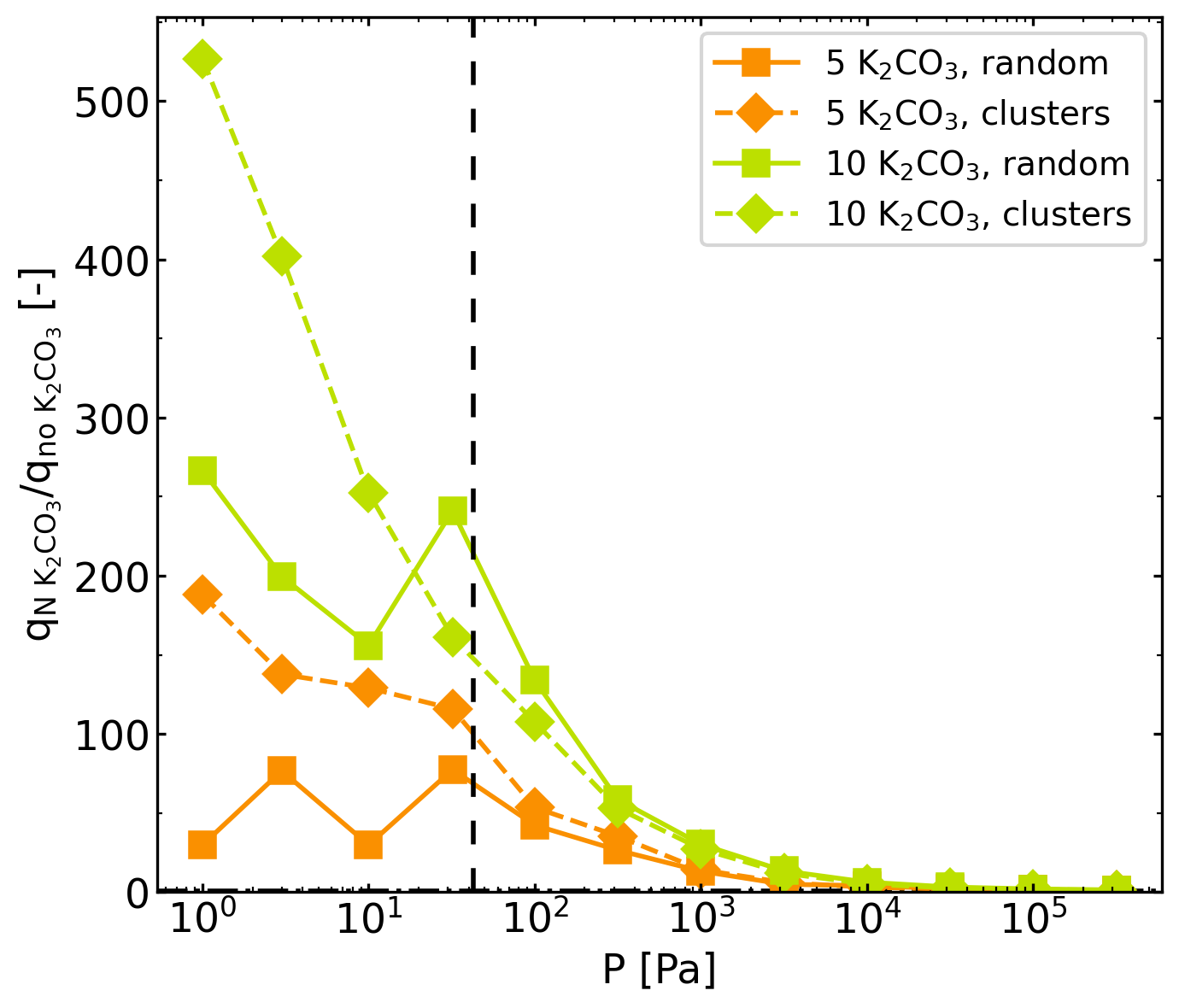}
        \caption{}\label{fig:CO2 nf 10 K2CO3 scaled}
    \end{subfigure}
    \begin{subfigure}[b]{0.325\textwidth}
        \includegraphics[width=\textwidth]{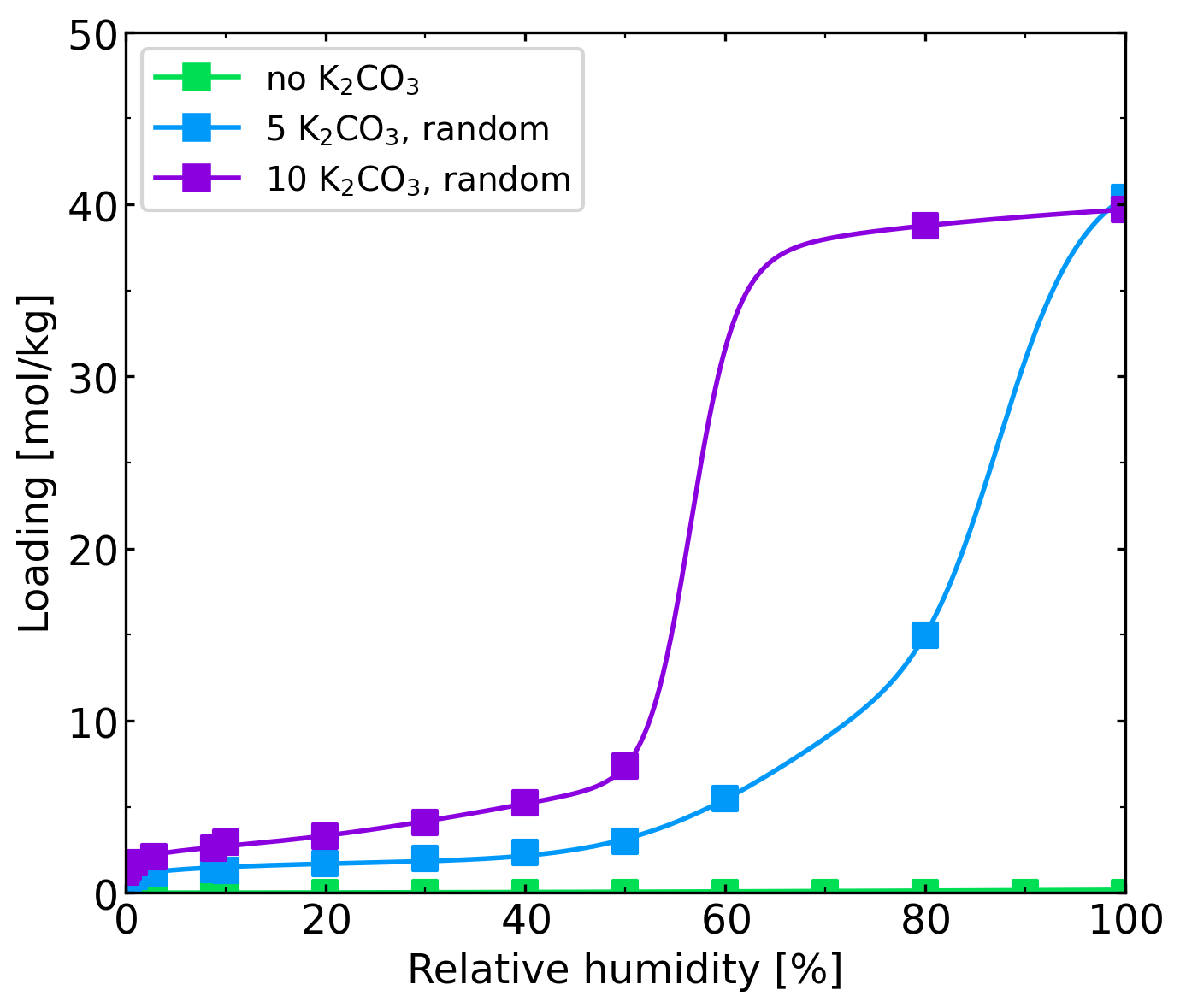}
        \caption{}\label{fig:H2O nf random}
    \end{subfigure}
    \begin{subfigure}[b]{0.325\textwidth}
        \includegraphics[width=\textwidth]{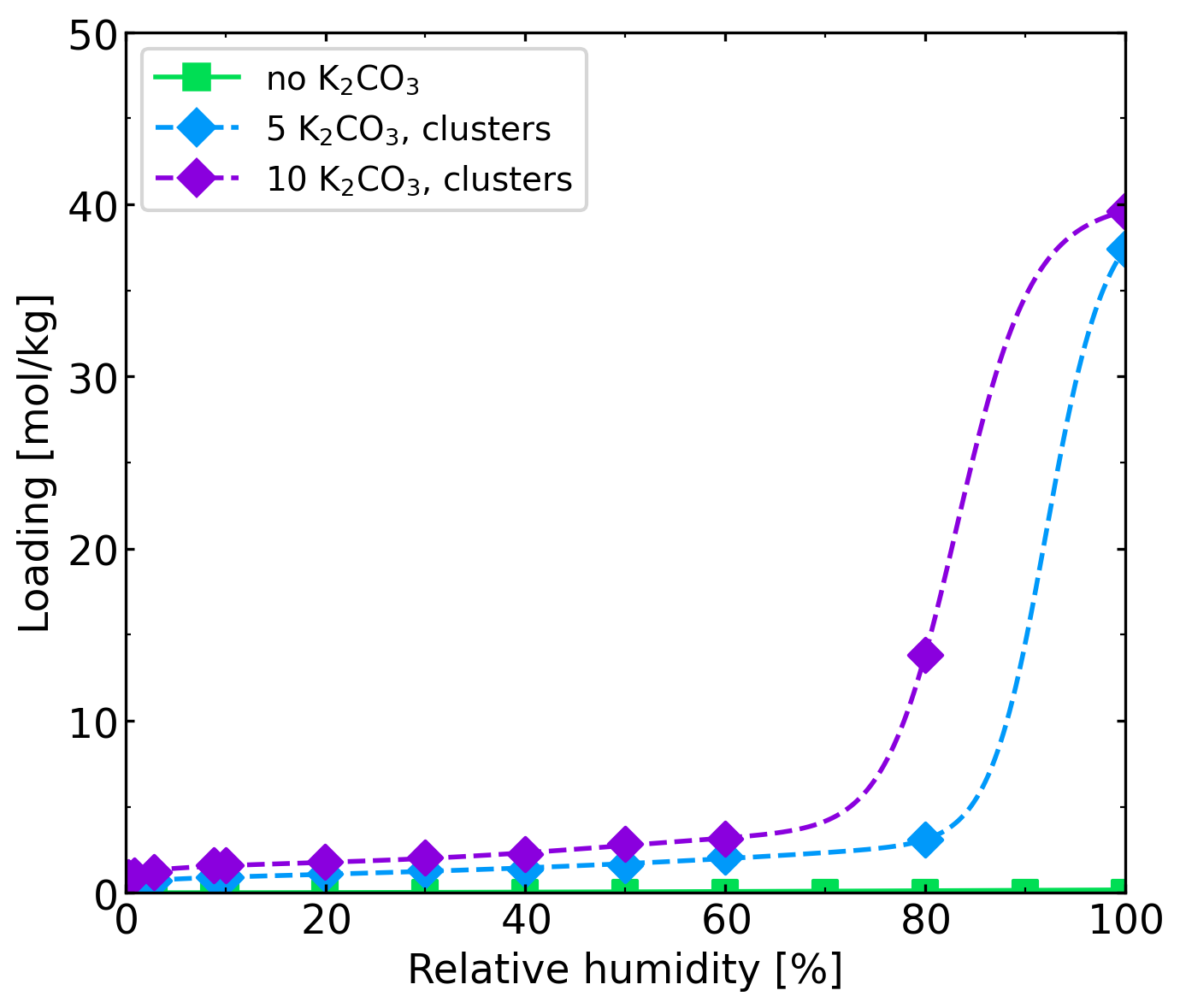}
        \caption{}\label{fig:H2O nf clusters}
    \end{subfigure}
    \begin{subfigure}[b]{0.325\textwidth}
        \includegraphics[width=\textwidth]{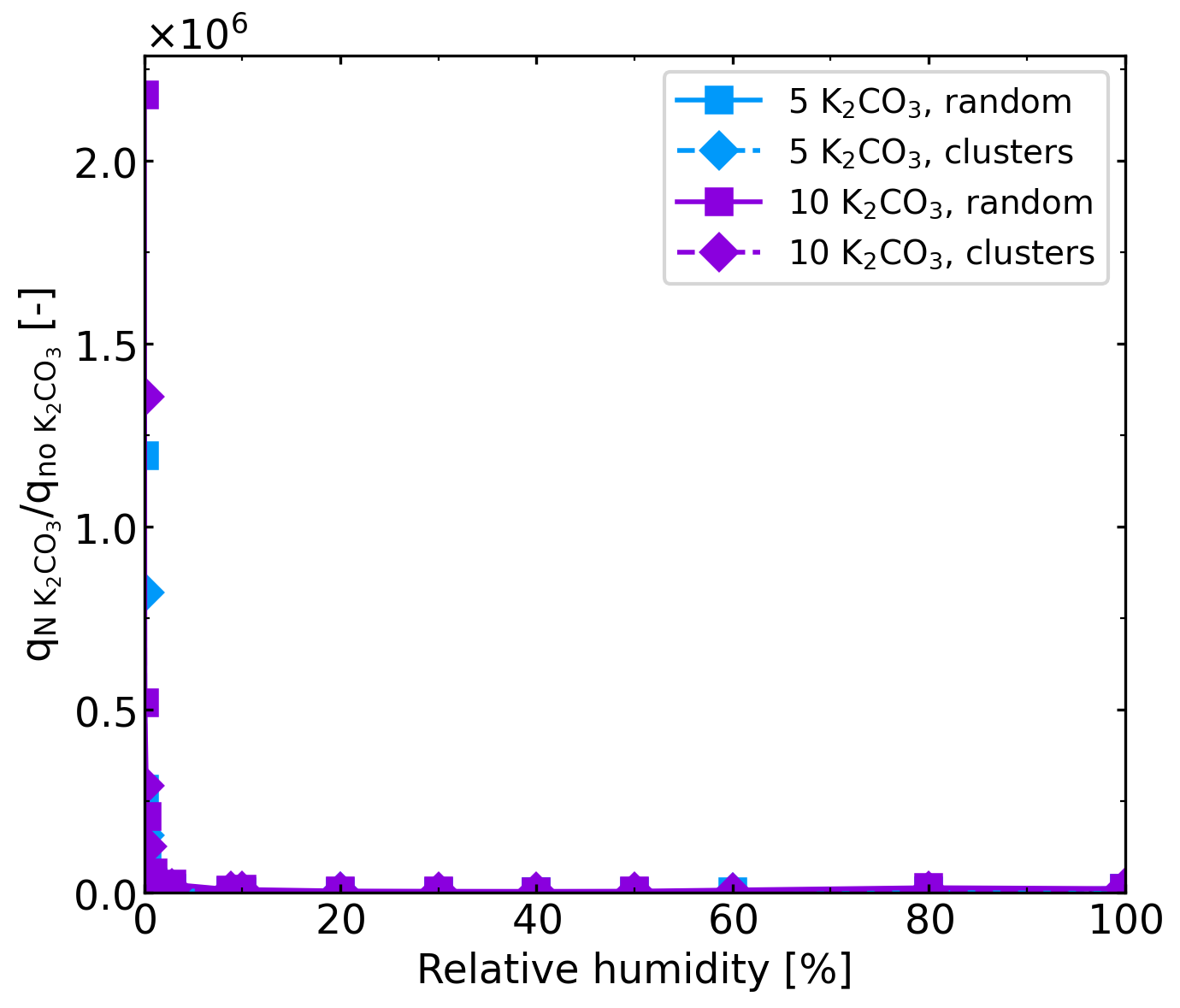}
        \caption{}\label{fig:H2O nf 10 K2CO3 scaled}
    \end{subfigure}
    \caption{Comparisons of computed adsorption isotherms of \carbon (a-c) and \water (d-f) in \nofunc at $300.15$ K for varying amounts of \ptc. Figures (a,d) show isotherms in which the \ptc was added randomly into the carbon, while (b,e) show isotherms in which the \ptc was added in clusters of $5$. Lines in these figures are fits through the data points obtained using RUPTURA.\cite{Sharma2023RUPTURA} In figures (c,f), isotherms using both methods of \ptc inclusion are shown, scaled by the isotherm without \ptc. The black vertical line in the (scaled) \carbon isotherms indicate the partial pressure equivalent to 420 ppm (at 1 atm), its current approximate atmospheric concentration.\cite{Lan2023}}\label{fig:iso nofunc K2CO3}
\end{figure*}

\begin{figure*}[!ht]
\centering
    \begin{subfigure}[b]{0.245\textwidth}
        \includegraphics[width=\textwidth]{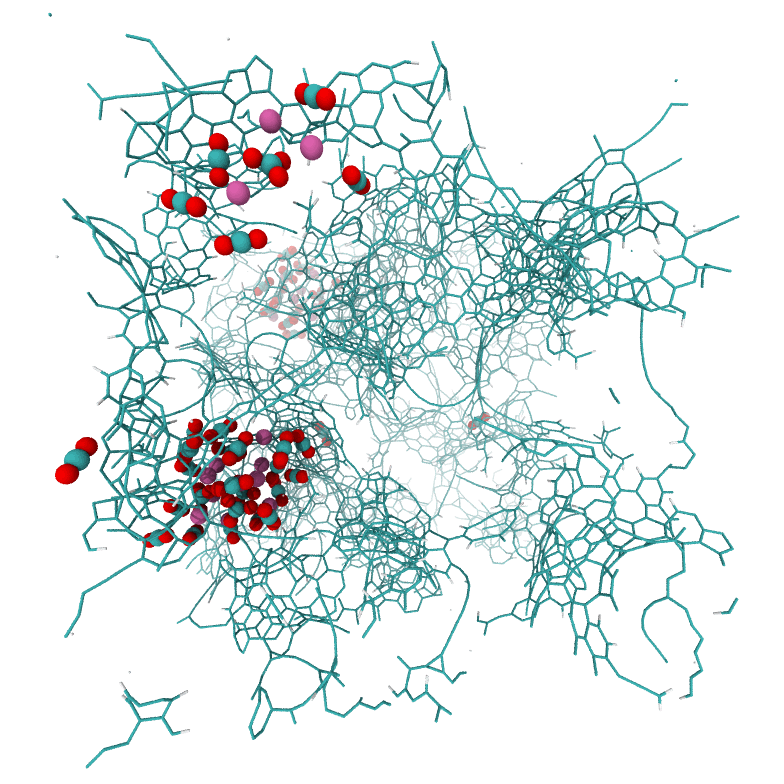}
        \caption{}\label{fig:CO2 nf 10000 cluster}
    \end{subfigure}
    \begin{subfigure}[b]{0.245\textwidth}
        \includegraphics[width=\textwidth]{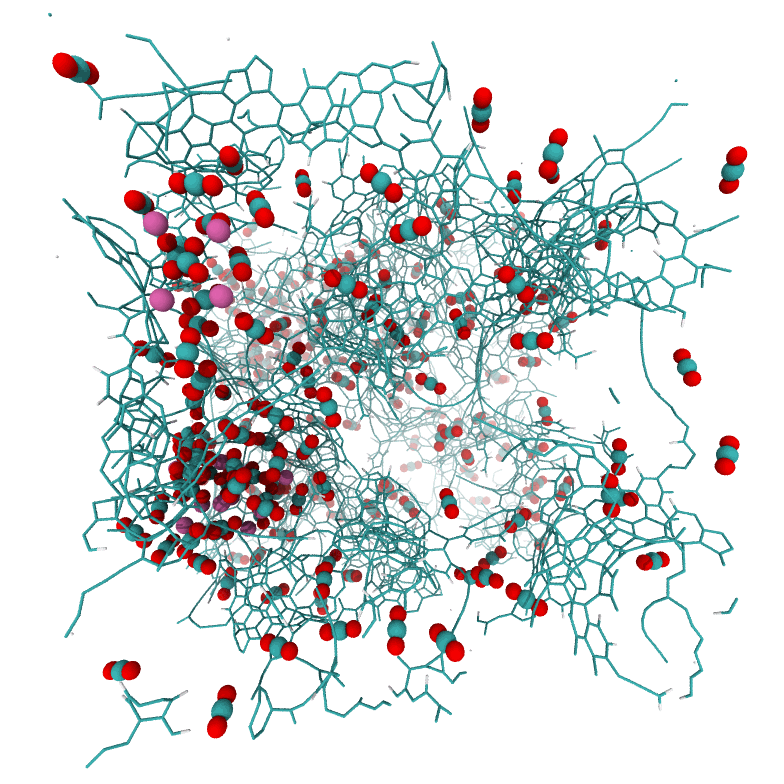}
        \caption{}\label{fig:CO2 nf 316228 cluster}
    \end{subfigure}
    \begin{subfigure}[b]{0.245\textwidth}
        \includegraphics[width=\textwidth]{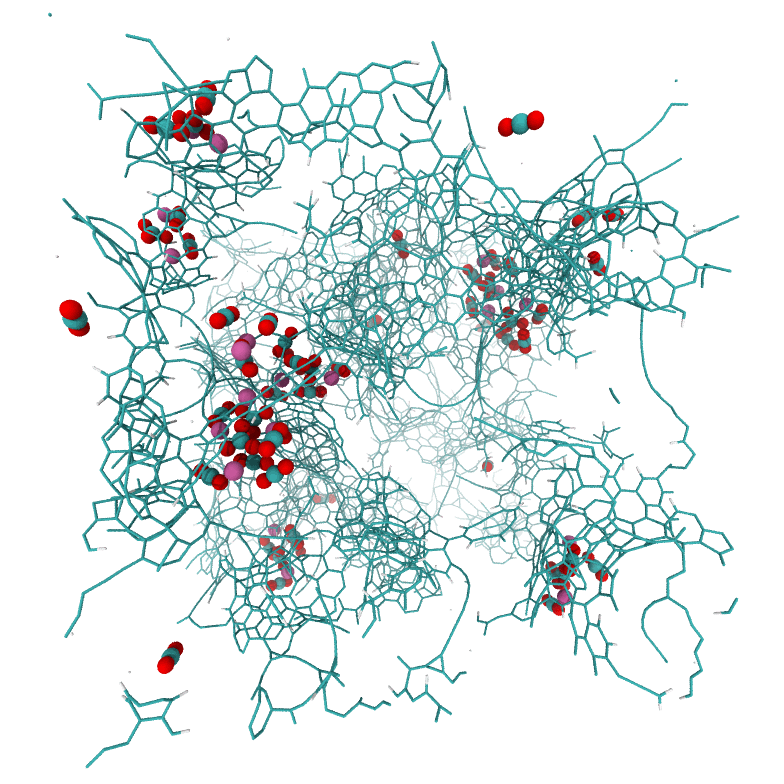}
        \caption{}\label{fig:CO2 nf 10000 random}
    \end{subfigure}
    \begin{subfigure}[b]{0.245\textwidth}
        \includegraphics[width=\textwidth]{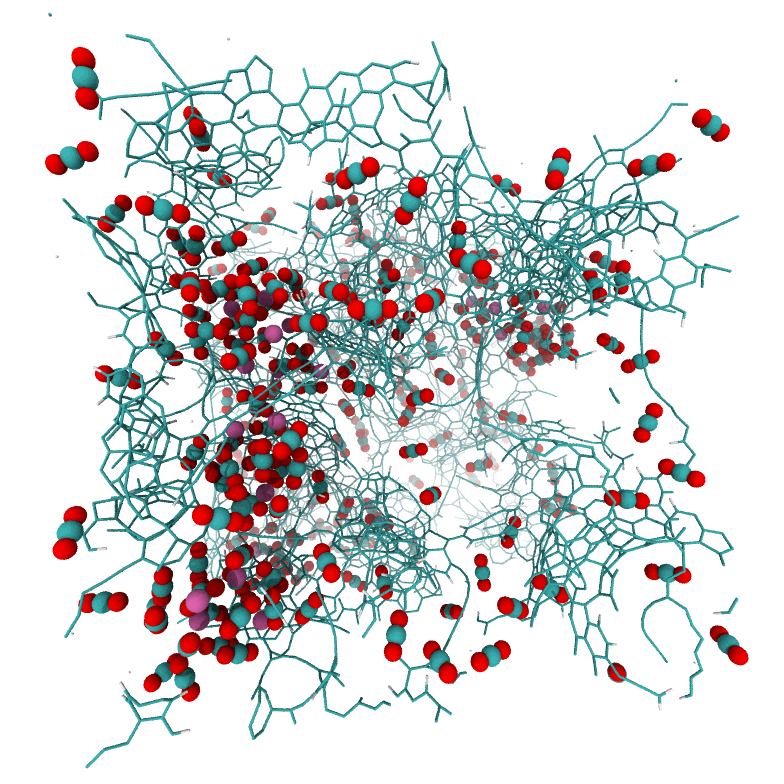}
        \caption{}\label{fig:CO2 nf 316228 random}
    \end{subfigure}
    \begin{subfigure}[b]{0.245\textwidth}
        \includegraphics[width=\textwidth]{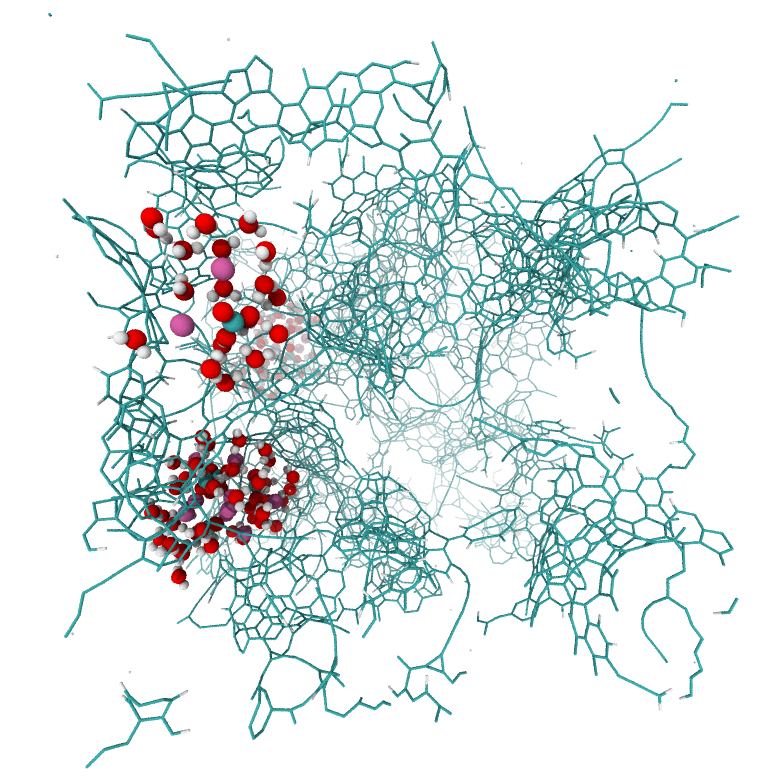}
        \caption{}\label{fig:H2O nf 357 cluster}
    \end{subfigure}
    \begin{subfigure}[b]{0.245\textwidth}
        \includegraphics[width=\textwidth]{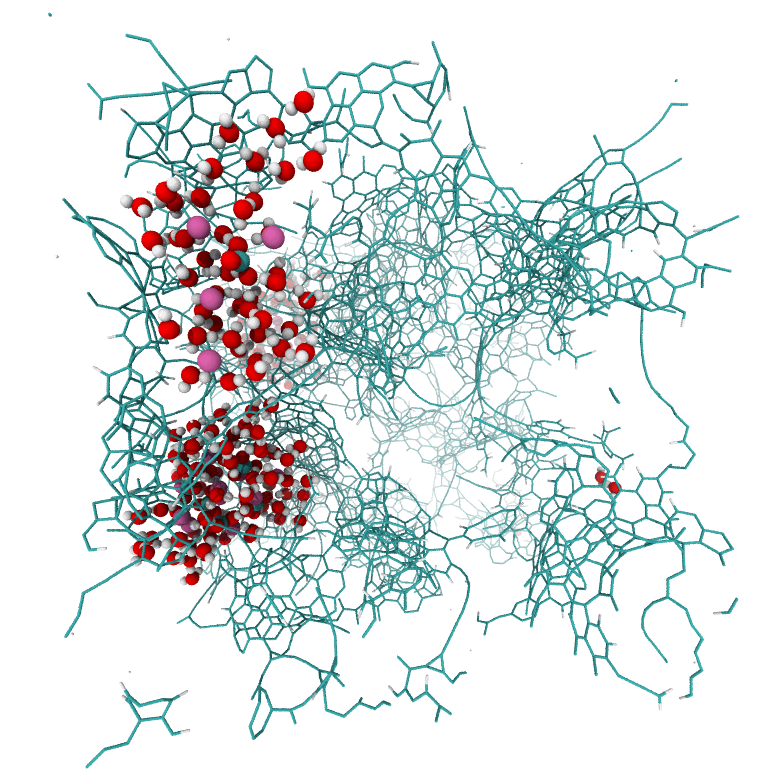}
        \caption{}\label{fig:H2O nf 1784 cluster}
    \end{subfigure}
    \begin{subfigure}[b]{0.245\textwidth}
        \includegraphics[width=\textwidth]{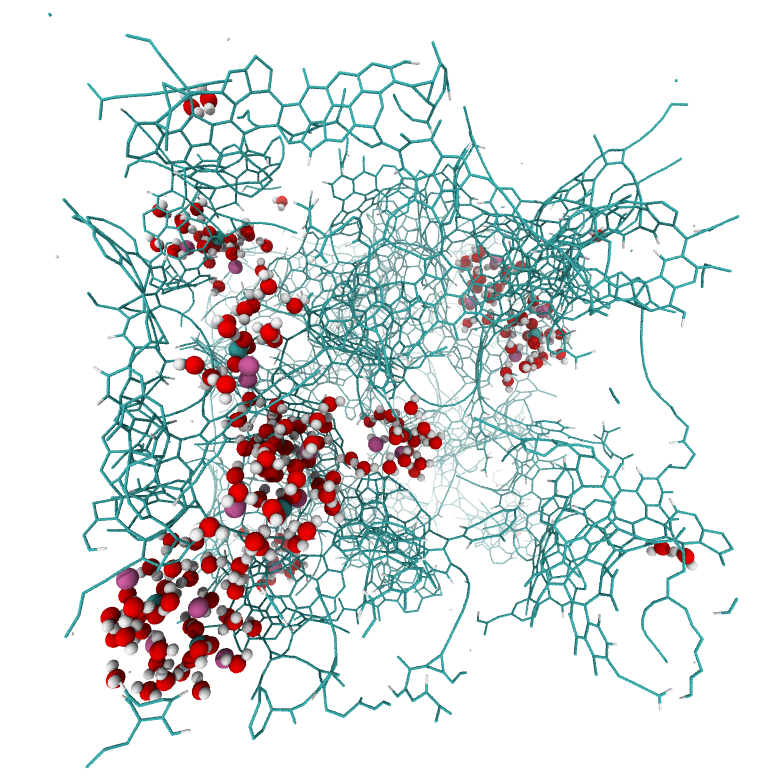}
        \caption{}\label{fig:H2O nf 357 random}
    \end{subfigure}
    \begin{subfigure}[b]{0.245\textwidth}
        \includegraphics[width=\textwidth]{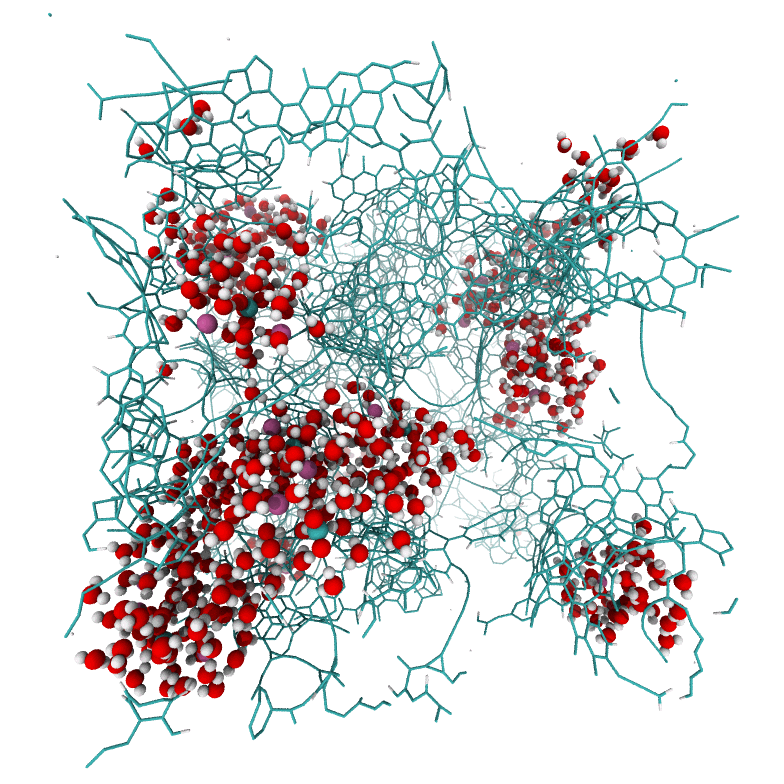}
        \caption{}\label{fig:H2O nf 1784 random}
    \end{subfigure}
    \caption{Snapshots showcasing adsorption of \carbon (a-d) and \water (e-h) in \nofunc at $300.15$ K with $10$ \ptc added in clusters (a,b,e,f) or randomly (c,d,g,h) under various partial pressure conditions. \carbon is shown at $10000$ Pa in (a,c) and at $316228$ Pa in (b,d). \water is shown at $357$ Pa ($10\%$ relative humidity) in (e,g) and at 1784 Pa ($50\%$ relative humidity) in (f,h).}\label{fig:renders nofunc 10 K2CO3}
\end{figure*}

\begin{figure*}[!ht]
\centering
    \begin{subfigure}[b]{0.245\textwidth}
        \includegraphics[width=\textwidth]{images/\rdfpth/rdf-f-CO2-C_co2_C_K2CO3.png}
        \caption{}\label{fig:rdf CO2 f C_CO2 - C}
    \end{subfigure}
    \begin{subfigure}[b]{0.245\textwidth}
        \includegraphics[width=\textwidth]{images/\rdfpth/rdf-f-CO2-C_co2_K_K2CO3.png}
        \caption{}\label{fig:rdf CO2 f C_CO2 - K}
    \end{subfigure}
    \begin{subfigure}[b]{0.245\textwidth}
        \includegraphics[width=\textwidth]{images/\rdfpth/rdf-f-CO2-O_co2_C_K2CO3.png}
        \caption{}\label{fig:rdf CO2 f O_CO2 - C}
    \end{subfigure}
    \begin{subfigure}[b]{0.245\textwidth}
        \includegraphics[width=\textwidth]{images/\rdfpth/rdf-f-CO2-O_co2_K_K2CO3.png}
        \caption{}\label{fig:rdf CO2 f O_CO2 - K}
    \end{subfigure}
    \begin{subfigure}[b]{0.245\textwidth}
        \includegraphics[width=\textwidth]{images/\rdfpth/rdf-f-H2O-Hw1_C_K2CO3.png}
        \caption{}\label{fig:rdf H2O f H_H2O - C}
    \end{subfigure}
    \begin{subfigure}[b]{0.245\textwidth}
        \includegraphics[width=\textwidth]{images/\rdfpth/rdf-f-H2O-Hw1_K_K2CO3.png}
        \caption{}\label{fig:rdf H2O f H_H2O - K}
    \end{subfigure}
    \begin{subfigure}[b]{0.245\textwidth}
        \includegraphics[width=\textwidth]{images/\rdfpth/rdf-f-H2O-Ow1_C_K2CO3.png}
        \caption{}\label{fig:rdf H2O f O_H2O - C}
    \end{subfigure}
    \begin{subfigure}[b]{0.245\textwidth}
        \includegraphics[width=\textwidth]{images/\rdfpth/rdf-f-H2O-Ow1_K_K2CO3.png}
        \caption{}\label{fig:rdf H2O f O_H2O - K}
    \end{subfigure}
    \caption{Computed radial distribution functions (RDFs) between \ptc and adsorbates in \func containing $2$ clusters of $5$ \ptc in the presence of \carbon (a-d) and water (e-h) at $300.15$ K. The RDFs compare the effects of increased pressure on the loading behavior. Under \carbon adsorption, RDFs are shown between $\text{\Ccarbon} - \text{\Cptc}$ (a), $\text{\Ccarbon} - \text{\Kptc}$ (b), $\text{\Ocarbon} - \text{\Cptc}$ (c), and $\text{\Ocarbon} - \text{\Kptc}$ (d). Under water adsorption, RDFs are shown between $\text{\Hwater} - \text{\Cptc}$ (e), $\text{\Hwater} - \text{\Kptc}$ (f), $\text{\Owater} - \text{\Cptc}$ (g), and $\text{\Owater} - \text{\Kptc}$ (h).}\label{fig:rdfs ptc adsorbates cont}
\end{figure*}

\begin{figure*}[!ht]
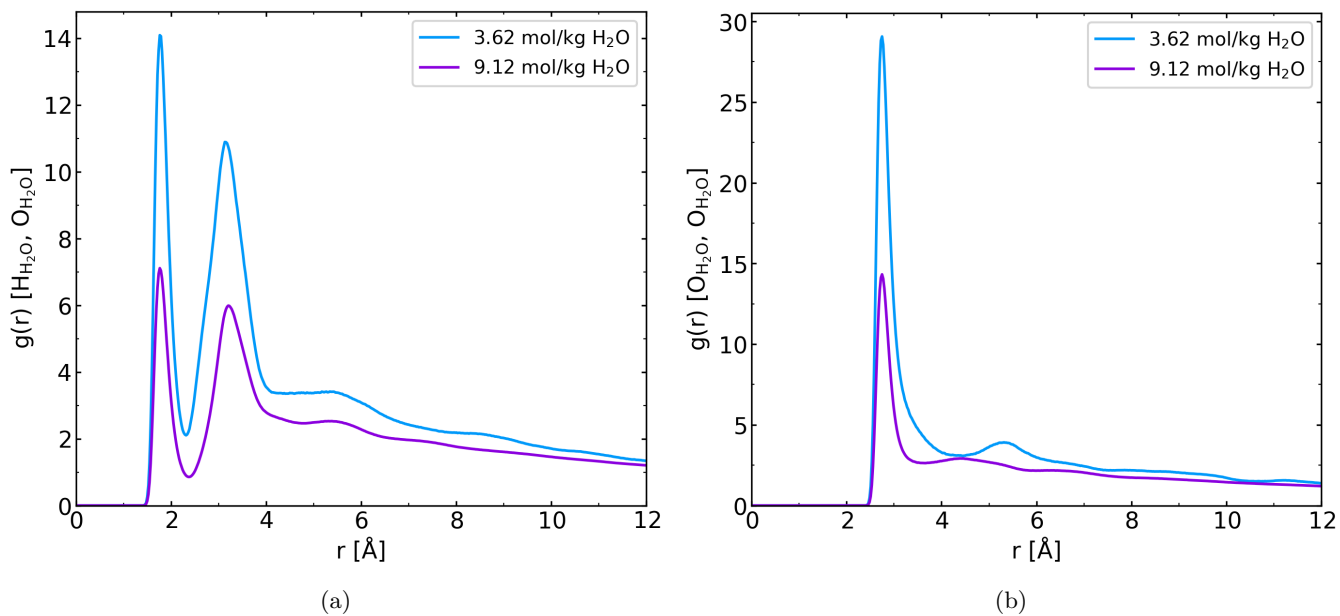

\centering
    \begin{subfigure}[b]{0.495\textwidth}
        \includegraphics[width=\textwidth]{images/\rdfpth/rdf-f-H2O-Hw1_Ow1.png}
        \caption{}\label{fig:rdf H H2O O H2O}
    \end{subfigure}
    \begin{subfigure}[b]{0.495\textwidth}
        \includegraphics[width=\textwidth]{images/\rdfpth/rdf-f-H2O-Ow1_Ow1.png}
        \caption{}\label{fig:rdf O H2O O H2O}
    \end{subfigure}
    \caption{Computed radial distribution functions (RDFs) between \water atoms in \func containing $2$ clusters of $5$ \ptc at $300.15$ K. These show hydrogen bonding between water molecules, evidencing the growth of water clusters through this mechanism.}\label{fig:H2O H2O mutual HB}
\end{figure*}

\end{document}